%% file: main.tex
\PassOptionsToPackage{comma,numbers,sort&compress,super}{natbib}

\documentclass[%
 onecolumn,  
superscriptaddress,
preprint,
nofootinbib,
 aps,
]{revtex4-2}

\usepackage{import}
\usepackage{usepackets/usepackets}
\usepackage{array, makecell} 
\renewcommand{\thetable}{\arabic{table}}
\usepackage{threeparttable}
\usepackage{booktabs,tabularx}
\usepackage{multirow}

\usepackage{adjustbox}

\usepackage{titletoc}
\dottedcontents{section}[2em]{\bfseries}{2.9em}{1pc}
\dottedcontents{subsection}[6em]{}{3.3em}{1pc}
\dottedcontents{subsubsection}[10em]{}{3.3em}{1pc}

\usepackage[articletitle=true, etalmode = truncate, maxauthors = 100, doi = true]{achemso}

\newcommand{\nocontentsline}[3]{}
\let\origcontentsline\addcontentsline
\newcommand\stoptoc{\let\addcontentsline\nocontentsline}
\newcommand\resumetoc{\let\addcontentsline\origcontentsline}

\newcommand{\revision}[1]{{\color{black}{#1}}}

\makeatletter
\def\p@subsection{}
\def\p@subsubsection{}
\makeatother

\begin{document}


\title{Hyperfine Coupling Constants on Quantum Computers: Performance, Errors, and Future Prospects}

\author{Phillip W. K. Jensen}
 \email{pwkj@chem.ku.dk}
\affiliation{Department of Chemistry, University of Copenhagen, Universitetsparken 5, DK-2100 Copenhagen \O, Denmark}

\author{Gustav Stausbøll Hedemark}
\affiliation{Department of Chemistry, University of Copenhagen, Universitetsparken 5, DK-2100 Copenhagen \O, Denmark}

\author{Karl Michael Ziems}
\affiliation{School of Chemistry, University of Southampton, 
SO17 1BJ Southampton, United Kingdom}
\affiliation{Department of Chemistry, Technical University of Denmark, Kemitorvet Building 207, DK-2800 Kongens Lyngby, Denmark.}

\author{Erik Rosendahl Kjellgren}
\affiliation{Department of Physics, Chemistry and Pharmacy,
University of Southern Denmark, Campusvej 55, 5230 Odense, Denmark. }

\author{Peter Reinholdt}
\affiliation{Department of Physics, Chemistry and Pharmacy,
University of Southern Denmark, Campusvej 55, 5230 Odense, Denmark. }

\author{Stefan Knecht}
\affiliation{Algorithmiq Ltd, Kanavakatu 3C, FI-00160 Helsinki, Finland}

\author{Sonia Coriani}
\affiliation{Department of Chemistry, Technical University of Denmark, Kemitorvet Building 207, DK-2800 Kongens Lyngby, Denmark.}

\author{Jacob Kongsted}
\affiliation{Department of Physics, Chemistry and Pharmacy,
University of Southern Denmark, Campusvej 55, 5230 Odense, Denmark. }

\author{Stephan P. A. Sauer}
 \email{sauer@chem.ku.dk}
\affiliation{Department of Chemistry, University of Copenhagen, Universitetsparken 5, DK-2100 Copenhagen \O, Denmark}

\date{\today}

\begin{abstract}

We present the first implementation and computation of electron spin resonance isotropic hyperfine coupling constants (HFCs) on quantum hardware. As illustrative test cases, we compute the HFCs for the hydroxyl radical (OH$^{\bullet}$), nitric oxide (NO$^{\bullet}$), and the triplet hydroxyl cation (OH$^{+}$). Our approach integrates the qubit-ADAPT method with unrestricted orbital optimization in an active space framework. To accurately measure the necessary spin one-electron reduced density matrices on current hardware, we employ a combination of error mitigation, error suppression, and post-selection, including our in-house developed ansatz-based readout and gate error mitigation. The HFCs obtained from the quantum hardware experiments align with results from unrestricted complete active space self-consistent field calculations on classical hardware. These results mark a significant step towards leveraging quantum computing for chemically relevant molecular properties and highlight the critical role of multi-method error strategies in the noisy intermediate-scale quantum era. 
%
%

\end{abstract}

\maketitle


\input{sections/introduction}

\input{sections/theory}

\input{sections/computational_details}
\input{sections/hardware_results}

\input{sections/conclusion}

\input{sections/acknowledgements}

\def\bibsection{}  
\section*{References}
\bigbreak

\bibliographystyle{achemso}
\bibliography{refs}

\input{sections/appendix}

\end{document}

%% file: sections/introduction.tex
\stoptoc

\section{Introduction}

The potential of quantum computers to simulate
molecules is promising for many applications in industry and research, notably within materials science and drug discovery~\cite{santagati_drug_2024}. One of the earliest demonstrations showcased the potential energy curve of molecular hydrogen using just two qubits~\cite{lanyon_towards_2010}. Since then, advancements in quantum hardware have enabled the simulation of increasingly large molecular systems. For example, case studies on quantum hardware include potential energy surfaces~\cite{kandala_hardware-efficient_2017, guo_experimental_2024} and absorption spectra~\cite{ziems_understanding_2025} of small molecules, the Diels-Alder reaction~\cite{liepuoniute_simulation_2024}, enzymatic reactions~\cite{ettenhuber_calculating_2024, ollitrault_estimation_2024}, Hartree-Fock and fermionic quantum Monte Carlo calculations~\cite{google_hartree_fock_2020, huggins_unbiasing_2022}. These case studies either involve small molecules in minimal basis sets, allowing the full space to be simulated on a quantum computer, or large molecules but in a small active space to be simulated on a quantum computer. Either way, these systems can easily be simulated on classical computers and, as such, do not provide new chemical insights. The main goal of today's quantum hardware simulations of molecular systems is, therefore, not to demonstrate quantum advantage but to explore the limits (in terms of qubits and gates) that current quantum computers can manage, as well as advance error mitigation techniques. The hope is that error-corrected quantum computers will eventually allow simulations of active spaces beyond the reach of classical methods, opening new frontiers in chemical discovery. Nevertheless, it is still an open question whether the variational quantum eigensolver (VQE)~\cite{peruzzo_variational_2014, mcclean_theory_2016} offers advantage over classical methods such as selected configuration interaction-based methods~\cite{holmes_heat-bath_2016,smith_cheap_2017,levine_casscf_2020}, density matrix renormalization group~\cite{wouters_density_2014, olivares-Amaya_ab-initio_2015, baiardi_density_2020}, \revision{quantum Monte Carlo methods~\cite{booth2009fermion,spencer2012sign,lee2022twenty}, and full coupled-cluster reduction approaches~\cite{xu2018full,xu2020towards},} which allow large active spaces to be simulated on classical computing architectures. 

VQE has become the primary approach for computing molecular properties on today's quantum computers. It can be executed on noisy quantum computers while still providing useful ground-state energies and electronic structures. However, due to the noise levels in current quantum hardware, VQE calculations often involve pre-optimized circuits and ideal variational parameters obtained through classical simulations, which are then executed on quantum hardware to evaluate the target properties. Since error-corrected quantum computers are not yet available, accumulated errors during computation restrict the depth of quantum circuits. Consequently, while current devices may support over 100 qubits, device noise often limits the practical use of all of them. One way to reduce the quantum workload is to utilize hybrid algorithms such as the orbital-optimized variational quantum eigensolver (oo-VQE)~\cite{sokolov_quantum_2020, mizukami_orbital_2020, bierman_improving_2023, fitzpatrick_self_consistent_2024}. This approach builds upon conventional VQE but restricts quantum computations to an active space that is within reach of current quantum hardware while treating the remaining space classically. By offloading a portion of the workload to classical computing, we can obtain properties on quantum hardware, even with small active spaces.

In this work, we present the first implementation and computation of electron spin resonance (ESR) isotropic hyperfine coupling constants (HFCs) on quantum hardware. ESR spectroscopy is a powerful tool for studying the geometric and electronic structures of free radicals and transition metal complexes~\cite{solomon_inorganic_2006}. ESR spectroscopy relies on the interaction between electron magnetic moments and nuclear spins. We compute HFCs, also known as the Fermi-contact interaction, for the hydroxyl radical (OH$^{\bullet}$), nitric oxide (NO$^{\bullet}$), and the triplet hydroxyl cation (OH$^{+}$). The implementation combines the oo-VQE method with unrestricted orbital optimization. For the radicals, we employ an active space of three electrons in three orbitals, and for the ion, the active space includes four electrons in four orbitals. We then provide a thorough proof-of-concept study on IBM's Torino device, measuring unrestricted one-electron reduced density matrix elements whose off-diagonal elements map to non-Ising Pauli strings. Applying a combination of  error mitigation, error suppression, and post-selection, we obtain results comparable to unrestricted complete active space self-consistent field (UCASSCF) calculations on classical hardware. Moreover, we provide a detailed analysis of our error strategies together with a direct comparison to the unmitigated result that showcases the necessity of employing such complex error management strategies in the noisy intermediate-scale quantum (NISQ) era. While the systems under investigation here are trivial to treat classically, our work serves as proof-of-concept and stepping stone toward more challenging systems. Hence, this work lays the foundation for calculating ESR spectroscopy properties using quantum hardware, with the goal of scaling these simulations to more complex systems in the future as the quantum hardware becomes more advanced.

The article is structured as follows: Section \ref{sec:theory} covers the theory behind HFC, followed by the unrestricted oo-qubit-ADAPT method. In Section \ref{sec:comp_details}, we summarize the computational details for the quantum hardware experiments, which include the optimization of the molecular spin-orbitals and the quantum circuits, and the error mitigation strategies. Finally, in Section \ref{section:hardware_results}, we present and discuss the spin one-electron reduced density matrix (1-RDM) and HFC results obtained from the quantum hardware experiments. We compare the data from our hardware experiments with unrestricted Hartree-Fock (UHF), UCASSCF and experimental data to assess the accuracy and reliability of the hardware results.

%% file: sections/theory.tex
\section{Theory}
\label{sec:theory}

In the following, we present the theoretical framework underlying this work. Section \ref{subsec:hyperfine} describes the theory of the HFC, a first-order property whose accuracy relies on the quality of the ground-state approximation. To treat molecular systems with large basis sets while limiting the quantum computation to a small space (few qubits), we employ the oo-VQE algorithm, detailed in Section \ref{sec:vqe_uscf}, to approximate ground states. Finally, in Section \ref{sec:ina_ac_ihfc}, we explain the splitting of the HFC into its inactive and active contributions, with the active contribution computed on the quantum hardware.

\subsection{Hyperfine interaction by first-order perturbation theory}
\label{subsec:hyperfine}

In perturbation theory, the  treatment of the magnetic interactions between the magnetic moments of the electron spin and the nuclear spin magnetic moments starts by decomposing the Hamiltonian into a zeroth-order term and a perturbation: 
\begin{align}
\hat{H} = \hat{H}^{(0)} + \hat{H}^{(\text{FC})}.  \label{eq:hamil}
\end{align}
The zeroth-order part $\hat{H}^{(0)}$ is the molecular spin- and field-free Born–Oppenheimer Hamiltonian, which in second quantized form is given by~\cite{helgaker_molecular_2000_ch1}:
\begin{align}
\hat{H}^{(0)} &=\sum^N_{PQ} h_{PQ} \hat{a}^{\dagger}_P \hat{a}_Q + \frac{1}{2}\sum^N_{PQRS} g_{PQRS}\hat{a}^{\dagger}_P  \hat{a}^{\dagger}_R  \hat{a}_S \hat{a}_Q  + h_{\text{nuc}},\label{eq:fermionic_H0}
\end{align}
where $\hat{a}_P$ $(\hat{a}^\dagger_P)$ are the annihilation (creation) operators and the indices $PQ,\hdots$ denote molecular spin-orbitals, $N$ is the number of spin-orbitals, and $h_{\text{nuc}}$ is the nuclear repulsion energy. The one- and two-electron integrals are given by
\begin{align}
h_{PQ}&= \int \phi^*_P \left(\vec{x}\right)\left( -\frac{\hbar^2}{2m_e} \hat{\vec{\nabla}}^2 - \frac{e^2}{4\pi \epsilon_0}\sum_K \frac{Z_K}{| \vec{R}_K - \vec{r}|} \right) ~\phi_Q\left(\vec{x}\right) \mathrm{d}\vec{x} \label{eq:h_PQ}\\[0.3cm]
g_{PQRS}&= \frac{e^2}{4\pi \epsilon_0}\int \frac{\phi^*_P\left(\vec{x}_1\right)  \phi^*_R \left(\vec{x}_2\right) \phi_Q\left(\vec{x}_1\right) \phi_S\left(\vec{x}_2\right)}{|\vec{r}_1 - \vec{r}_2|} \mathrm{d}\vec{x}_1 \mathrm{d}\vec{x}_2  \label{eq:g_PQRS}
\end{align}
with $e$ being the electron charge, $\epsilon_0$ the vacuum permittivity, and $m_e$ the electron mass. The integration over $\vec{x}$ signifies integration on both the spatial coordinates $\vec{r}$ and the spin coordinate of the electron. The set of functions $\{\phi_P\}$ represents an orthonormal molecular spin-orbital basis, i.e., $\braket{\phi_P|\phi_Q} = \delta_{PQ}$. The perturbation term $\hat{H}^{(\text{FC})}$ describes the interaction of the magnetic moments of the electron spins with the magnetic field generated from the nuclear magnetic spins inside the nucleus, known as the \emph{Fermi-contact} (FC) interaction. Another first-order contribution to the hyperfine interaction is the spin-dipolar interaction. However, this interaction is anisotropic and averages to zero for molecules in the gas or liquid phase and is therefore not considered~\cite{sauer_molecular_2011_ch5}. \revision{A second-order contribution involving spin-orbit coupling operators exists too. However, for atoms with nuclear charges smaller than 18 these are generally negligible and first-order perturbation theory is sufficient for accurate calculations}. The FC Hamiltonian in second quantization is given by 
\begin{align}
\hat{H}^{(\text{FC})} &= \sum_K f_K \sum^n_{pq} \left\{ A^{(K)}_{p\beta,q\alpha } \hat{T}^{1,-1}_{pq} \hat{I}^{(K)}_+ - A^{(K)}_{p\alpha, q\beta}  \hat{T}^{1,1}_{pq} \hat{I}^{(K)}_-  + \left( A^{(K)}_{p\alpha,q\alpha} \hat{a}^{\dagger}_{p\alpha} \hat{a}_{q\alpha} -  A^{(K)}_{p\beta,q\beta} \hat{a}^{\dagger}_{p\beta} \hat{a}_{q\beta} \right) \hat{I}^{(K)}_z \right\},\label{eq:hamil_fc}
\end{align}
where $\hat{T}^{1,-1}_{pq} = \hat{a}^\dagger_{p\beta} \hat{a}_{q \alpha}$ and $\hat{T}^{1,1}_{pq} = -\hat{a}^\dagger_{p\alpha} \hat{a}_{q \beta}$ are the usual spin-adapted triplet one-electron excitation operators that act on the electronic wave function, $\hat{I}^{(K)}_-$ and $ \hat{I}^{(K)}_+ $ are the nuclear spin step-down and step-up operators, respectively, while $\hat{I}^{(K)}_z$ is the nuclear spin-projection operator which all act on the nuclear spin wave function.
The amplitudes are given by $A^{(K)}_{p\sigma, q\tau} = \phi^*_{p\sigma}(\vec{R}_K) \phi_{q\tau}(\vec{R}_K)$ where $\phi_{p\sigma}(\vec{R}_K) $ is the amplitude of the molecular spin-orbital at the \emph{K}-th nucleus.
The indices $p,q, \hdots$ denote molecular spatial orbitals, and $\alpha$ and $\beta$ indicate spin orientation.
The prefactor $f_K$ accounts for the strength of the electron-nuclear magnetic interaction and is given as $f_K = \frac{g_K  g_e  e \mu_0\mu_N}{6m_e} $, where $g_K$ is the g-factor of the \emph{K}-th nucleus, $g_e$ is the g-factor of the electron ($g_e = 2.0023 \hdots$),  $\mu_0$ is the vacuum permeability, $\mu_N = \frac{e\hbar}{2m_p}$ is the nuclear magneton and $m_p$ is the proton mass. The 
Hamiltonian~\eqref{eq:hamil} can be derived from the elimination of the small component approach of the Dirac equation, referred to as the non-relativistic Schr\"odinger-Pauli Hamiltonian~\cite{sauer_molecular_2011_ch2}.
We refer to  Refs.~\cite{sauer_molecular_2011_ch5, kutzelnigg_origin_1988, miller_interpretation_2004} for a more detailed explanation of the derivation of the hyperfine interaction. 

The first-order energy is given by the expectation value of the FC Hamiltonian with respect to the ground-state of the zeroth-order Hamiltonian, $\ket{\Psi^{(0)}_0}$. Assuming the nuclear spins do not interact, the  wave function is given by $\ket{\Psi^{(0)}_0, \boldsymbol{M}^{(I)}} \equiv \ket{\Psi^{(0)}_0} \otimes \prod_K \ket{I^{(K)} M^{(K)}}$. Here, the nuclear spin wave function of the \emph{K}-th nucleus, $\ket{I^{(K)} M^{(K)}}$, is written in terms of its spin quantum numbers $ I^{(K)}$ and $M^{(K)}$, corresponding to the total and projected spin quantum numbers, respectively. The first-order energy evaluates to
\begin{align}
\left<\Psi^{(0)}_0, \boldsymbol{M}^{(I)}\right|  \hat{H}^{(\text{FC})}\left|\Psi^{(0)}_0, \boldsymbol{M}^{(I)} \right> &=  \sum_K f_K \hbar M^{(K)} \text{tr}\left[ \boldsymbol{A}^{(K)}_{\alpha} \boldsymbol{D}_{\alpha} - \boldsymbol{A}^{(K)}_{\beta} \boldsymbol{D}_{\beta}\right], \label{eq:fc_energy}
\end{align}
where the matrix $\boldsymbol{D}_{\sigma}$ is the spin-density 1-RDM with elements $D_{p\sigma,q \sigma} = \braket{\Psi^{(0)}_0| \hat{a}^\dagger_{p\sigma} \hat{a}_{q\sigma} |\Psi^{(0)}_0 }$, and the amplitude distribution matrix $\boldsymbol{A}^{(K)}_{\sigma} $ consists of the elements $ A^{(K)}_{p\sigma, q\sigma}$ explained in Eq. \eqref{eq:hamil_fc}. Note that the expectation value of the first two terms in \eqref{eq:hamil_fc} vanish from spin-symmetry of the electronic wave function since $\ket{\Psi^{(0)}_0}$ is an eigenstate of the spin-projection operator $\hat{S}_z$.

When we want to compare with experimental results, a new Hamiltonian is defined, known as the \emph{spin-Hamiltonian} (SH), which is given by~\cite{mcweeny_origin_1965, miller_interpretation_2004}
\begin{align}
\hat{H}_{\text{spin}} = \frac{2\pi}{\hbar} \sum_K \hat{\vec{S}} \boldsymbol{\alpha}^{(K)} \hat{\vec{I}}^{\hspace{0.05cm}(K)}, \label{eq:spin_hamil}
\end{align}
where $\boldsymbol{\alpha}^{(K)} $ is the hyperfine coupling matrix that couples the magnetic moment of electrons with total spin $\hat{\vec{S}}$ to the magnetic moments of nuclei with a nuclear spin $\hat{\vec{I}}^{\hspace{0.05cm}(K)}$.
The SH provides a simple way to parameterize experimental results without explicitly referencing the molecular geometry or electronic structure. Instead, it introduces free parameters $(\boldsymbol{\alpha}^{(K)} )$ that can be fitted to experimental data.
The SH parameters are extracted from experimental data by fitting procedures, which can be compared with theoretical results~\cite{miller_interpretation_2004}. 
Our focus is the isotropic FC interaction, which depends only on the relative positions of the nuclei and not on the molecular orientation relative to the external magnetic field. It is referred to as the \emph{isotropic hyperfine coupling constant} (HFC), denoted as $\boldsymbol{\alpha}^{(K)} = \alpha_{\text{iso}}^{(K)} \mathbb{1}$.

The working equation for the theoretical prediction of the HFC can be obtained by comparing  Eq.~\eqref{eq:fc_energy} with the expectation value of Eq.~\eqref{eq:spin_hamil}, which gives 
\begin{align}
\alpha_{\text{iso}}^{(K)} &=   \frac{f_K}{2\pi M}    \text{tr}
\left[ \boldsymbol{A}^{(K)}_{\alpha} \boldsymbol{D}_{\alpha} - \boldsymbol{A}^{(K)}_{\beta} \boldsymbol{D}_{\beta}
\right], \label{eq:iso_a}
\end{align}
where \emph{M} is the projected spin quantum number for the electronic wave function. The isotropic hyperfine coupling constant $\alpha_{\text{iso}}^{(K)}$ is usually expressed in units of MHz (frequency). Note that only the  \emph{z}-component term in Eq. \eqref{eq:spin_hamil}, i.e., $\hat{S}_z \hat{I}^{(K)}_z$, remains, as the electronic wave function is an eigenfunction of the $\hat{S}_z$ operator. 

\subsection{Orbital-optimized variational quantum eigensolver}
\label{sec:vqe_uscf}

We will now briefly explain the orbital-optimized variational quantum eigensolver (oo-VQE) algorithm, motivated by similar schemes in Refs.~\cite{sokolov_quantum_2020, mizukami_orbital_2020, bierman_improving_2023, fitzpatrick_self_consistent_2024}. oo-VQE is a multiconfigurational self-consistent field (MCSCF) approach~\cite{helgaker_molecular_2000_ch12}. We will use the unrestricted version of the oo-VQE algorithm 
to approximate the ground state of the zeroth-order Hamiltonian, $\ket{\Psi^{(0)}_0}$, as it includes the contributions from the spin polarization of the electrons in the inactive doubly-occupied orbitals to the HFC. A key advantage of using oo-VQE is its ability to treat molecular systems with large basis set expansions while restricting quantum computations to a small set of qubits to be simulated on a quantum computer, making the algorithm more NISQ friendly. \revision{Moreover, orbital optimization improves the quality of the approximate zeroth-order ground state. Since the HFC is computed as an expectation value with respect to the zeroth-order ground state, we expect that a better approximation of $\ket{\Psi^{(0)}_0}$ leads to more accurate HFCs.}


The first step in oo-VQE is to partition the molecular orbitals into inactive, active, and virtual spaces: 
\begin{align}
\left|\Psi^{(0)}_0 (\vec{\theta})\right> &= \ket{I}\otimes  \ket{A(\vec{\theta})}  \otimes  \ket{V}, \label{eq:wf}
\end{align}
where $\ket{I}$ denotes the inactive spin-orbitals and can be expanded as $\ket{I}=\ket{1}\otimes \ket{1} \otimes \cdots \ket{1}$, $\ket{V}$ denotes the virtual spin-orbitals, $\ket{V}=\ket{0}\otimes \ket{0} \otimes \cdots \ket{0}$, and the active part, $\ket{A(\vec{\theta})}$, is given as
\begin{align}
\left|A(\vec{\theta})\right> &= \hat{U}(\vec{\theta}) \left( \left|1\right>\otimes\left|1\right> \otimes \cdots \otimes \left|0\right>\otimes \left|0\right>\right) \label{eq:active_space_wf}
\end{align}
where $\vec{\theta}$ are the quantum circuit parameters. Unlike conventional MCSCF, where the active space wave function is usually parameterized linearly, we employ an exponential parametrization of the form
\begin{align}
\hat{U}(\vec{\theta}) &= \prod_i \mathrm{exp}\left(\text{i} \theta_i \hat{P}_i \right). \label{eq:ansatz_circuit}
\end{align}
Here, $\hat{P}_i $ are strings of Pauli operators, composed of Pauli-\emph{X} and -\emph{Y} operators, selected using the qubit-ADAPT method~\cite{tang_qubit_ADAPT_VQE_2021}. The size of the Pauli strings is either 2 for single excitations or 4 for double excitations, since the Pauli-\emph{Z} strings are removed from the operators in this framework. Compared to ansatzes such as factorized UCCSD~\cite{romero_strategies_2019} and fermionic-ADAPT~\cite{grimsley_adaptive_2019}, the qubit-ADAPT ansatz allows for relatively shallow circuit depths to converge toward the UCASSCF solution. \revision{However, the shallow circuit can come at the cost of potentially breaking symmetries such as spin and particle number. These symmetries can be restored by enforcing a strict convergence criterion in the optimization, although this increases the circuit depth.}

In oo-VQE, the minimization problem is both over the space of circuit parameters $\vec{\theta}$  and the space of orbital rotations $\vec{\kappa}$, 

\begin{align}
 E\left(\vec{\kappa}_{\text{opt}},\vec{\theta}_{\text{opt}} \right)  &=  \min_{\vec{\kappa},\vec{\theta}}\left<\Psi^{(0)}_0 (\vec{\theta})\right|  \mathrm{e}^{-\hat{\kappa}\left(\vec{\kappa} \right)} \hat{H}^{(0)} \mathrm{e}^{\hat{\kappa}\left(\vec{\kappa} \right)}\left|\Psi^{(0)}_0 (\vec{\theta})\right>  \nonumber \\[0.3cm]
 &= \min_{\vec{\kappa},\vec{\theta}}\sum_i c_i\left(\vec{\kappa}\right) \left<A(\vec{\theta})\right|  \hat{\mathcal{P}}_i \left|A(\vec{\theta})\right>.
 \label{eq:energy_function}
\end{align}
Here, $\vec{\kappa}_{\text{opt}}$ and $\vec{\theta}_{\text{opt}} $ denote the optimal parameters, $\hat{\mathcal{P}}_i$ are the Pauli strings from the mapping of the fermionic Hamiltonian to Pauli operators, $\hat{\kappa}$ is the generator of the orbital rotation operator, and the coefficients $c_i\left(\vec{\kappa}\right)$ are determined by the one- and two-electron integrals used to construct the fermionic Hamiltonian Eq. \eqref{eq:fermionic_H0}. The latter also depend on the coefficients from the mapping of fermionic to Pauli operators with additional sign corrections arising from the number of Pauli-\emph{Z} operators in the inactive space to obtain the proper sign. The minimization with respect to orbital and circuit parameters can be done on a classical device, and the expectation value in Eq.~\eqref{eq:energy_function} is computed on a quantum device. We use the unrestricted form of the generator of the orbital rotation operator, where the wave function is optimized with respect to both singlet and triplet orbital variations,  defined by~\cite{szabo_modern_1996_ch2, helgaker_molecular_2000_ch10}

\begin{align}
 \hat{\kappa}\left(\vec{\kappa} \right)=\sum_{p>q} \kappa^{(0,0)}_{pq} \left( \hat{E}_{pq} - \hat{E}_{qp}\right) + \sum_{p>q} \kappa^{(1,0)}_{pq} \left( \hat{T}^{1,0}_{pq} - \hat{T}^{1,0}_{qp}\right), \quad pq\in\{vi,ai,av\} \label{eq:unrestricted_oo}
\end{align}
where $\hat{E}_{pq} = \hat{a}^\dagger_{p\alpha} \hat{a}_{q \alpha} + \hat{a}^\dagger_{p\beta} \hat{a}_{q \beta}$ and $\hat{T}^{1,0}_{pq} = \frac{1}{\sqrt{2}} ( \hat{a}^\dagger_{p\alpha} \hat{a}_{q \alpha} - \hat{a}^\dagger_{p\beta} \hat{a}_{q \beta})  $ are the spin-adapted singlet and triplet one-electron excitation operators, respectively, and the vector $\vec{\kappa}$ consists of the set of orbital-rotation parameters $\{\kappa^{(0,0)}_{pq}\}$ and $\{\kappa^{(1,0)}_{pq}\}$. The index \emph{i} is used for occupied (inactive) orbitals, the index \emph{v} is used for the active orbitals, and the index \emph{a} is used for the unoccupied (virtual) orbitals. The orbital-rotation operator, $\mathrm{e}^{\hat{\kappa}\left(\vec{\kappa} \right)}$, therefore, acts between the spaces where $\hat{U}(\vec{\theta})$ acts within the active space. We include the triplet operators in the generator \eqref{eq:unrestricted_oo}  to account for spin polarization of the wave function to the HFC, as will be explained in the next section. The effect of the orbital-rotation operator is a rotation of the molecular spin-orbitals, which modifies the one- and two-electron integrals in Eqs.~\eqref{eq:h_PQ} and \eqref{eq:g_PQRS} according to 
\begin{align}
\tilde{h}_{P'Q'} &= \sum_{PQ}  \left[ \mathrm{e}^{\hat{\kappa}}\right]_{P'P} h_{PQ} \left[ \mathrm{e}^{-\hat{\kappa}}\right]_{Q'Q} \label{eq:new_h_PQ}\\[0.3cm]
\tilde{g}_{P'Q'R'S'} &= \sum_{PQRS}  \left[ \mathrm{e}^{\hat{\kappa}}\right]_{P'P} \left[ \mathrm{e}^{\hat{\kappa}}\right]_{Q'Q}  g_{PQRS} \left[ \mathrm{e}^{-\hat{\kappa}}\right]_{R'R}\left[ \mathrm{e}^{-\hat{\kappa}}\right]_{S'S},\label{eq:new_g_PQRS}
\end{align}
and $ \mathrm{e}^{-\hat{\kappa}\left(\vec{\kappa} \right)} \hat{H}^{(0)} \mathrm{e}^{\hat{\kappa}\left(\vec{\kappa} \right)} $ in Eq. \eqref{eq:energy_function} is expressed in the spin-orbital basis defined from $\mathrm{e}^{\hat{\kappa}\left(\vec{\kappa} \right)}$ by using the integrals in \eqref{eq:new_h_PQ} and \eqref{eq:new_g_PQRS}. 

The advantage of using the active space approximation is that we can remove the qubits corresponding to the inactive and virtual parts, which leaves only the active space to be simulated on a quantum computer. Note that in conventional VQE, the optimization is only over the space of circuit parameters~\cite{mcclean_theory_2016}. For example, ADAPT-based methods converge to the complete active-space configuration interaction (CASCI) solution, whereas oo-ADAPT-based methods converge to the CASSCF solution.

\subsection{Inactive and active space contributions to the HFC}
\label{sec:ina_ac_ihfc}

The HFC can be split into its inactive and active contributions
\begin{align}
\alpha^{(K)}_{\text{iso}} & = \left[\alpha^{(K)}_{\text{iso}}\right]_{I} + \left[\alpha^{(K)}_{\text{iso}}\right]_{A} \label{eq:HFC_split}
\end{align}
where 
\begin{align}
\left[\alpha^{(K)}_{\text{iso}}\right]_{I} &= \frac{f_K}{2\pi M} \text{tr}\left[  \big[\boldsymbol{A}^{(K)}_{\alpha} \big]_I - \big[\boldsymbol{A}^{(K)}_{\beta} \big]_I \right] \label{eq:HFC_inactive} \\[0.3cm]
\left[\alpha^{(K)}_{\text{iso}}\right]_{A} &= \frac{f_K}{2\pi M} \text{tr}\left[ \big[\boldsymbol{A}^{(K)}_{\alpha}\big]_A   \big[\boldsymbol{D}_{\alpha} \big]_{A} -\big[\boldsymbol{A}^{(K)}_{\beta}\big]_A   \big[\boldsymbol{D}_{\beta} \big]_{A}\right].\label{eq:HFC_active}
\end{align}
The subscripts \emph{I} and \emph{A} denote the inactive and active parts, respectively. Note that the virtual space does not contribute to the HFC because $[\boldsymbol{D}_{\sigma} ]_{V}=\boldsymbol{0}$. Equation \eqref{eq:HFC_inactive} shows that when using unrestricted orbitals there might be a contribution from the inactive space, which will not be the case using restricted orbitals because $ \big[\boldsymbol{A}^{(K)}_{\alpha} \big]_I =  \big[\boldsymbol{A}^{(K)}_{\beta} \big]_I$ when $\left\{\ket{\phi_{p\alpha}}\right\}=\left\{\ket{\phi_{p\beta}}\right\}$. While using a spin-unrestricted wave function introduces spin contamination, and thus can lead to inaccurate HFC calculations, the contribution from the inactive space, Eq. \eqref{eq:HFC_inactive}, may be important or even dominant to the HFC, which justifies the use of a spin-unrestricted wave function. An alternative approach is the restricted-unrestricted method proposed by Fernandez \emph{et al}.~\cite{fernandez_spin_1992}.

%% file: sections/computational_details.tex
\section{Computational details}
\label{sec:comp_details}

We calculate the HFCs via computing the spin-density 1-RDMs on the IBM Torino device on classically pre-optimized circuits. The \emph{SlowQuant} software~\cite{slowquant_2024} is used to interface with Qiskit~\cite{javadiabhari_quantum_2024} and IBM devices~\cite{ibm_quantum}, as well as to provide the error mitigation and post-processing tools. 
In terms of error mitigation strategies, we use ansatz-based readout and gate error mitigation and an RDM purification method. Each of these methods will be explained in this section. 

\subsection{Molecular systems}

The molecules we consider are hydroxyl radical (OH$^{\bullet}$), nitric oxide (NO$^{\bullet}$), and hydroxyl cation (OH$^{+}$) at their experimental bond lengths of 0.9697 $\text{\r{A}}$, 1.1508 $\text{\r{A}}$, and 1.0289 $\text{\r{A}}$, respectively.  The 6-311++G**-J basis set~\cite{kjaer_pople_2011}, which is  optimized for ESR properties, was adopted in all calculations. We use the experimental nuclear g-factors, $g_K$, for hydrogen, oxygen, and nitrogen from Ref. \cite{stone_table_2005}, and, in cases where experimental data is available, our results are compared with the experimental HFC values from Ref. \cite{kossmann_performance_2007}.  

The molecules OH$^{\bullet}$ and NO$^{\bullet}$ are free radicals with one unpaired electron with spin $S=\tfrac{1}{2}$, and we treat them with an active space of three electrons in three orbitals, corresponding to a 6-qubit system using the Jordan-Wigner mapping. The cation OH$^{+}$ has spin $S=1$, and is treated with an active space of four electrons in four orbitals, corresponding to an 8-qubit system.

\revision{The selection of active spaces was guided by two main considerations:  constraints on circuit depth (and thus number of qubits) on current noisy hardware and having some chemically foundation behind the choice of the active space. For the latter, our choice was motivated by the conventional molecular orbital diagrams. For OH$^{\bullet}$, the active space was chosen to include the $2p_x$ and $2p_y$ orbitals from oxygen and the antibonding $\sigma^*$ orbital. In the case of NO$^{\bullet}$, the expected orbitals are the bonding $\sigma$ orbital and the two antibonding $\pi^*$ orbitals. For OH$^{+}$, the active space was chosen to include the bonding $\sigma$ and antibonding  $\sigma^*$ orbitals, and the $2p_x$ and $2p_y$ orbitals from oxygen.}


\subsection{Circuits and spin-orbitals}
\label{subsec:class_process}

\begin{figure}[t]
\centering  
\includegraphics[width=1.0\textwidth]{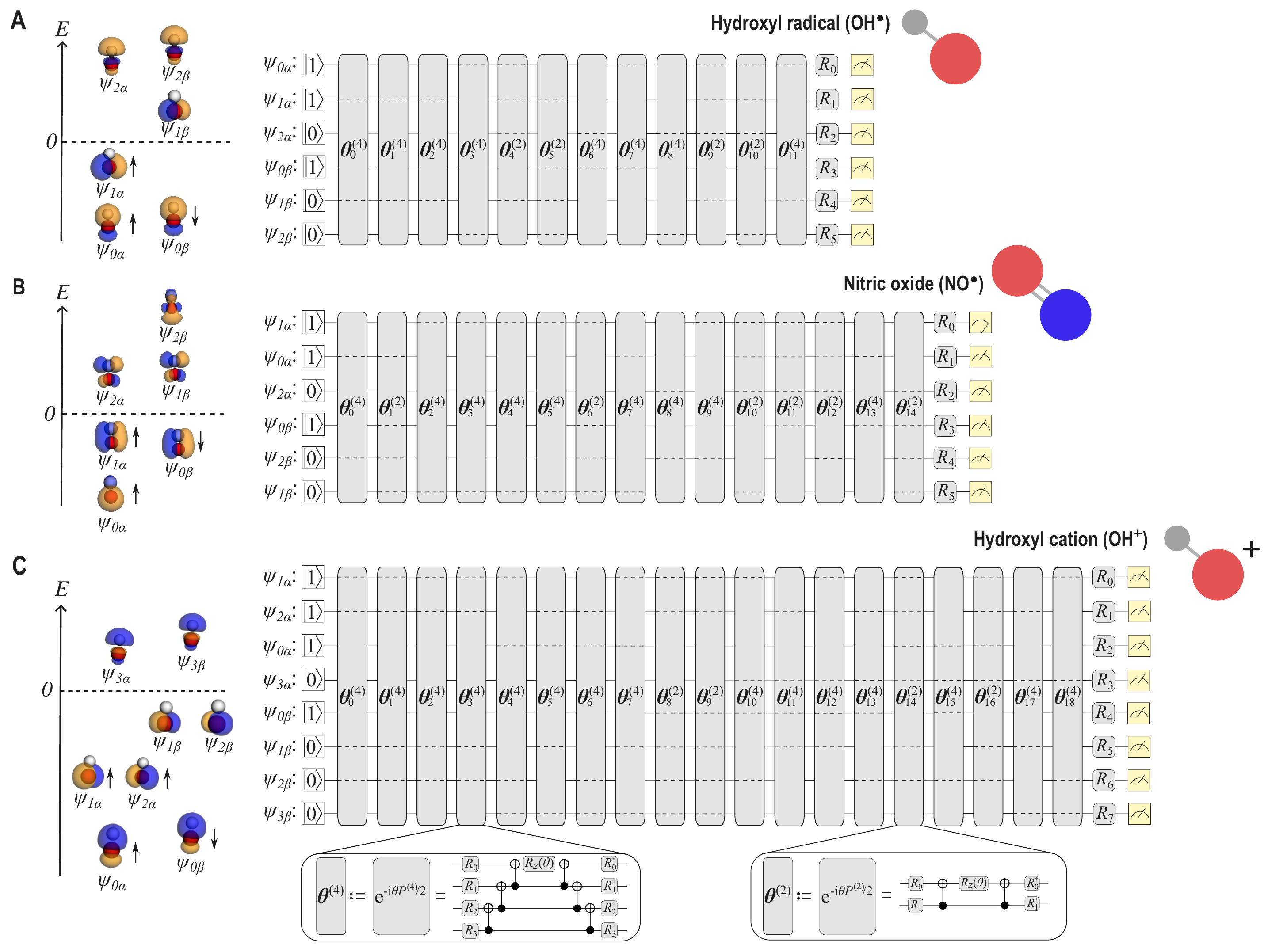}
\caption{Circuits and spin-orbitals for OH$^{\bullet}$ (A), NO$^{\bullet}$ (B), and OH$^{+}$ (C) to approximate the zeroth-order ground state $\ket{\Psi^{(0)}_0}$. The circuits are derived from the qubit-ADAPT method~\cite{tang_qubit_ADAPT_VQE_2021}. The basis rotation gates used are $R_i = \{H, HS^\dagger\}$. The circuits are executed on the IBM Torino device (Heron r1 chip, 133 qubits). The transpiled circuit for OH$^{\bullet}$ involves 77 CZ two-qubit gates, and 136 $\sqrt{X}$, 118 $R_z(\theta)$, 2 $X$ single-qubit gates with a total of 12 variational parameters. The transpiled circuit for NO$^{\bullet}$ involves 78 CZ two-qubit gates, and 127 $\sqrt{X}$, 131 $R_z(\theta)$, 7~$X$ single-qubit gates with a total of 15 variational parameters. The transpiled circuit for OH$^{+}$ involves 144 CZ two-qubit gates, and 242 $\sqrt{X}$, 182 $R_z(\theta)$, 4 $X$ single-qubit gates with a total of 19 variational parameters. }
\label{fig:main_circuits}
\end{figure}

The ansatz circuit in Eq.~\eqref{eq:ansatz_circuit} is obtained from the qubit-ADAPT approach~\cite{tang_qubit_ADAPT_VQE_2021}, combined with unrestricted orbital optimization in Eq.~\eqref{eq:unrestricted_oo} to include the contributions to the HFC from the spin polarization of the electrons in the doubly-occupied orbitals (inactive space). This method is referred to as oo-qubit-ADAPT(e,o), where ``(e,o)'' specifies the active space, with  ``e'' and  ``o'' representing the number of electrons and molecular orbitals, respectively, in the active space.

The oo-qubit-ADAPT(e,o) optimization was performed classically using the state-vector simulator of the  \emph{Aurora} software platform~\cite{aurora_2023}. The resulting optimized qubit-ADAPT circuits are shown in Figure \ref{fig:main_circuits}, and the optimized variational parameters can be found in the Supporting Information (SI) Section \ref{subsec:app:circuits}. The circuits in Figure \ref{fig:main_circuits} are then executed on IBM's Torino device to calculate the active space 1-RDMs, $[\boldsymbol{D}_{\sigma}]_{A}$, and from the 1-RDMs we can compute the HFCs. Ideally, the oo-qubit-ADAPT(e,o) ansatz would be optimized on a quantum computer, and the resulting 1-RDMs would be stored and reused in property calculations.  However, this was not done due to device noise, shot noise, and restricted QPU time on the quantum hardware. Details on the classically oo-qubit-ADAPT(e,o) optimization can be found in  Section \ref{app:sec:spin_orbitals} of the SI.  Figure \ref{fig:main_circuits} shows the classically optimized oo-qubit-ADAPT(3,3) spin-orbitals for OH$^{\bullet}$ and NO$^{\bullet}$ and oo-qubit-ADAPT(4,4) spin-orbitals for OH$^{+}$.

The amplitude distribution matrices, $\boldsymbol{A}^{(K)}_{\sigma}$, in Eqs. \eqref{eq:HFC_inactive} and \eqref{eq:HFC_active} are computed classically using the oo-qubit-ADAPT(e,o) spin-orbitals. These matrices can be found in the SI Section \ref{app:sec:spin_orbitals}. The error from the quantum hardware experiments, therefore, comes entirely from calculating the active space 1-RDMs which are obtained using quantum hardware. If $\tilde{\alpha}^{(K)}_{\text{iso}}$ represents the HFC obtained from a quantum hardware experiment, then the error becomes (SI Section \ref{app:error_analysis})
\begin{align}
\Delta\alpha^{(K)}_{\text{iso}} =\alpha^{(K)}_{\text{iso}}-\tilde{\alpha}^{(K)}_{\text{iso}} = \frac{f_K}{2\pi M} \sum^{n_A}_{vw}\left( A^{(K)}_{v\alpha, w\alpha}  \Delta D_{v\alpha, w\alpha} -A^{(K)}_{v\beta, w\beta} \Delta  D_{v\beta, w\beta}  \right), \label{eq:error_formula}
\end{align}
where the indices \emph{v} and \emph{w} denote the molecular spatial orbitals in the active space, and $\Delta D_{v\sigma, w\sigma}  =  D_{v\sigma, w\sigma} - \tilde{D}_{v\sigma, w\sigma}$, where $\tilde{D}_{v\sigma, w\sigma}$ is the estimate obtained from a quantum hardware experiment. Equation \eqref{eq:error_formula} can lead to both error cancellation and error amplification. For example, if the errors in the $\alpha$-spin $(\Delta D_{v\alpha, w\alpha})$ and the errors in the $\beta$-spin $(\Delta  D_{v\beta, w\beta})$ are both large, the total error may still be small if these errors cancel each other out. Another situation of error cancellation occurs when the corresponding amplitudes $ A^{(K)}_{v\sigma, w\sigma} $ are very small. In this case, a large error in $\Delta D_{v\sigma, w\sigma}$ would have minimal impact, as the vanishing amplitude $ A^{(K)}_{v\sigma, w\sigma} $ removes this error.  In the hardware results, we will clarify whether a small error in HFC is due to error cancellation or if the errors in the hardware experiments were indeed small.

\subsection{Operators}
\label{subsec:operator}

The 1-RDM number operators, $\hat{a}^\dagger_{p\sigma} \hat{a}_{p\sigma} \mapsto \frac{1}{2}(I-Z)_{p\sigma}$, qubit-wise commute with the Pauli-\emph{Z} string, allowing the diagonal elements of the 1-RDM to be obtained from a single hardware call, i.e., using the same probability distribution. Since the Pauli-\emph{Z} string does not change the electron number, post-selection is applied for the diagonal elements of the 1-RDM, that is on the spin $\alpha$ and $\beta$ part separately. The off-diagonal elements of the 1-RDM, which correspond to one-electron excitation operators, are measured sequentially on the quantum hardware. These operators contain Pauli-\emph{X} and Pauli-\emph{Y}  operators, and thus post-selection cannot be applied for the off-diagonal elements.

\subsection{Ansatz-based readout and gate error mitigation}
\label{subsec:M0}

We employ the error mitigation technique introduced by co-authors in Ref.~\cite{ziems_understanding_2025}. It builds upon the readout error mitigation (REM) method~\cite{maciejewski_mitigation_2020, bravyi_mitigating_2021} by constructing a confusion matrix,  $\boldsymbol{M}^{-1}_{U_0} $, that mitigates error as
\begin{align}
\vec{p}_{\text{mitigated}} = \boldsymbol{M}^{-1}_{U_0} \vec{p}_{\text{raw}}. \label{eq:Mrem_U}
\end{align}
Here, $\vec{p}_{\text{raw}}$ denotes the bit-string probability vector before error mitigation, and  $\vec{p}_{\text{mitigated}}$ denotes the quasi-probability vector after error mitigation. To encode for both readout errors and gate errors in the confusion matrix, Ref.~\cite{ziems_understanding_2025} proposed to incorporate the ansatz circuit (Eq. \eqref{eq:ansatz_circuit}) with $\hat{U}(\vec{\theta} = \vec{0})$ into the confusion matrix. The key part of this approach is that setting the circuit parameters to zero gives, in a noiseless case, $\hat{U}(\vec{\theta} = \vec{0}) = I$, but allows the accumulation of gate errors in a noisy setting that are subsequently removed by inverting the confusion matrix. \revision{Note that this error mitigation technique is not tied to a specific hardware architecture and only requires that the circuit evaluates to the identity when the parameters are set to zero.}

We use the fully correlated confusion matrix, which requires 64 circuits for the 6-qubit systems (OH$^{\bullet}$ and NO$^{\bullet}$) and 256 circuits for the 8-qubit system (OH$^{+}$). Note that the fully correlated confusion matrix scales exponentially with the size of the active space. For the quantum hardware calculations in this work, the majority of the QPU time was indeed spent on constructing the fully correlated confusion matrix, as described in Table \ref{tbl:overview_QPU_time}.

\begin{table}[t]
\begin{threeparttable}
\centering\renewcommand\cellalign{lc}
\setcellgapes{3pt}\makegapedcells
\caption{Overview of the tools used in the quantum hardware experiments. The \emph{SlowQuant} software~\cite{slowquant_2024} is used to interface with Qiskit~\cite{javadiabhari_quantum_2024} and IBM devices~\cite{ibm_quantum} and apply the tools outlined in the table.}
\centering\renewcommand\cellalign{lc}
\setcellgapes{3pt}\makegapedcells
\begin{tabularx}{\linewidth}{l|X}
\hline\hline  \\[-0.4cm]
\makecell[cl]{\multirow{2}{*}{Post-selection}} & Post-selection is on the spin $\alpha$ and $\beta$ part separately for the diagonal elements of the 1-RDM (Section \ref{subsec:operator}).     \\[0.2cm]
\makecell[cl]{\multirow{1}{*}{Error mitigation }}& Ansatz-based readout and gate error mitigation, $\boldsymbol{M}_{U_0}$~\cite{ziems_understanding_2025} (Section \ref{subsec:M0}).  \\[0.2cm]
\makecell[cl]{\multirow{2}{*}{Removal of 1-RDM}}& If any occupation number is outside the range $[-\epsilon,1+\epsilon]$ where $\epsilon\geq 0$, we discard the 1-RDM in the HFC calculations (Section \ref{subsec:removal_of_RDMs}).  \\[0.2cm]
\makecell[cl]{\multirow{1}{*}{RDM purification}} &  1-RDM is reconstructed to ensure trace condition tr$[\overline{\boldsymbol{D}}_\sigma] = N_\sigma$ (Section \ref{subsec:rdm_puri}).  \\ 
\makecell[cl]{\multirow{2}{*}{Error suppression}}&  Pauli-Twirling~\cite{dankert_exact_2009} to reduce the impact of coherent noise, and dynamical decoupling~\cite{viola_dynamical_1998, paolo_symmetrizing_1999, vitali_parity_1999, lu-ming_suppressing_1999, viola_dynamical_1999, ezzell_dynamical_2023} to reduce the effects of cross-talk.  \\ 
\bottomrule
\end{tabularx}
\label{tbl:overview_processing}
\end{threeparttable}
\end{table}

\begin{table}[t]
\begin{threeparttable}
\centering\renewcommand\cellalign{lc}
\setcellgapes{3pt}\makegapedcells
\caption{Overview of the number of measurements (shots) and quantum processing unit (QPU) times for computing the 1-RDMs, $[\boldsymbol{D}_{\alpha}]_{A}$ and $[\boldsymbol{D}_{\beta}]_{A}$,  on the IBM Torino device (Heron r1 chip, 133 qubits) using the circuits in Figure \ref{fig:main_circuits}.  }
\centering\renewcommand\cellalign{lcc}
\setcellgapes{3pt}\makegapedcells
\begin{tabularx}{\linewidth}{l|XXX}
\hline
Molecule & OH$^{\bullet}$  &  NO$^{\bullet}$   &OH$^{+}$   \\  \hline  \hline 
Shots per Pauli string & 56k shots &  56k shots &  16.3k shots \\ 
Total shots\tnote{\textcolor{blue}{$*$}} &  4.984M shots  & 4.984M shots &  4.9715M shots  \\ 
\makecell[cl]{QPU time for $\boldsymbol{M}_{U_0}$} & $\sim$17 minutes  & $\sim$17 minutes  & $\sim$25 minutes   \\ 
\makecell[cl]{Total QPU time\tnote{\textcolor{blue}{$*$}}} & $\sim$24 minutes  &  $\sim$24 minutes  &  $\sim$30 minutes  \\ 
\bottomrule
\end{tabularx}
\begin{tablenotes}\footnotesize
\item[\textcolor{blue}{$*$}] Include the construction of the $\boldsymbol{M}_{U_0}$ matrix.
\end{tablenotes}
 \label{tbl:overview_QPU_time}
\end{threeparttable}
\end{table}

\subsection{Removal of unphysical 1-RDMs}
\label{subsec:removal_of_RDMs}

For all hardware results, we remove 1-RDMs that are unphysical when computing the HFCs.  If any occupation number (eigenvalues of the 1-RDMs) is outside the range $[-\epsilon,1+\epsilon]$ where $\epsilon\geq 0$, we discard the 1-RDM. Ideally, we set $\epsilon = 0$; however, when the occupation number is close to either zero or one, statistical errors in the measurements and device noise may move the occupation number slightly outside the range [0,1]. The 1-RDM sample is only removed if there is a significant error, as defined by $\epsilon$. In this work, we set $\epsilon=0.05$.

\subsection{RDM purification}
\label{subsec:rdm_puri}

On the quantum hardware, we prepare the circuits shown in Figure \ref{fig:main_circuits} and calculate the 1-RDM elements as

\begin{align}
D_{v\sigma, w\sigma}=\left< A (\vec{\theta}_{\text{opt}}) \right|\hat{a}^\dagger_{v\sigma}\hat{a}_{w\sigma} \left|A (\vec{\theta}_{\text{opt}})\right> \label{eq:D_psqs_hardware}
\end{align}
where $\ket{A (\vec{\theta}_{\text{opt}})}$, defined in Eq.~\eqref{eq:active_space_wf}, is prepared using the qubit-ADAPT circuits in Figure \ref{fig:main_circuits}, and the fermionic excitation operators are mapped into qubit operators via the Jordan-Wigner transformation. The transpiled circuits are shown in the SI Section \ref{subsec:app:circuits}. Once the 1-RDMs are computed, they are reconstructed as: $\overline{\boldsymbol{D}}_\sigma = \boldsymbol{V}_\sigma \overline{\boldsymbol{\lambda}}_\sigma \boldsymbol{V}_\sigma^\text{T}$ where $\overline{\boldsymbol{\lambda}}_\sigma = \left(N_{\sigma}/\sum_i \lambda_i \right)\boldsymbol{\lambda}_\sigma$ are the scaled eigenvalues with $N_{\sigma}$ being the number of $\sigma\in\{\alpha,\beta\}$ electrons, and  $\boldsymbol{V}_\sigma$ are the eigenstates of the original spin-density 1-RDM. This ensures that the trace condition tr$[\overline{\boldsymbol{D}}_\sigma] = N_\sigma$ is fulfilled. The impact of RDM purification on the HFC is a multiplication of $N_{\sigma}/\sum_i \lambda_i $ on each corresponding (spin) term in the active space contribution to the HFC, Eq. \eqref{eq:HFC_active}. Other works have also performed RDM purification~\cite{smart_quantum-classical_2019, tilly_reduced_2021, deGraciaTrivio_complete_2023,Skogh_the_2024}. Note that the RDM purification method will not impact results obtained with post-selection, as post-selection already ensures the correct number of electrons when estimating the diagonal elements of the 1-RDMs. Therefore, we will evaluate the performance of the RDM-purification method without post-selection. 


%% file: sections/hardware_results.tex
\section{Hardware results}
\label{section:hardware_results}

The HFCs are computed on the IBM Torino device, applying the tools outlined in Table \ref{tbl:overview_processing}. We compare the performance of error mitigation, post-selection, and error suppression against results obtained without error mitigation and post-selection but using 1-RDM purification. In Section \ref{subsec:rdm}, we present the 1-RDM quantum hardware results, and in Section \ref{subsec:ihfc}, we show the HFCs results. 

\subsection{Reduced density matrices}
\label{subsec:rdm}

\begin{figure}[t]
    \centering
   \includegraphics[width=1.0\textwidth]{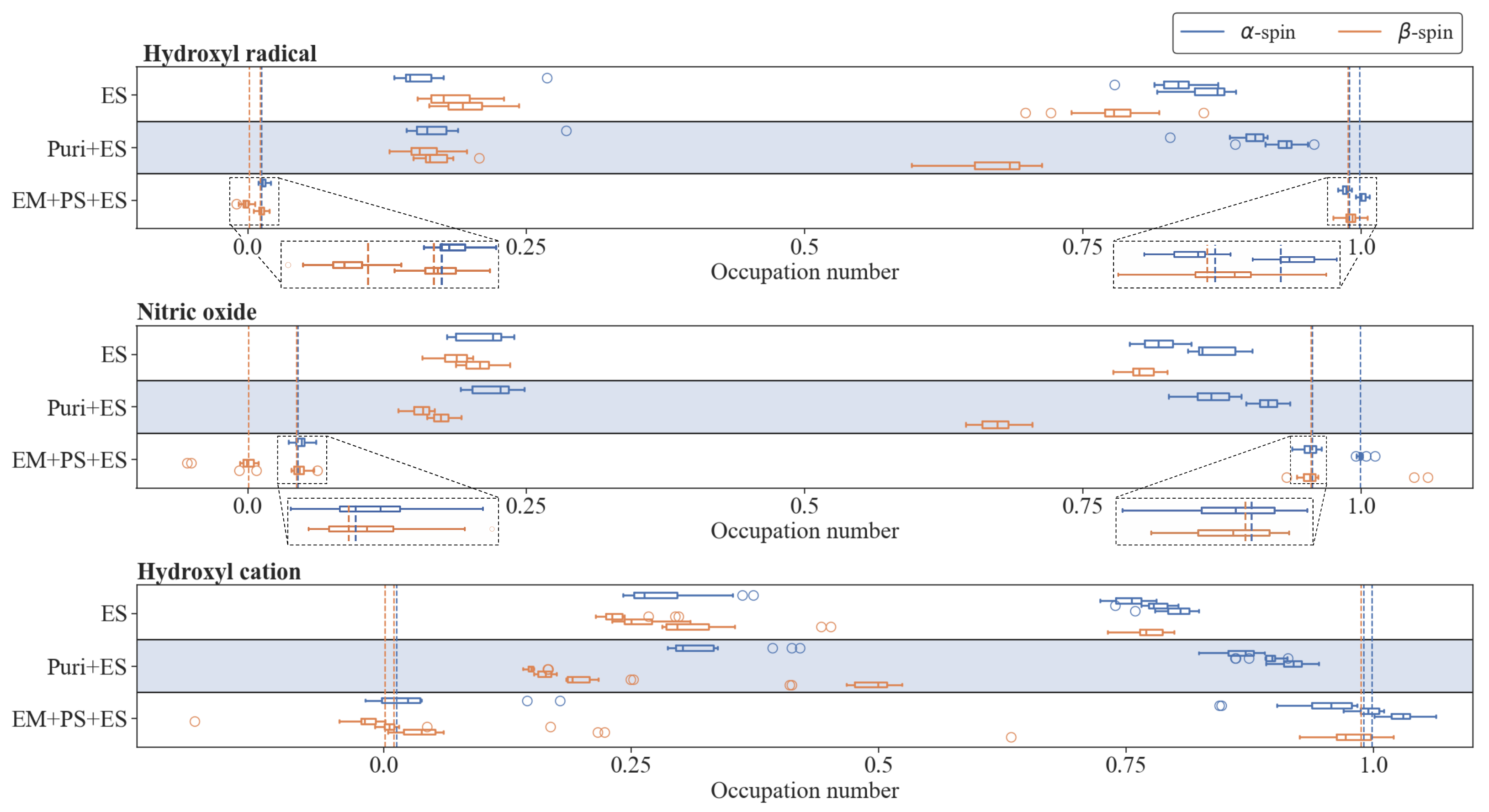}
    \caption{\revision{Occupation numbers obtained from quantum hardware experiments on the IBM Torino device for hydroxyl radical (OH$^{\bullet}$),   nitric oxide (NO$^{\bullet}$), and hydroxyl cation (OH$^{+}$)  using the circuits shown in Figure \ref{fig:main_circuits}. Each box plot  represents data from 15 independent hardware runs which show quartiles of the dataset, except for points that are determined to be outliers (empty circles). An overview of the tools used in hardware experiments is provided in Table \ref{tbl:overview_processing}, while Table \ref{tbl:overview_QPU_time} summarizes the number of measurements and QPU times. The EM+PS+ES method includes error mitigation ($\boldsymbol{M}_{U_0}$), post-selection, and error suppression. The Puri+ES method includes the 1-RDM purification method and error suppression, while the ES method includes only error suppression. The vertical dashed lines correspond to the exact (UCASSCF) occupation numbers.}}
    \label{fig:occupation_numbers}
\end{figure}

Figure \ref{fig:occupation_numbers} shows the occupation numbers, which correspond to the eigenvalues of the 1-RDMs, obtained from the quantum hardware experiments for 15 independent hardware runs for OH$^{\bullet}$, NO$^{\bullet}$,  OH$^{+}$, using the number of measurements listed in Table \ref{tbl:overview_QPU_time}. For OH$^{\bullet}$ and NO$^{\bullet}$, an active space of three electrons in three orbitals is used (Figure \ref{fig:main_circuits}), corresponding to three $\alpha$-spin and three $\beta$-spin occupation numbers. For OH$^{+}$, an active space of four electrons in four orbitals is used, corresponding to four occupation numbers for each spin state. We compare results from error mitigation, post-selection, and error suppression (EM+PS+ES) to that of without error mitigation and post-selection but with 1-RDM purification (Puri+ES). Additionally, we show the impact of 1-RDM purification by comparing results with only error suppression (ES). These methods are listed in Table \ref{tbl:overview_processing}. The EM+PS+ES results show a significantly smaller bias compared to Puri+ES and ES results. For OH$^{\bullet}$ and NO$^{\bullet}$, the difference between EM+PS+ES and the exact occupation numbers is, on average, on the order of $10^{-3}$, whereas for Puri+ES and ES, the difference is on the order of $10^{-1}$.  This shows a reduction in error by a factor of roughly 100 when additionally employing error mitigation and post-selection. For OH$^{+}$, the error increases due to the deeper circuit depth, with an average difference between EM+PS+ES and the exact occupation numbers on the order of $10^{-2}$, while the error for Puri+ES and ES also increases accordingly. The comparison of Puri+ES with ES show that purification does not consistently improve the occupation numbers. For example, while purification improves the occupation numbers near one for the $\alpha$-spin, it leads to worse occupation numbers for the corresponding $\beta$-spin near one for all systems.  Puri+ES and ES differ only by the scaling constant, $N_\sigma / \sum_i \lambda_i$, multiplied on the ES occupation numbers. For  OH$^{\bullet}$,  NO$^{\bullet}$, and OH$^{+}$, the average scaling constants for the 15 runs of the ES method are, for the $\alpha$-spin, $1.08 \pm 1.68 \cdot 10^{-2}$, $1.06 \pm 2.08 \cdot 10^{-2}$, and $1.15 \pm 2.58 \cdot 10^{-2}$, respectively, and for the $\beta$-spin $0.866  \pm 3.15  \cdot 10^{-2}$, $0.837 \pm 1.49 \cdot 10^{-2}$, and $0.631 \pm 3.06 \cdot 10^{-2}$, respectively. The scaling constant is greater than one for $\alpha$-spin and smaller than one for $\beta$-spin for all systems, which explains why the occupation numbers near one for the $\alpha$-spin improve with Puri+ES but worsen for the $\beta$-spin.

To remove unphysical 1-RDMs when computing the HFCs, we disregard the 1-RDM if any occupation number falls outside the range $[-0.05,1.05]$, i.e., setting $\epsilon=0.05$. This error tolerance is higher than the standard deviation of the statistical error due to shot noise alone, which is $\sim 10^{-3}$, and ensures that only significant deviations, likely caused by device noise rather than statistical uncertainty, are removed. Note that all Puri+ES and ES runs fall within the acceptable range in Figure \ref{fig:occupation_numbers}, even though they are far from the noiseless solution. We will therefore only remove 1-RDMs obtained from the EM+PS+ES results. With this error tolerance, the following 1-RDMs in the HFC calculations (discussed in the next section) are removed: For OH$^{\bullet}$, none of the runs are removed. In Figure \ref{fig:occupation_numbers} (top plot), we observe for OH$^{\bullet}$ that all the EM+PS+ES runs closely match the exact occupation numbers, and thus all the runs are used in the HFC calculations. For the NO$^{\bullet}$ molecule, two out of the 15 total runs are removed in the HFC calculations,  which correspond to the two outliers in Figure \ref{fig:occupation_numbers} (middle plot). \revision{In SI Section \ref{sec:app:occ_nums}, we show the occupation numbers for each individual run, where the  two outliers correspond to runs 1 and 7.} The occupation numbers that led to the removal of the 1-RDM are $\beta$-spin  occupation numbers: the two negative value outliers in Figure \ref{fig:occupation_numbers}, given by $-0.05119$ and $-0.05471$, fall below the lower limit of $-0.05$, and only the maximum outlier, given by $1.059$, exceeds the upper limit of $1.05$. For the OH$^{+}$ molecule,  we remove three runs in the HFC calculations.  \revision{These correspond to runs 1, 4, and 12.} The large negative $\beta$-spin outlier is given by $-0.1913$ in Figure \ref{fig:occupation_numbers} (bottom plot).  The other two occupation numbers which led to the removal of the 1-RDM  are $\alpha$-spin occupation numbers given by $1.062$ and $1.063$. Note that these $\alpha$-spin occupation numbers are not identified as outliers in Figure \ref{fig:occupation_numbers}, however, they still lie outside the acceptable range. For OH$^{+}$, the two occupation numbers close to one for the $\alpha$-spin are nearly identical, making them indistinguishable on the plot. Similarly, for the occupation numbers close to zero for the $\beta$-spin. 

\subsection{Isotropic hyperfine coupling constants}
\label{subsec:ihfc}

\begin{figure}[t]
\centering  
\includegraphics[width=1.0\textwidth]{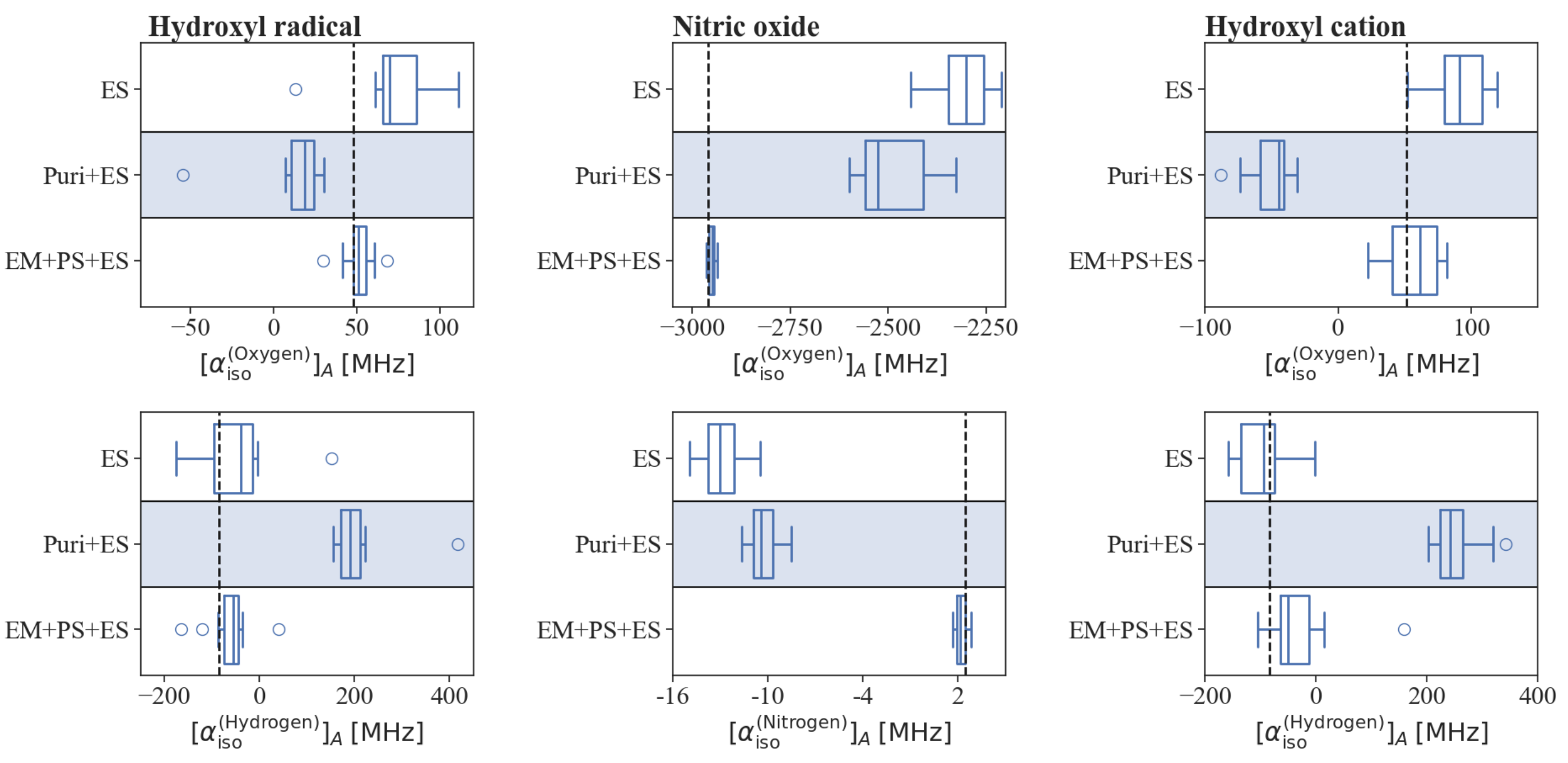}
\caption{\revision{Active space contribution to the isotropic hyperfine coupling constant (HFC), $[\alpha^{(K)}_{\text{iso}}]_{A}$, obtained from quantum hardware experiments on the IBM Torino device for hydroxyl radical (OH$^{\bullet}$),   nitric oxide  (NO$^{\bullet}$), and hydroxyl cation (OH$^{+}$) using the circuits shown in Figure \ref{fig:main_circuits}.  An overview of the tools used in hardware experiments is provided in Table \ref{tbl:overview_processing}, while Table \ref{tbl:overview_QPU_time} summarizes the number of measurements and QPU times. The EM+PS+ES method includes error mitigation ($\boldsymbol{M}_{U_0}$), post-selection, and error suppression. The Puri+ES method includes the 1-RDM purification method and error suppression, while the ES method includes only error suppression. The vertical dashed lines correspond to the exact (UCASSCF) active space contribution to the HFC.}}
\label{fig:main_hfc}
\end{figure}

We focus on the active space contribution to the HFC, $[\alpha^{(K)}_{\text{iso}}]_{A}$, given by Eq. \eqref{eq:HFC_active}. The inactive part (Eq. \eqref{eq:HFC_inactive}) is computed classically, so any errors in the quantum hardware experiments originate from the active space contribution, as shown in Eq. \eqref{eq:error_formula}. The HFC results are plotted in  Figure \ref{fig:main_hfc}, calculated from the 1-RDM runs in Section \ref{subsec:rdm}.  We compare the HFCs obtained with EM+PS+ES to that of without error mitigation and post-selection but with 1-RDM purification, Puri+ES, and only with error suppression, ES. The vertical dashed lines correspond to the exact (UCASSCF) active space contribution to
the HFC. Note that in the noiseless case, where neither device nor sampling noise is presented, the oo-qubit-ADAPT method gives the same result as UCASSCF. For all systems, we observe that the EM+PS+ES results are in good agreement with UCASSCF and improve, in some cases significantly, upon the Puri+ES and ES results. Notably, for NO$^{\bullet}$, the EM+PS+ES results show a substantial improvement over Puri+ES and ES compared to the other systems, despite having a larger circuit depth than OH$^{\bullet}$ (see Figure \ref{fig:main_circuits}).  This is due to the fact that Puri+ES and ES do not benefit from error cancellation. The errors in the HFCs, given in Eq. \eqref{eq:error_formula}, for NO$^{\bullet}$ can be approximated as (SI Section \ref{app:error_analysis})

\begin{align}
\Delta\alpha^{(N)}_{\text{iso}}  &\approx -323\cdot 0.25 \cdot  \Delta D_{1\beta, 1\beta}~\text{MHz} \label{eq:error_no_N} \\ 
\Delta\alpha^{(O)}_{\text{iso}}  &\approx -606 \cdot  \left( 4.88 \cdot \Delta D_{1\alpha, 1\alpha}  - 2.00\cdot  \Delta D_{1\beta, 1\beta}  \right)~\text{MHz}. \label{eq:error_no_O} 
\end{align}
Here, the error in the nitrogen HFC, Eq. \eqref{eq:error_no_N}, is proportional to the error in the  1-RDM diagonal element $\Delta D_{1\beta, 1\beta} = D_{1\beta, 1\beta} - \tilde{D}_{1\beta, 1\beta}$, so the errors from the hardware experiments directly translate into errors in the HFC. For oxygen,  Eq. \eqref{eq:error_no_O}, the error also depends on the $\alpha$-spin element, which allows for potential error cancellation. For the hardware experiments, the average errors in these 1-RDM diagonal elements are: $\braket{\Delta D_{1\alpha, 1\alpha}} = 1.86 \cdot 10^{-3} \pm 1.92  \cdot 10^{-3} $,  $\braket{\Delta D_{1\alpha, 1\alpha}} = 9.70 \cdot 10^{-2} \pm 2.25  \cdot 10^{-3} $, and  $\braket{\Delta D_{1\alpha, 1\alpha}} = 0.144 \pm 2.25  \cdot 10^{-2} $ for EM+PS+ES, Puri+ES, and ES, respectively, and $\braket{\Delta D_{1\beta, 1\beta}} = - 3.37 \cdot 10^{-3} \pm 4.24 \cdot 10^{-3}$, $\braket{\Delta D_{1\beta, 1\beta}} = -0.155 \pm 1.57 \cdot 10^{-2}$, and $\braket{\Delta D_{1\beta, 1\beta}} = -0.186 \pm 1.57 \cdot 10^{-2}$  for EM+PS+ES, Puri+ES, and ES, respectively. Thus, the impact of error mitigation and post-selection significantly reduced the errors in the 1-RDM diagonal elements.  Moreover, none of the methods benefit from error cancellation, due to the sign on the errors in the density matrix, so the improvement observed for NO$^{\bullet}$ reflects the direct impact of error mitigation and post-selection on the hardware results. However, for OH$^{\bullet}$ and OH$^{+}$, error cancellation is present. As a result, the closer agreement with UCASSCF observed for Puri+ES and ES is not due to improved hardware results but rather the influence of error cancellation.

\begin{table}[t]
\centering\renewcommand\cellalign{lc}
\setcellgapes{3pt}\makegapedcells
\caption{The total isotropic hyperfine coupling constant (HFC), including both inactive and active contributions, for the quantum hardware experiments (EM+PS+ES, Puri+ES, ES), zero device noise (shot-noise limit), the classical methods UHF and UCASSCF, and, when available, experimental data  (Experiment).  In each cell, except for the UHF method, the active space is denoted as “(e,o)”. } \vspace{0.1cm}
\begin{adjustbox}{width=17.0cm,center}
\begin{tabular}{|l|c|c|c|c|c|c|c}
\hline 
Molecule &  \multicolumn{2}{c|}{OH$^{\bullet}$}   &   \multicolumn{2}{c|}{NO$^{\bullet}$}    &  \multicolumn{2}{c|}{OH$^{+}$}  \\  \hline 
HFC [MHz] & $\alpha^{(\text{Oxygen})}_{\text{iso}}$ &  $\alpha^{(\text{Hydrogen})}_{\text{iso}}$   &  $\alpha^{(\text{Oxygen})}_{\text{iso}}$  &$\alpha^{(\text{Nitrogen})}_{\text{iso}}$ &$\alpha^{(\text{Oxygen})}_{\text{iso}}$ &  $\alpha^{(\text{Hydrogen})}_{\text{iso}}$ \\  \hline  \hline 
UHF &  $-99.2$  & $-112.7$& $-40.9$ & $62.5$ &$-101.7$ & $-113.3$ \\ 
\hline 
\makecell[cl]{ES} & $\makecell[c]{-65.9\pm
22.4\\ (3,3)}$  & $\makecell[c]{-48.6\pm
78.4\\ (3,3)}$ &  $\makecell[c]{577.5\pm
70.1\\ (3,3) }$  & $\makecell[c]{40.7 \pm
1.27\\ (3,3) }$ &  $\makecell[c]{-50.8\pm
19.3 \\ (4,4) } $   & $\makecell[c]{-97.0\pm
44.4 \\ (4,4)} $ \\ 
\hline 
\makecell[cl]{Puri+ES} & $\makecell[c]{-126.4\pm
19.6\\ (3,3)}$  & $\makecell[c]{206.0\pm
60.2\\ (3,3)}$ &  $\makecell[c]{400.0\pm
92.0\\ (3,3) }$  & $\makecell[c]{43.3
\pm 0.94\\ (3,3) }$ &  $\makecell[c]{-194.2\pm
15.0 \\ (4,4) } $   & $\makecell[c]{250.0\pm
38.5 \\ (4,4)} $ \\ 
\hline 
\makecell[cl]{EM+PS+ES} & $\makecell[c]{-89.5\pm 8.7\\(3,3)}$  &  $\makecell[c]{-60.7\pm 43.1 \\(3,3)}$&  $\makecell[c]{-65.0\pm
7.8 \\(3,3)}$ & $\makecell[c]{55.8
\pm 0.34 \\ (3,3)}$ & $\makecell[c]{-85.9\pm 
20.2\\ (4,4)}$
&  $\makecell[c]{-31.3 \pm 
66.4\\ (4,4)} $ \\ 
\hline 
\makecell[cl]{Shot-noise limit} &$\makecell[c]{-92.3\pm 1.58\\ (3,3)} $   &  $\makecell[c]{-83.1 \pm
6.1\\ (3,3)} $ & $\makecell[c]{-74.6 \pm
0.51\\ (3,3)} $  &   $\makecell[c]{56.1 \pm
0.010\\ (3,3)} $  & $\makecell[c]{-91.6 \pm
2.1\\ (4,4)}$ & $\makecell[c]{-83.4 \pm
5.2\\ (4,4)}$  \\ 
\hline 
  \makecell[cl]{\multirow{2}{*}{UCASSCF}} & $\makecell[c]{-92.3~  (3,3)}$   &  $\makecell[c]{-83.1~(3,3)}$ & $\makecell[c]{-74.6~(3,3)}$ &  $\makecell[c]{56.1~(3,3)}$ & $\makecell[c]{-91.6~(4,4)}$ &$\makecell[c]{-83.3~(4,4)}$  \\ 
& $\makecell[c]{-44.6~(7,22)}$ & $\makecell[c]{-77.9~(7,22)}$ & $\makecell[c]{-42.2~(11,16)}$ & $\makecell[c]{27.2~(11,16)}$& $-47.5~ (6,22)$& $-81.3~(6,22)$  \\ 
\hline 
Experiment &  $-51$ Ref.~\cite{engels_advances_1996}  & $\sim -61$ Ref.~\cite{engels_advances_1996}&  No exp. value & 22  Ref.~\cite{weltner_magnetic_1989} &No exp. value  &No exp. value  \\ 
\bottomrule
\end{tabular}
\end{adjustbox}
\label{tbl:main_hfc}
\end{table}

In Table \ref{tbl:main_hfc} we show the total HFC, which includes both the inactive and the active contributions,  for the quantum hardware experiments (EM+PS+ES,  Puri+ES, ES), in the shot-noise limit, and compare them with the classical methods UHF and UCASSCF. Further, in cases where experimental data is available, we compare with the experimental HFC values from Ref.~\cite{kossmann_performance_2007}. For the quantum hardware experiments, we report the average HFC values and standard deviations. For the shot-noise limit simulations, these values are obtained from 1000 independent runs using the number of measurements listed in Table \ref{tbl:overview_QPU_time}. Note that the standard deviations come entirely from the active space part, as the inactive part is computed classically and remains error-free. The shot-noise simulations provide a benchmark for the performance of future fault-tolerant quantum computers where device noise is eliminated and only shot-noise remains. In the shot-noise limit, the HFC values closely match UCASSCF. The standard deviations could be further reduced by increasing the number of measurements. For NO$^{\bullet}$, the EM+PS+ES values match closer with the shot-noise limit than the corresponding Puri+ES and ES values. As explained in the previous paragraph, this is because the errors in the diagonal elements of the 1-RDM are significantly smaller for EM+PS+ES, and error cancellation is not occurring in any of the methods. In contrast, for OH$^{+}$, we observe that ES outperforms EM+PS+ES for the hydrogen HFC, which is unexpected (this is also observed in Figure \ref{fig:main_hfc}). The errors are $-52$ MHz for EM+PS+ES and $13.7$ MHz for ES, which can be extracted from Table \ref{tbl:main_hfc}. This contrasts with Figure \ref{fig:occupation_numbers}, which shows that the hardware experiments for EM+PS+ES outperforms ES. The  error in the hydrogen HFC for OH$^{+}$ can be approximated as (SI Section \ref{app:error_analysis})

\begin{align}
\Delta\alpha^{(H)}_{\text{iso}}  \approx &2235\cdot  \left( 0.150 \cdot \Delta D_{2\alpha, 2\alpha}+ 0.518 \cdot \Delta D_{3\alpha, 3\alpha} + 0.557\cdot \Delta D_{2\alpha, 3\alpha} \right. \nonumber   \\
&\left. - 0.202 \cdot  \Delta D_{0\beta, 0\beta} - 0.496\cdot  \Delta D_{3\beta, 3\beta}  - 0.632 \cdot  \Delta D_{0\beta, 3\beta}\right)~\text{MHz}.
\label{eq:error_HFC_OH+_H}
\end{align}
Equation \eqref{eq:error_HFC_OH+_H} gives the errors of $-52$ MHz for EM+PS+ES and $13.7$ MHz for ES.  Of these errors,  the average $\alpha$-spin error ($0.150 \cdot \Delta D_{2\alpha, 2\alpha}+ 0.518 \cdot \Delta D_{3\alpha, 3\alpha} + 0.557\cdot \Delta D_{2\alpha, 3\alpha}$) is $-40.3$ MHz and the average $\beta$-spin error ($- 0.202 \cdot  \Delta D_{0\beta, 0\beta} - 0.496\cdot  \Delta D_{3\beta, 3\beta}  - 0.632 \cdot  \Delta D_{0\beta, 3\beta}$) is $-11.7$ MHz for EM+PS+ES. The sum of these contributions result in the total observed error. In contrast, for ES, the $\alpha$-spin error for hydrogen is $-231$ MHz and the $\beta$-spin error is $245.2$ MHz, leading to error cancellation. That is, for the ES method, even though the individual errors are large, due to large errors in the hardware experiments, they largely cancel each other out, resulting in a much smaller overall error which we are observing in Table \ref{tbl:main_hfc}. Error cancellation is also observed for OH$^{\bullet}$. For a more detailed explanation of error cancellation for the systems, we refer to SI Section \ref{app:error_analysis}. We also compare to larger active spaces using classic UCASSCF calculations that include all valence electrons for the molecular systems, i.e., 6, 7, and 11 electrons for OH$^{+}$, OH$^{\bullet}$, NO$^{\bullet}$, respectively. The UCASSCF calculations with large active space show better agreement with the experimental values when available, and an active space of these sizes is necessary for computing reliable HFCs. \revision{Comparing with the current state-of-the-art methods for calculations of HFCs on classical computers, i.e. DFT for larger molecules and metal complexes and Coupled Cluster methods for small molecules~\cite{bruder_application_2024}, a new approach for calculating HFCs needs to achieve an accuracy of 10 MHz or less in its mean absolute errors in order to be competitive.  The results from our study show that the active spaces currently possible on NISQ devices are too small for obtaining such an accuracy but with larger active spaces it would have been possible. However, there is also the question of the standard deviation in the results from quantum computations, which also has to be within this accuracy of 10 MHz or less in order to be compatible. And in respect to this, we could show in the present study that for some HFCs this is already possible with the current NISQ devices.}

%% file: sections/conclusion.tex
\section{Conclusion and Outlook}

In this work, we have implemented and demonstrated the computation of isotropic hyperfine coupling constants (HFCs) on quantum computers for hydroxyl radical (OH$^{\bullet}$), nitric oxide (NO$^{\bullet}$), and hydroxyl cation (OH$^{+}$) using a hardware-efficient quantum algorithm, oo-qubit-ADAPT, applied within an active space framework and unrestricted orbital optimization. Our results highlight the feasibility of computing HFCs on current quantum hardware, the challenges posed by device noise, and the limitations of error mitigation techniques.

\revision{The advantage of using a hybrid quantum algorithm to compute HFCs lies in employing the VQE algorithm to estimate the zeroth-order ground state in the active space approximation. Furthermore, since we are dealing with open-shell systems, which is known to be challenging for accurate HFC calculations~\cite{vogler_important_2020,windom_benchmarking_2022, jaworski_electron_2022}, expensive classical methods are required such as coupled-cluster or MCSCF approaches due to strong electron correlation. The hope is that VQE, and in particular the oo-qubit-ADAPT algorithm used in our study, can obtain results of similar quality as the expensive classical methods while reducing the computational cost, e.g., in terms of runtime and memory. Note that the case studies presented in this study serve as proof-of-concept demonstrations and  quantum advantage was not observed. It remains, however, an open question whether VQE can offer an advantage over state-of-the-art classical methods, e.g., the overhead from the number of measurements which increases the runtime. Significant improvements in quantum hardware will be necessary before VQE may become competitive to classical methods.}

We computed the spin-density 1-RDMs using six qubits for OH$^{\bullet}$ and NO$^{\bullet}$, and eight qubits for OH$^{+}$. Our analysis compared results obtained from the quantum hardware experiments with error mitigation, post-selection, and error suppression (EM+PS+ES) against those obtained without error mitigation and post-selection but with 1-RDM purification (Puri+ES). Additionally, we evaluated the impact of 1-RDM purification by comparing results using only error suppression (ES). We showed that employing our in-house developed ansatz-based readout and gate error mitigation with post-selection significantly reduced the bias in the computed 1-RDMs compared to the unmitigated results. For OH$^{\bullet}$ and NO$^{\bullet}$, the average difference between EM+PS+ES and the exact (UCASSCF) occupation numbers was on the order of $10^{-3}$, whereas for Puri+ES and ES, the difference was on the order of $10^{-1}$. For OH$^{+}$, the deeper circuit depth led to an increase in error, and the average difference between EM+PS+ES and the exact occupation numbers was on the order of $10^{-2}$, with errors for Puri+ES and ES also increasing accordingly. Furthermore, a comparison of Puri+ES with ES revealed that purification did not consistently improve the accuracy of the occupation numbers.

The HFCs obtained using EM+PS+ES produced results comparable to those from UCASSCF. For NO$^{\bullet}$, EM+PS+ES closely matched UCASSCF and were significantly more accurate than those from Puri+ES and ES. This improvement reflects the direct impact of error mitigation and post-selection, as no error cancellation occurred for this system. For OH$^{\bullet}$ and OH$^{+}$, we observed error cancellation, and in one case, ES even outperformed EM+PS+ES. However, this was not due to improved hardware performance but rather the effect of error cancellation.

Expanding the active space in UCASSCF revealed that larger active spaces, including all valence electrons, would be necessary to achieve accurate HFC values. While these calculations remain feasible classically for small molecules, they become computationally demanding as the system size increases. oo-qubit-ADAPT has the potential to scale beyond classical capabilities, making it a promising approach for future quantum simulations of HFCs in molecular systems.

Our study lays the groundwork for future quantum computing applications in ESR spectroscopy and electronic structure theory. As quantum hardware advances, with improved qubit quality, error correction, and larger circuit depths, we anticipate that quantum algorithms such as oo-qubit-ADAPT will play an increasingly important role in computing spectroscopic properties.

%% file: sections/acknowledgements.tex
\section{Acknowledgements}

We acknowledge the financial support of the Novo Nordisk
Foundation to the focused research project Hybrid Quantum
Chemistry on Hybrid Quantum Computers $(\text{HQC})^2$ (grant
number: NNFSA220080996).

%% file: sections/appendix.tex
\newpage
\appendix
\appendixpage

\tableofcontents

\setcounter{table}{0}
\setcounter{figure}{0}

\renewcommand{\thetable}{S\arabic{table}}

\renewcommand{\thefigure}{S\arabic{figure}}

\renewcommand{\thesection}{\Alph{section}}
\renewcommand{\thesubsection}{\Alph{section}.\arabic{subsection}}
\renewcommand{\thesubsubsection}{\Alph{section}.\arabic{subsection}.\arabic{subsubsection}}


\resumetoc
\newpage
\section{Error analysis}
\label{app:error_analysis}

In the active space approximation the spin-density one-electron reduced density matrix (1-RDM) can be written in the block form 
\begin{align}
\boldsymbol{D}_{\sigma} = \begin{pmatrix}
      [\boldsymbol{D}_{\sigma} ]_{I}& [\boldsymbol{D}_{\sigma} ]_{IA} &[\boldsymbol{D}_{\sigma} ]_{IV}           \\[0.3em]
     [\boldsymbol{D}_{\sigma} ]_{AI} &[\boldsymbol{D}_{\sigma} ]_{A}         &  [\boldsymbol{D}_{\sigma} ]_{AV}\\[0.3em]
     [\boldsymbol{D}_{\sigma} ]_{VI}     & [\boldsymbol{D}_{\sigma} ]_{VA} & [\boldsymbol{D}_{\sigma} ]_{V}
     \end{pmatrix}  =  \begin{pmatrix}
       \mathbf{1} & \mathbf{0} & \mathbf{0}          \\[0.3em]
     \mathbf{0}& [\boldsymbol{D}_{\sigma} ]_{A}          & \mathbf{0}\\[0.3em]
      \mathbf{0}      & \mathbf{0}& \mathbf{0}
     \end{pmatrix},
\end{align}
where $[\boldsymbol{D}_{\sigma} ]_{I}$ is the block with the one-electron excitations within the inactive space, i.e., the elements $ D_{i\sigma,j\sigma}$ where the indices \emph{i} and \emph{j}  are used for occupied (inactive) orbitals, $[\boldsymbol{D}_{\sigma} ]_{A}$ is the block with the one-electron excitations within the active space, and $[\boldsymbol{D}_{\sigma} ]_{V}$ is the block with the one-electron excitations within the virtual space. The off-diagonal blocks are the one-electron excitations that occur between these spaces. The trace in Eq.~\eqref{eq:iso_a} in the main text can then be divided into its inactive and active contribution:
\begin{align}
\text{tr}\left[ \boldsymbol{A}^{(K)}_{\alpha} \boldsymbol{D}_{\alpha} - \boldsymbol{A}^{(K)}_{\beta} \boldsymbol{D}_{\beta}\right] &= \sum^{n_I}_{i}\left( A^{(K)}_{i\alpha, i\alpha}  -A^{(K)}_{i\beta, i\beta}  \right) +\sum^{n_A}_{vw}\left( A^{(K)}_{v\alpha, w\alpha}  D_{v\alpha, w\alpha} -A^{(K)}_{v\beta, w\beta} D_{v\beta, w\beta}  \right),
\end{align}
where the indices \emph{v} and \emph{w}  denote active molecular orbitals. The isotropic hyperfine coupling constant  (HFC) therefore takes the form: 
\begin{align}
\alpha^{(K)}_{\text{iso}} & = \left[\alpha^{(K)}_{\text{iso}}\right]_{I} + \left[\alpha^{(K)}_{\text{iso}}\right]_{A} \\[0.3cm]
\left[\alpha^{(K)}_{\text{iso}}\right]_{I} &= \frac{f_K}{2\pi M} \text{tr}\left[  \big[\boldsymbol{A}^{(K)}_{\alpha} \big]_I - \big[\boldsymbol{A}^{(K)}_{\beta} \big]_I \right]  \\[0.3cm]
\left[\alpha^{(K)}_{\text{iso}}\right]_{A} &= \frac{f_K}{2\pi M} \text{tr}\left[ \big[\boldsymbol{A}^{(K)}_{\alpha}\big]_A   \big[\boldsymbol{D}_{\alpha} \big]_{A} -\big[\boldsymbol{A}^{(K)}_{\beta}\big]_A   \big[\boldsymbol{D}_{\beta} \big]_{A}\right].
\end{align}
Note that the virtual space does not contribute to the HFC because $[\boldsymbol{D}_{\sigma} ]_{V}=\textbf{0}$.

Let $\tilde{\alpha}^{(K)}_{\text{iso}}$ represents the HFC obtained from a quantum hardware experiment. Since we are not doing oo-qubit-ADAPT on the quantum hardware, the molecular spin-orbitals are instead obtained from a oo-qubit-ADAPT state-vector simulator~\cite{aurora_2023}, which are then used to construct the  $\boldsymbol{A}^{(K)}_{\sigma}$  matrices.  As a result, the error in this case originates entirely from the active space 1-RDMs, $[\boldsymbol{D}_{\sigma} ]_{A}$, which is obtained using quantum hardware. The error therefore becomes
\begin{align}
\Delta\alpha^{(K)}_{\text{iso}}  &=\alpha^{(K)}_{\text{iso}} - \tilde{\alpha}^{(K)}_{\text{iso}}\nonumber  \\[0.3cm]
&=    \left[\alpha^{(K)}_{\text{iso}}\right]_{A} - \left[\tilde{\alpha}^{(K)}_{\text{iso}}\right]_{A} \nonumber \\[0.3cm]
&= \frac{f_K}{2\pi M} \sum^{n_A}_{vw}\left( A^{(K)}_{v\alpha, w\alpha}  \Delta D_{v\alpha, w\alpha} -A^{(K)}_{v\beta, w\beta} \Delta  D_{v\beta, w\beta}  \right), \label{eq:app:error_formula}
\end{align}
where 

\begin{align}
\Delta D_{v\sigma, w\sigma}  =  D_{v\sigma, w\sigma} - \tilde{D}_{v\sigma, w\sigma}
\end{align}
and $ D_{v\sigma, w\sigma}$ being the exact value and $\tilde{D}_{v\sigma, w\sigma}$ the estimate obtained from a quantum hardware experiment. The constant is given by 
\begin{align}
\frac{f_K}{2\pi M} =  400.12~\text{MHz}~\frac{g_K}{ M}. \label{eq:app:error_constant}
\end{align}
For example, the error $\Delta\alpha^{(K)}_{\text{iso}}$ may be small even when the error in $\tilde{D}_{v\sigma, w\sigma}$ is large if the corresponding element $A^{(K)}_{v\sigma, w\sigma}$ is small. The nuclear g-factors, $g_K$, for hydrogen $\prescript{1}{}{\text{H}}$, oxygen $\prescript{17}{}{\text{O}}$, and nitrogen $\prescript{14}{}{\text{N}}$ are~\cite{stone_table_2005}

\begin{align}
g_H &= 5.58569468 \nonumber \\
g_O &=-0.757516\nonumber \\
g_N &= 0.40376100\nonumber
\end{align}
and the spin projection $M$ takes values of $\frac{1}{2}$ for radicals and 1 for triplets.

\hfill \break
\textbf{Error for nitric oxide}
\hfill \break
The amplitude distribution matrices, $\boldsymbol{A}^{(K)}_{\sigma}$, for nitric oxide (NO$^{\bullet}$) are given in Eqs. \eqref{eq:app:A_NO_1}-\eqref{eq:app:A_NO_4}. We observe that only the diagonal elements $A^{(K)}_{1\alpha, 1\alpha}$ and $A^{(K)}_{1\beta, 1\beta}$ have significant amplitudes compared to the other elements. The errors in the HFCs for NO$^{\bullet}$ can therefore be approximated as

\begin{align}
\Delta\alpha^{(O)}_{\text{iso}}  &\approx -606 \cdot  \left( 4.88 \cdot \Delta D_{1\alpha, 1\alpha}  - 2.00\cdot  \Delta D_{1\beta, 1\beta}  \right)~\text{MHz} \label{eq:app:error_HFC_NO_O} \\
\Delta\alpha^{(N)}_{\text{iso}}  &\approx -323\cdot 0.25 \cdot  \Delta D_{1\beta, 1\beta}~\text{MHz}. \label{eq:app:error_HFC_NO_N}
\end{align}
Figures \ref{fig:app:NO_rdms_alpha} and \ref{fig:app:NO_rdms_beta} show the $\alpha$ and $\beta$ spin-density 1-RDM elements, respectively, obtained from the quantum experiments for 15 independent runs. The average errors of $ \Delta D_{1\alpha, 1\alpha} $ and $\Delta D_{1\beta, 1\beta} $ are calculated over these runs, excluding runs 1 and 7 which are removed from the HFC calculations in the error-mitigation, post-selection, and error-suppression (EM+PS+ES) results (Section \eqref{subsec:rdm} in the main text). For EM+PS+ES, the errors are:  $\braket{\Delta D_{1\alpha, 1\alpha}} = 1.86 \cdot 10^{-3} \pm 1.92  \cdot 10^{-3} $ and $\braket{\Delta D_{1\beta, 1\beta}} =  - 3.37 \cdot 10^{-3} \pm 4.24 \cdot 10^{-3}$. For the unmitigated and without post-selection (ES) results, the errors are: $\braket{\Delta D_{1\alpha, 1\alpha}} = 0.144 \pm 2.25  \cdot 10^{-2} $ and $\braket{\Delta D_{1\beta, 1\beta}} =  -0.186 \pm 1.57 \cdot 10^{-2}$. Thus, we achieve a factor of 77 and 55 reduction in error for the $\alpha$ and $\beta$ spin diagonal elements, respectively, compared to the ES results. For the RDM purification method (Puri+ES) results (Section \ref{subsec:rdm_puri} in main text), the error is given by  $ D_{1\sigma, 1\sigma} - (N_{\sigma}/\sum_i \lambda_i) \tilde{D}_{1\sigma, 1\sigma} $. The average scaling constants, $\braket{N_{\sigma}/\sum_i \lambda_i}$, for the 15 runs of the ES method are  $1.055 \pm 2.08 \cdot 10^{-2}$ and $0.8369 \pm 1.49 \cdot 10^{-2}$ for the $\alpha$ and $\beta$ spin, respectively. Using these average scaling constants, the errors for Puri+ES results are $\braket{\Delta D_{1\alpha, 1\alpha}} = 9.70 \cdot 10^{-2} \pm 2.25  \cdot 10^{-3} $ and $\braket{\Delta D_{1\beta, 1\beta}} = -0.155 \pm 1.57 \cdot 10^{-2} $. Comparing to EM+PS+ES, we achieve a factor of 52 and 50 reduction in error for the $\alpha$ and $\beta$ spin diagonal elements, respectively, and Puri+ES is therefore performing better than ES. Plugging the average errors into Eqs. \eqref{eq:app:error_HFC_NO_O} and \eqref{eq:app:error_HFC_NO_N}, we obtain the errors for the HFCs

\begin{align}
&\left[\Delta\alpha^{(O)}_{\text{iso}}\right]_{\text{EM+PS+ES}}  \approx -9.6 ~\text{MHz} && \left[\Delta\alpha^{(N)}_{\text{iso}}\right]_{\text{EM+PS+ES}}   \approx 0.27~\text{MHz} \label{eq:app:error_HFC_NO_1} \\[0.3cm]
&\left[\Delta\alpha^{(O)}_{\text{iso}}\right]_{\text{Puri+ES}}  \approx -475~\text{MHz}  && \left[\Delta\alpha^{(N)}_{\text{iso}}\right]_{\text{Puri+ES}}  \approx 13~\text{MHz} \\[0.3cm]
&\left[\Delta\alpha^{(O)}_{\text{iso}}\right]_{\text{ES}}  \approx -651~\text{MHz} && \left[\Delta\alpha^{(N)}_{\text{iso}}\right]_{\text{ES}}  \approx 15~\text{MHz} \label{eq:app:error_HFC_NO_3} 
\end{align}
The exact errors can be extracted from Table \ref{tbl:main_hfc} in the main text which  closely match those in Eqs. \eqref{eq:app:error_HFC_NO_1}-\eqref{eq:app:error_HFC_NO_3}. We achieve a factor of 50 and 68 reduction in the error for the HFC of oxygen between  EM+PS+ES and Puri+ES and EM+PS+ES and ES, respectively, which correspond roughly to the reduction of errors in the diagonal elements between the methods. For the nitrogen atom, the error in $D_{1\beta, 1\beta}$ directly translates into error in the HFC.
\hfill \break
\hfill \break
\textbf{Error for hydroxyl radical}
\hfill \break
The amplitude distribution matrices, $\boldsymbol{A}^{(K)}_{\sigma}$, for hydroxyl radical (OH$^{\bullet}$) are given in Eqs. \eqref{eq:app:A_OH_1}-\eqref{eq:app:A_OH_4}. The diagonal elements $A^{(K)}_{0\sigma, 0\sigma}$ and $A^{(K)}_{2\sigma, 2\sigma}$ and the off-diagonal elements, $A^{(K)}_{0\sigma, 2\sigma}=A^{(K)}_{2\sigma, 0\sigma}$, have significant amplitudes compared to the other elements. The errors in the HFCs for OH$^{\bullet}$ can therefore be approximated as

\begin{align}
\Delta\alpha^{(O)}_{\text{iso}}  \approx &-606 \cdot  \left( 0.240 \cdot \Delta D_{0\alpha, 0\alpha}+ 1.15 \cdot \Delta D_{2\alpha, 2\alpha} - 1.05 \cdot \Delta D_{0\alpha, 2\alpha} \right. \nonumber   \\
&\left. - 0.357\cdot  \Delta D_{0\beta, 0\beta}- 1.29\cdot  \Delta D_{2\beta, 2\beta}  + 1.36\cdot  \Delta D_{0\beta, 2\beta}\right)~\text{MHz} \label{eq:app:error_HFC_OH_O} \\[0.3cm]
\Delta\alpha^{(H)}_{\text{iso}}  \approx &4470\cdot  \left( 0.227 \cdot \Delta D_{0\alpha, 0\alpha}+ 0.447 \cdot \Delta D_{2\alpha, 2\alpha} + 0.637\cdot \Delta D_{0\alpha, 2\alpha} \right. \nonumber   \\
&\left. - 0.228\cdot  \Delta D_{0\beta, 0\beta}- 0.460\cdot  \Delta D_{2\beta, 2\beta}  - 0.647\cdot  \Delta D_{0\beta, 2\beta}\right)~\text{MHz}.
\label{eq:app:error_HFC_OH_H}
\end{align}
Figures \ref{fig:app:OH_rdms_alpha} and \ref{fig:app:OH_rdms_beta} show the $\alpha$ and $\beta$ spin-density 1-RDM elements, respectively, obtained from the quantum experiments for 15 independent runs, and Table \ref{app:tbl:rdm_errors_oh} shows the errors based on the average values for these runs. Plugging the values from Table \ref{app:tbl:rdm_errors_oh} into Eqs. \eqref{eq:app:error_HFC_OH_O} and \eqref{eq:app:error_HFC_OH_H}, we obtain the HFC errors

\begin{table}[t]
\centering\renewcommand\cellalign{lc}
\setcellgapes{3pt}\makegapedcells
\caption{Average errors in the 1-RDM elements  for OH$^{\bullet}$ obtained from the quantum experiments for the 15 independent hardware runs shown in Figs. \ref{fig:app:OH_rdms_alpha} and \ref{fig:app:OH_rdms_beta}. The table only includes 1-RDM errors that impact the HFCs, as the remaining 1-RDM elements have negligible effects due to their correspondingly small amplitude distribution matrix elements $(A^{(K)}_{p\sigma, q\sigma})$. An overview of the tools, measurements, and QPU times are provided in the main text in Tables \ref{tbl:overview_processing} and \ref{tbl:overview_QPU_time}. } \vspace{0.1cm}
\begin{adjustbox}{width=16.0cm,center}
\begin{tabular}{ l | c | c | c | c | c }
Elements &  $\braket{\Delta D_{p\sigma, q\sigma}}_{\text{EM+PS+ES}}$ & $\braket{\Delta D_{p\sigma, q\sigma}}_{\text{Puri+ES}}$ & $\braket{\Delta D_{p\sigma, q\sigma}}_{\text{ES}}$& $\left|\frac{\braket{\Delta D_{p\sigma, q\sigma}}_{\text{Puri+ES}}}{\braket{\Delta D_{p\sigma, q\sigma}}_{\text{EM+PS+ES}}}\right|$ & $\left|\frac{\braket{\Delta D_{p\sigma, q\sigma}}_{\text{ES}}}{\braket{\Delta D_{p\sigma, q\sigma}}_{\text{EM+PS+ES}}}\right|$  \\
 \hline  \hline
$(0\alpha, 0\alpha)$ & $1.22 \cdot 10^{-3} \pm 1.48  \cdot 10^{-3}$  & $7.63 \cdot 10^{-2} \pm 2.66 \cdot 10^{-2}$  & $ 0.142 \pm 2.47 \cdot 10^{-2} $ & 63& 116  \\
 $(2\alpha, 2\alpha)$ & $ -1.78 \cdot 10^{-3} \pm 2.96  \cdot 10^{-3}$  & $-0.158 \pm 3.47\cdot 10^{-2}$ & $-0.146 \pm 3.22 \cdot 10^{-2}$ & 89 & 82 \\
  $(0\alpha, 2\alpha)$ & $-2.26 \cdot 10^{-3} \pm 9.41  \cdot 10^{-3}$   & $-5.27 \cdot 10^{-3} \pm 3.14 \cdot 10^{-3} $ &$-5.60 \cdot 10^{-3} \pm 2.92 \cdot 10^{-3}$ & 2.3 & 2.5 \\
$(0\beta, 0\beta)$ & $ -1.93 \cdot 10^{-3} \pm 6.93  \cdot 10^{-3}$  & $0.314 \pm 3.25\cdot 10^{-2} $ & $ 0.210 \pm 3.75 \cdot 10^{-2}$  &  163 & 109  \\
$(2\beta, 2\beta)$ & $ 1.76 \cdot 10^{-3} \pm 3.71  \cdot 10^{-3} $  &  $-0.155 \pm 2.34 \cdot 10^{-2} $ & $-0.181 \pm 2.34 \cdot 10^{-2}$ & 88 & 103  \\
$(0\beta, 2\beta)$ & $4.13 \cdot 10^{-3} \pm 8.70  \cdot 10^{-3}$  & $ 1.12 \cdot 10^{-2} \pm 2.03 \cdot 10^{-2}$ & $1.00 \cdot 10^{-2} \pm 3.97 \cdot 10^{-3}$ & 2.7 & 2.4  \\
\bottomrule
\end{tabular}
\end{adjustbox}
\label{app:tbl:rdm_errors_oh}
\end{table}

\begin{align}
&\left[\Delta\alpha^{(O)}_{\text{iso}}\right]_{\text{EM+PS+ES}}  \approx-2.8 ~\text{MHz} && \left[\Delta\alpha^{(H)}_{\text{iso}}\right]_{\text{EM+PS+ES}}   \approx -22.4 ~\text{MHz} \label{eq:app:error_HFC_OH_O_numbers_1}  \\
&\left[\Delta\alpha^{(O)}_{\text{iso}}\right]_{\text{Puri+ES}}  \approx 33.2  ~\text{MHz}  &&\left[\Delta\alpha^{(H)}_{\text{iso}}\right]_{\text{Puri+ES}}  \approx -287~\text{MHz}. \\
&\left[\Delta\alpha^{(O)}_{\text{iso}}\right]_{\text{ES}}  \approx -26.8 ~\text{MHz}  &&\left[\Delta\alpha^{(H)}_{\text{iso}}\right]_{\text{ES}}  \approx-34.4 ~\text{MHz}. \label{eq:app:error_HFC_OH_H_numbers_3}
\end{align}
The exact errors can be extracted from Table \ref{tbl:main_hfc} in the main text which closely match those in Eqs. \eqref{eq:app:error_HFC_OH_O_numbers_1}-\eqref{eq:app:error_HFC_OH_H_numbers_3}. To calculate the errors for the Puri+ES method in  Table \ref{app:tbl:rdm_errors_oh},  we used the average scaling constants for the 15 hardware runs, $\braket{N_{\sigma}/\sum_i \lambda_i}$, given by $1.078 \pm 1.68 \cdot 10^{-2}$ and $0.8662 \pm 3.15 \cdot 10^{-2}$ for the $\alpha$ and $\beta$ spin, respectively. The error difference between EM+PS+ES and ES is not as large as  compared to NO$^{\bullet}$ (Eqs. \eqref{eq:app:error_HFC_NO_1}-\eqref{eq:app:error_HFC_NO_3}), which is due to a combination of error cancellation in ES and error amplification in EM+PS+ES. For ES, the $\alpha$-spin errors in Eqs. \eqref{eq:app:error_HFC_OH_O} and \eqref{eq:app:error_HFC_OH_H} are $77.5$ MHz and $-163.6$ MHz for oxygen and hydrogen, respectively, while the $\beta$-spin errors are $-104.3$ MHz and $129.2$ MHz. Despite these large individual errors, their sum significantly reduces the total error. In contrast, for EM+PS+ES, the $\alpha$-spin errors are $-0.4$ MHz (oxygen) and $-8.8$ MHz (hydrogen), while the $\beta$-spin errors are $-2.4$ MHz (oxygen) and $-13.6$ MHz (hydrogen). Although these individual errors are much smaller than those in ES, their sum increases the total error. We observe a large error of $-287$ MHz for Puri+ES in the case of hydrogen, compared to the smaller error of $-34.4$ MHz obtained with the ES method. It is not because the 1-RDM elements obtained from the ES method is more accurate than those from Puri+ES. Instead, it arises from a combination of error cancellation and error amplification. In Puri+ES for hydrogen, error amplification occurs because the $\alpha$-spin error is $-253$ MHz and the $\beta$-spin error is $-33.7$ MHz. In contrast, for the ES method, although the individual errors are large, they largely cancel each other out, leading to a much smaller overall error.
\hfill \break
\hfill \break
\textbf{Error for hydroxyl cation}
\hfill \break
The amplitude distribution matrices, $\boldsymbol{A}^{(K)}_{\sigma}$, for hydroxyl cation (OH$^{+}$) are given in Eqs. \eqref{eq:app:A_OH+_1}-\eqref{eq:app:A_OH+_4}. For the $\alpha$-spin amplitudes, only the lower $2 \times 2$ blocks contain non-vanishing amplitudes. For the $\beta$-spin amplitudes, only $A^{(K)}_{0\beta, 0\beta}, A^{(K)}_{3\beta, 3\beta}$, and the off-diagonal elements $A^{(K)}_{0\beta, 3\beta} = A^{(K)}_{3\beta, 0\beta}$ have non-vanishing amplitudes. The errors in the HFCs for OH$^{+}$ can therefore be approximated as

\begin{align}
\Delta\alpha^{(O)}_{\text{iso}}  \approx &-303 \cdot  \left( 0.436 \cdot \Delta D_{2\alpha, 2\alpha}+ 1.32 \cdot \Delta D_{3\alpha, 3\alpha} - 1.52 \cdot \Delta D_{2\alpha, 3\alpha} \right. \nonumber   \\
&\left. - 0.576 \cdot  \Delta D_{0\beta, 0\beta}- 1.73\cdot  \Delta D_{3\beta, 3\beta}  + 2.00 \cdot  \Delta D_{0\beta, 3\beta}\right)~\text{MHz} \label{eq:app:error_HFC_OH+_O} \\[0.3cm]
\Delta\alpha^{(H)}_{\text{iso}}  \approx &2235\cdot  \left( 0.150 \cdot \Delta D_{2\alpha, 2\alpha}+ 0.518 \cdot \Delta D_{3\alpha, 3\alpha} + 0.557\cdot \Delta D_{2\alpha, 3\alpha} \right. \nonumber   \\
&\left. - 0.202 \cdot  \Delta D_{0\beta, 0\beta} - 0.496\cdot  \Delta D_{3\beta, 3\beta}  - 0.632 \cdot  \Delta D_{0\beta, 3\beta}\right)~\text{MHz}.
\label{eq:app:error_HFC_OH+_H}
\end{align}
Figures \ref{fig:app:OH+_rdms_alpha} and  \ref{fig:app:OH+_rdms_beta}  show the $\alpha$ and $\beta$ spin-density 1-RDM elements, respectively, obtained from the quantum experiments for 15 independent runs, and Table \ref{app:tbl:rdm_errors_oh+} shows the errors based on the average values for these runs. Plugging the values from Table \ref{app:tbl:rdm_errors_oh+} into Eqs. \eqref{eq:app:error_HFC_OH+_O} and \eqref{eq:app:error_HFC_OH+_H}, we obtain the HFC errors

\begin{table}[t]
\centering\renewcommand\cellalign{lc}
\setcellgapes{3pt}\makegapedcells
\caption{ Average errors in the 1-RDM elements  for OH$^{+}$ obtained from the quantum experiments for the 15 independent hardware runs shown in Figs. \ref{fig:app:OH+_rdms_alpha} and  \ref{fig:app:OH+_rdms_beta}. For the EM+PS+ES method, runs 1, 4, and 12 are excluded from the average errors, as they are also removed from the HFC calculations (Section \ref{subsec:removal_of_RDMs} in the main text). The table only includes 1-RDM errors that impact the HFCs, as the remaining 1-RDM elements have negligible effects due to their correspondingly small amplitude distribution matrix elements $(A^{(K)}_{p\sigma, q\sigma})$. An overview of the tools, measurements, and QPU times are provided in the main text in Tables \ref{tbl:overview_processing} and \ref{tbl:overview_QPU_time}. } \vspace{0.1cm}
\begin{adjustbox}{width=16.0cm,center}
\begin{tabular}{ l | c | c | c | c | c }
Elements &  $\braket{\Delta D_{p\sigma, q\sigma}}_{\text{EM+PS+ES}}$ & $\braket{\Delta D_{p\sigma, q\sigma}}_{\text{Puri+ES}}$ & $\braket{\Delta D_{p\sigma, q\sigma}}_{\text{ES}}$& $\left|\frac{\braket{\Delta D_{p\sigma, q\sigma}}_{\text{Puri+ES}}}{\braket{\Delta D_{p\sigma, q\sigma}}_{\text{EM+PS+ES}}}\right|$ & $\left|\frac{\braket{\Delta D_{p\sigma, q\sigma}}_{\text{ES}}}{\braket{\Delta D_{p\sigma, q\sigma}}_{\text{EM+PS+ES}}}\right|$  \\
 \hline  \hline
$(2\alpha, 2\alpha)$ &   $-5.26 \cdot 10^{-3} \pm 1.31 \cdot 10^{-2} $ & $0.104 \pm 2.22 \cdot 10^{-2} $ & $0.217 \pm 1.94 \cdot 10^{-2} $  & 20 & 41  \\
 $(3\alpha, 3\alpha)$ &  $-1.74 \cdot 10^{-2} \pm 4.80 \cdot 10^{-2} $  &  $-0.314 \pm 4.83\cdot 10^{-2} $ & $-0.272\pm 4.21\cdot 10^{-2} $   & 18 &  16 \\
  $(2\alpha, 3\alpha)$ &  $-1.48 \cdot 10^{-2} \pm 2.62 \cdot 10^{-2} $ & $7.06 \cdot 10^{-3} \pm  6.36 \cdot 10^{-3}  $  &  $8.60 \cdot 10^{-3} \pm  5.55 \cdot 10^{-3}  $ & 0.48 & 0.58  \\
$(0\beta, 0\beta)$ &  $3.16 \cdot 10^{-2} \pm 0.101 $  & $0.501 \pm 1.26 \cdot 10^{-2} $  &  $0.217 \pm  1.99 \cdot 10^{-2} $  &16 & 6.9    \\
$(3\beta, 3\beta)$ & $-1.32 \cdot 10^{-2} \pm  4.15 \cdot 10^{-2}$   & $-0.192  \pm  3.39 \cdot 10^{-2}$   &   $-0.310  \pm  5.37 \cdot 10^{-2}$ & 15 & 24   \\
$(0\beta, 3\beta)$& $8.54 \cdot 10^{-3} \pm  1.59 \cdot 10^{-2}$   & $-4.09  \cdot 10^{-4} \pm   1.55 \cdot 10^{-3}$  & $3.04 \cdot 10^{-4} \pm  2.45\cdot 10^{-3}$  & 0.048 & 0.036   \\
\bottomrule
\end{tabular}
\end{adjustbox}
\label{app:tbl:rdm_errors_oh+}
\end{table}

\begin{align}
&\left[\Delta\alpha^{(O)}_{\text{iso}}\right]_{\text{EM+PS+ES}}  \approx -5.7 ~\text{MHz} && \left[\Delta\alpha^{(H)}_{\text{iso}}\right]_{\text{EM+PS+ES}}   \approx -52.0 ~\text{MHz} \label{eq:app:error_HFC_OH_O+_numbers_1}  \\
&\left[\Delta\alpha^{(O)}_{\text{iso}}\right]_{\text{Puri+ES}}  \approx 102  ~\text{MHz}  &&\left[\Delta\alpha^{(H)}_{\text{iso}}\right]_{\text{Puri+ES}}  \approx -333 ~\text{MHz}. \\
&\left[\Delta\alpha^{(O)}_{\text{iso}}\right]_{\text{ES}}  \approx -40.7 ~\text{MHz}  &&\left[\Delta\alpha^{(H)}_{\text{iso}}\right]_{\text{ES}}  \approx 13.8 ~\text{MHz}. \label{eq:app:error_HFC_OH_H+_numbers_3}
\end{align}
The exact errors can be extracted from Table \ref{tbl:main_hfc} in the main text which closely match those in Eqs. \eqref{eq:app:error_HFC_OH_O+_numbers_1}-\eqref{eq:app:error_HFC_OH_H+_numbers_3}.  To calculate the errors for the Puri+ES method in  Table \ref{app:tbl:rdm_errors_oh},  we used the average scaling constants for the 15 hardware runs, $\braket{N_{\sigma}/\sum_i \lambda_i}$, given by $1.147 \pm 2.58 \cdot 10^{-2}$ and $0.6308 \pm 3.06 \cdot 10^{-2}$ for the $\alpha$ and $\beta$ spin, respectively. Unlike the other molecules, we observe that the error in the ES method for hydrogen is lower than that of the EM+PS+ES method. Comparing this to the errors in the 1-RDM elements in Table \ref{app:tbl:rdm_errors_oh+}, the diagonal elements show significantly smaller errors in the EM+PS+ES method. For the off-diagonal elements, both methods yield small errors on the order of $10^{-4}-10^{-3}$. Thus, the reason ES performs better is due to a combination of error cancellation and amplification. For EM+PS+ES, the $\alpha$-spin error for hydrogen is $-40.3$ MHz and the $\beta$-spin error is $-11.7$ MHz, and their sum increases the total error $-52.0$ MHz. For ES, the $\alpha$-spin error for hydrogen is $-231$ MHz and the $\beta$-spin error is $245.2$ MHz, and their sum decreases the total error $13.8$ MHz. That is, the individual errors are much large for ES, but they largely cancel each other out, leading to a smaller overall error.

\section{Spin-orbitals}
\label{app:sec:spin_orbitals}

In this section, we will review the spin-orbitals for the hydroxyl radical (OH$^{\bullet}$), nitric oxide (NO$^{\bullet}$), and hydroxyl cation (OH$^{+}$) molecules, which are used as the computational basis for the quantum hardware calculations. The active space for the radicals, OH$^{\bullet}$ and NO$^{\bullet}$, consists of three electrons in three orbitals (a 6 qubit system), and for the cation, OH$^{+}$, is four electrons in four orbitals (a 8 qubit system). Since the oo-qubit-ADAPT method is not being executed on the quantum hardware, the molecular spin-orbitals are instead obtained from an oo-qubit-ADAPT state-vector simulator~\cite{aurora_2023}, and the resulting spin-orbitals are then used to construct the $\boldsymbol{A}^{(K)}_{\sigma}$ matrices.   

\subsection{Hydroxyl radical}
\label{sec:app:spin_orbs_OH}
\begin{figure}[t]
\centering  
\includegraphics[width=0.3\textwidth]{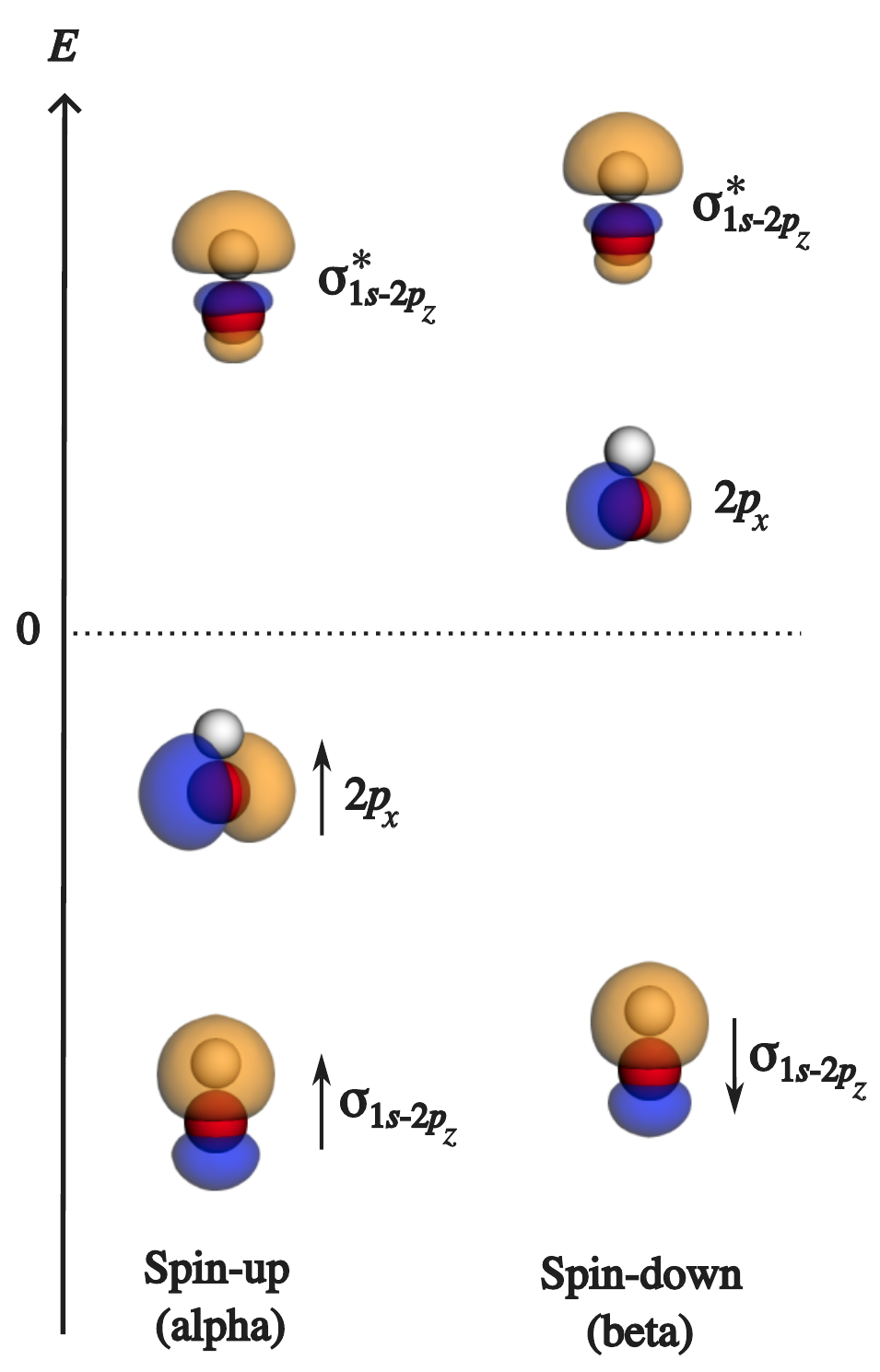}
\caption{Spin-orbitals for the hydroxyl radical molecule in an active space of three orbitals and three electrons, with an experimental internuclear distance of 0.9697 $\text{\AA}$, employing the 6-311++G**-J basis set~\cite{kjaer_pople_2011}. The orbitals were obtained using a oo-qubit-ADAPT(3,3) state-vector simulator~\cite{aurora_2023}. The spin-orbital energies are $(-0.829191, -0.657719 ,  0.694536)E_H$ and  $(-0.809123,  0.463627,   0.740324)E_H$ for the $\alpha$ and $\beta$ spin-orbitals, respectively, which correspond to the diagonal elements of the Fock matrix in the oo-qubit-ADAPT(3,3) spin-orbital basis.  The isovalue is set to $\pm 0.05$, where the blue and orange indicate the positive and negative sign of the wave functions, respectively. }
\label{fig:app:OH_MOs}
\end{figure}

The $\boldsymbol{A}^{(K)}_{\sigma}$  matrices for the OH$^{\bullet}$ molecule are computed using the molecular spin-orbitals obtained from an oo-qubit-ADAPT(3,3) state-vector simulator~\cite{aurora_2023}, starting from the unrestricted MP2 (UMP2) natural orbitals. The ``(3,3)" refers to an active space of three electrons in three orbitals. The resulting qubit-ADAPT circuit is shown in Figure \ref{fig:app:OH_circuit}. The calculation was performed with the experimental internuclear distance of 0.9697 $\text{\AA}$ using the 6-311++G**-J basis set~\cite{kjaer_pople_2011} which is optimized for ESR properties. The oo-qubit-ADAPT(3,3) energy converged to $-75.43387721$ $E_H$. The resulting matrices are 
\begin{align}
\left[\boldsymbol{A}^{(O)}_{\alpha}\right]_A &= \begin{pmatrix}
 2.39998196\cdot 10^{-1} &-3.06329871\cdot 10^{-4}&-5.26467324\cdot 10^{-1} \\[0.3em]
-3.06329871\cdot 10^{-4}& 3.90994564\cdot 10^{-7}&  6.71974499\cdot 10^{-4} \\[0.3em]
-5.26467324\cdot 10^{-1} & 6.71974499\cdot 10^{-4} & 1.15487469
     \end{pmatrix}  \label{eq:app:A_OH_1} \\[0.3cm]
\left[\boldsymbol{A}^{(O)}_{\beta}\right]_A &= \begin{pmatrix}
 3.56536063\cdot 10^{-1} &-2.18648977\cdot 10^{-4} &-6.78741132\cdot 10^{-1} \\[0.3cm]
-2.18648977\cdot 10^{-4} & 1.34088469\cdot 10^{-7}  &4.16244160\cdot 10^{-4} \\[0.3cm]
-6.78741132\cdot 10^{-1}&  4.16244160\cdot 10^{-4}&  1.29212602
     \end{pmatrix} \\[0.3cm]
\left[\boldsymbol{A}^{(H)}_{\alpha}\right]_A &= \begin{pmatrix}
 2.26860384\cdot 10^{-1} &-7.84048776\cdot 10^{-5}&  3.18271006\cdot 10^{-1}\\[0.3cm]
-7.84048776\cdot 10^{-5} & 2.70973924\cdot 10^{-8}& -1.09997166\cdot 10^{-4}\\[0.3cm]
 3.18271006\cdot 10^{-1} &-1.09997166\cdot 10^{-4}&  4.46514421\cdot 10^{-1}
     \end{pmatrix} \\[0.3cm]
\left[\boldsymbol{A}^{(H)}_{\beta}\right]_A &= \begin{pmatrix}
2.27833989\cdot 10^{-1}& 1.79852270\cdot 10^{-4}& 3.23714906\cdot 10^{-1}\\[0.3cm]
1.79852270\cdot 10^{-4}& 1.41975476\cdot 10^{-7}& 2.55540716\cdot 10^{-4}\\[0.3cm]
3.23714906\cdot 10^{-1}& 2.55540716\cdot 10^{-4}& 4.59946037\cdot 10^{-1}
     \end{pmatrix},  \label{eq:app:A_OH_4}
\end{align}
where the subscript \emph{A} indicates the active space part of the  $\boldsymbol{A}^{(K)}_{\sigma}$  matrices.  Figure \ref{fig:app:OH_MOs} shows the oo-qubit-ADAPT(3,3) spin-orbitals used to compute the $\boldsymbol{A}_\sigma$ matrices \eqref{eq:app:A_OH_1}-\eqref{eq:app:A_OH_4}. The
molecular orbital diagram is to be expected  for OH$^{\bullet}$ from a (3,3) active space calculation. The lowest energy orbital is the bonding sigma orbital, $\sigma_{1s-2p_z}$, formed by a $1s$ orbital of hydrogen and a $2p_z$ orbital of oxygen. Next is the $2p_x$ orbital of oxygen, and finally the highest energy orbital is the anti-bonding sigma orbital, $\sigma^*_{1s-2p_z}$.

\clearpage

\subsection{Nitric oxide}
\label{sec:app:spin_orbs_no}
\begin{figure}[t]
\centering  
\includegraphics[width=0.3\textwidth]{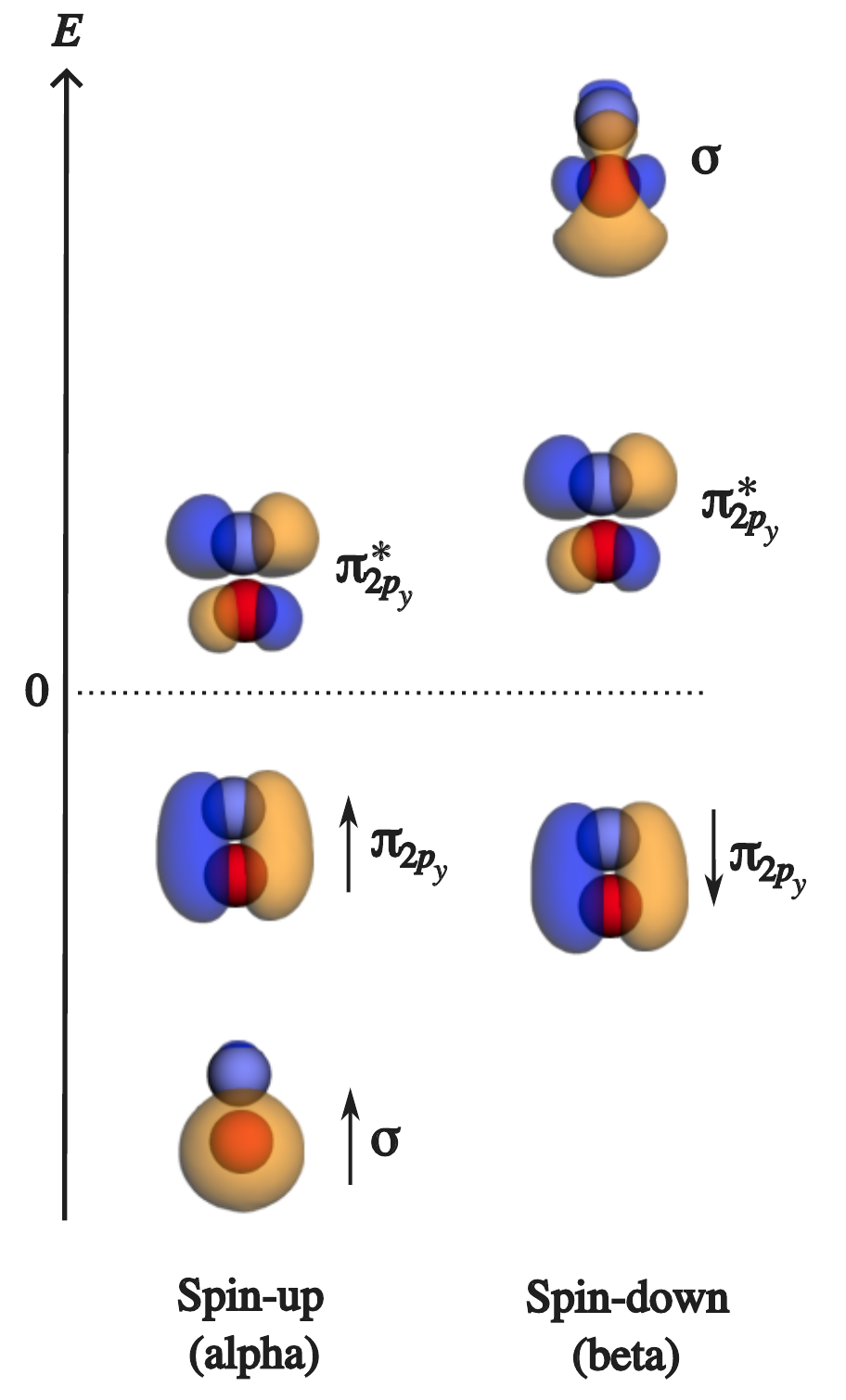}
\caption{Spin-orbitals for the nitric oxide molecule in an active space
of three orbitals and three electrons, with an experimental internuclear distance of 1.1508 $\text{\AA}$, employing the 6-311++G**-J basis set~\cite{kjaer_pople_2011}. The orbitals were obtained using a oo-qubit-ADAPT(3,3) state-vector simulator~\cite{aurora_2023}. The spin-orbital energies are  $(-1.325911,  -0.644536 ,    0.136797)E_H$ and  $( -0.654843,  0.195640,  1.917405)E_H$ for the $\alpha$ and $\beta$  spin-orbitals, respectively, which correspond to the diagonal elements of the Fock matrix in the oo-qubit-ADAPT(3,3) spin-orbital basis. The isovalue is set to $\pm 0.05$, where the blue and orange indicate the positive and negative sign of the orbitals, respectively.  }
\label{fig:app:NO_MOs}
\end{figure}

The $\boldsymbol{A}^{(K)}_{\sigma}$  matrices for the NO$^{\bullet}$ molecule are computed using the molecular spin-orbitals obtained  from a oo-qubit-ADAPT(3,3) state-vector simulator~\cite{aurora_2023}, starting from the UMP2 natural orbitals, but rotated by multiplying with the block-diagonal rotation matrix diag$(R_y(\theta),\hdots,R_y(\theta))$ setting $\theta=\pi/8$. The resulting qubit-ADAPT circuit is shown in Figure \ref{fig:app:NO_circuit}. The calculation was performed at the experimental internuclear distance of 1.1508 $\text{\AA}$ using the 6-311++G**-J basis set~\cite{kjaer_pople_2011}. The oo-qubit-ADAPT(3,3) energy converged to $-129.31953239$ $E_H$. The resulting matrices are 
\begin{align}
\left[\boldsymbol{A}^{(O)}_{\alpha}\right]_A &= \begin{pmatrix}
5.58658180\cdot 10^{-5} & -1.65263012\cdot 10^{-2} & 2.76627739\cdot 10^{-5}\\[0.3cm]
-1.65263012\cdot 10^{-2}&  4.88883257
& -8.18323887\cdot 10^{-3}\\[0.3cm]
2.76627739\cdot 10^{-5} &-8.18323887\cdot 10^{-3}  &1.36976257\cdot 10^{-5}
     \end{pmatrix} \label{eq:app:A_NO_1} \\[0.3cm]
\left[\boldsymbol{A}^{(O)}_{\beta}\right]_A &= \begin{pmatrix}
7.45097744\cdot 10^{-7}&  1.22090554\cdot 10^{-3}& -4.40679659\cdot 10^{-6}\\[0.3cm]
1.22090554\cdot 10^{-3}&  2.00055677& -7.22090815\cdot 10^{-3}\\[0.3cm]
-4.40679659\cdot 10^{-6}& -7.22090815\cdot 10^{-3}&  2.60635016\cdot 10^{-5}
     \end{pmatrix} \\[0.3cm]
\left[\boldsymbol{A}^{(N)}_{\alpha}\right]_A &= \begin{pmatrix}
1.64938828\cdot 10^{-7}& -3.60797431\cdot 10^{-5}&  8.14245987\cdot 10^{-8}\\[0.3cm]
-3.60797431\cdot 10^{-5}&  7.89230697\cdot 10^{-3}& -1.78113222\cdot 10^{-5}\\[0.3cm]
8.14245987\cdot 10^{-8}& -1.78113222\cdot 10^{-5}&  4.01965101\cdot 10^{-8}
     \end{pmatrix} \label{eq:app:A_NO_3} \\[0.3cm]
\left[\boldsymbol{A}^{(N)}_{\beta}\right]_A &= \begin{pmatrix}
1.40783480\cdot 10^{-7}&  1.87674238\cdot 10^{-4}& -6.49108572\cdot 10^{-7}\\[0.3cm]
1.87674238\cdot 10^{-4}&  2.50182901\cdot 10^{-1}& -8.65307181\cdot 10^{-4}\\[0.3cm]
-6.49108572\cdot 10^{-7}& -8.65307181\cdot 10^{-4}&  2.99283650\cdot 10^{-6}
     \end{pmatrix} \label{eq:app:A_NO_4} 
\end{align}
where the subscript \emph{A} indicates the active space part of the  $\boldsymbol{A}^{(K)}_{\sigma}$  matrices. Figure \ref{fig:app:NO_MOs} shows the oo-qubit-ADAPT(3,3) spin-orbitals used to compute the  $\boldsymbol{A}^{(K)}_{\sigma}$  matrices \eqref{eq:app:A_NO_1}-\eqref{eq:app:A_NO_4}. 
The lowest energy $\alpha$-spin orbital is a bonding sigma orbital, which seems to be formed by a $2s$ orbital of oxygen and a $2p_z$ orbital of nitrogen.  Next is the $\pi$ bonding, $\pi_{2p_y}$, and anti-bonding, $\pi^*_{2p_y}$, orbitals. The highest energy spin-orbital, denoted as a bonding sigma orbital, resembles that of a $\sigma_{dz^2-2p_z}$ sigma bonding orbital. This molecular diagram appears somewhat unexpected when compared to the textbook NO$^{\bullet}$ molecular orbital diagram. In an active space consisting of three electrons in three orbitals, we would expect seeing the bonding sigma orbital, along with the two anti-bonding $\pi^*_{2p_x}$ and $\pi^*_{2p_y}$ orbitals. The reason the orbitals seem unusual is due to the small active space.

Note that the spin-orbitals in the $\boldsymbol{A}^{(K)}_{\sigma}$  matrices are not ordered by increasing energy (as in Figure \ref{fig:app:NO_MOs}). The first, second, and third row in the  $\boldsymbol{A}^{(K)}_{\alpha}$ matrices correspond to the $\pi_{2p_y}$, $\sigma_{2s-2p_z}$, and $\pi^*_{2p_y}$ spin-orbitals, respectively, and the first, second, and third row in the  $\boldsymbol{A}^{(K)}_{\beta}$ matrices correspond to the $\pi_{2p_y}$, $\sigma_{dz^2-2p_z}$, and $\pi^*_{2p_y}$ spin-orbitals, respectively.

\subsection{Hydroxyl cation}
\label{sec:app:spin_orbs_oh+}
\begin{figure}[t]
\centering  
\includegraphics[width=0.4\textwidth]{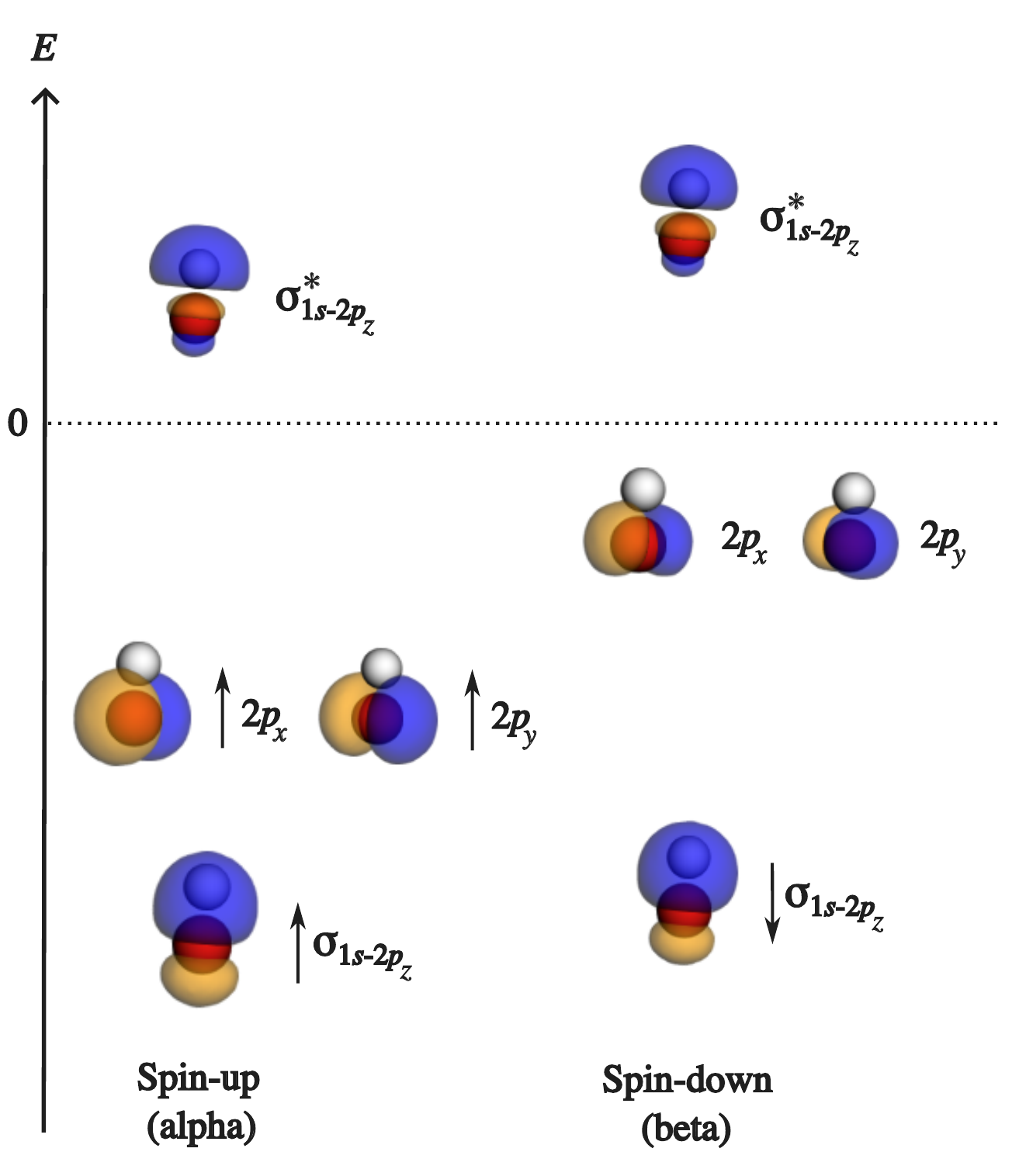}
\caption{Spin-orbitals for the hydroxyl cation molecule in an active space of four orbitals and four electrons, with an experimental internuclear distance of 1.0289 $\text{\AA}$, employing the 6-311++G**-J basis set~\cite{kjaer_pople_2011}. The orbitals were obtained using a oo-qubit-ADAPT(4,4) state-vector simulator~\cite{aurora_2023}. The spin-orbital energies are  $( -1.359513, -1.214864,   -1.214864 ,  0.239389)E_H$ and  $( -1.300019,  -0.003197,  -0.003142,  0.332230)E_H$ for the $\alpha$ and $\beta$  spin-orbitals, respectively, which correspond to the diagonal elements of the Fock matrix in the oo-qubit-ADAPT(4,4) spin-orbital basis. The isovalue is set to $\pm 0.05$, where the blue and orange indicate the positive and negative sign of the orbitals, respectively.  }
\label{fig:app:OH+_MOs}
\end{figure}

The $\boldsymbol{A}^{(K)}_{\sigma}$  matrices for the OH$^{+}$ molecule are computed using the molecular spin-orbitals obtained from a oo-qubit-ADAPT(4,4) state-vector simulator~\cite{aurora_2023}, starting from the unrestricted Hartree-Fock orbitals. The resulting qubit-ADAPT circuit is shown in Figure \ref{fig:app:OH+_circuit}. The calculation was performed with the experimental internuclear distance of 1.0289 $\text{\AA}$ using the 6-311++G**-J basis set~\cite{kjaer_pople_2011}. The oo-qubit-ADAPT(4,4) energy converged to $-75.0178478249$ $E_H$. The resulting matrices are 
\begin{align}
\left[\boldsymbol{A}^{(O)}_{\alpha}\right]_A &= \begin{pmatrix}
1.659290 \cdot 10^{-10} & 1.20077 \cdot 10^{-10}& -8.510219 \cdot 10^{-6} & 1.477859 \cdot 10^{-5}\\[0.3em]
 1.200771 \cdot 10^{-10}&  8.689561 \cdot 10^{-11} &-6.158551 \cdot 10^{-6} &1.069475 \cdot 10^{-5}\\[0.3em]
-8.510219 \cdot 10^{-6}& -6.158551 \cdot 10^{-6} & 4.364749 \cdot 10^{-1} &-7.579690 \cdot 10^{-1}\\[0.3em]
1.477859 \cdot 10^{-5} & 1.069475 \cdot 10^{-5} &-7.579690 \cdot 10^{-1} & 1.316265
     \end{pmatrix}  \label{eq:app:A_OH+_1} \\[0.3cm]
\left[\boldsymbol{A}^{(O)}_{\beta}\right]_A &= \begin{pmatrix}
5.761173 \cdot 10^{-1} & -2.918772 \cdot 10^{-6}& -1.226419 \cdot 10^{-5} & -9.979111 \cdot 10^{-1}\\[0.3em]
-2.918772 \cdot 10^{-6}&  1.478732 \cdot 10^{-11} &6.213381 \cdot 10^{-11} &5.055697 \cdot 10^{-6}\\[0.3em]
-1.226419\cdot 10^{-5}& 6.213381 \cdot 10^{-11} & 2.610758 \cdot 10^{-10} & 2.124319 \cdot 10^{-5}\\[0.3em]
-9.979111 \cdot 10^{-1} & 5.055697 \cdot 10^{-6} &2.124319 \cdot 10^{-5} & 1.728513
     \end{pmatrix}  \\[0.3cm]
\left[\boldsymbol{A}^{(H)}_{\alpha}\right]_A &= \begin{pmatrix}
2.117710\cdot 10^{-11} &-4.052889\cdot 10^{-12}&  1.780855\cdot 10^{-6}  &3.310612\cdot 10^{-6}\\[0.3cm]
-4.052889\cdot 10^{-12} & 7.756446\cdot 10^{-13} &-3.408212\cdot 10^{-7} &-6.335872\cdot 10^{-7}\\[0.3cm]
 1.780855\cdot 10^{-6} &-3.408212\cdot 10^{-7} & 1.497581\cdot 10^{-1}  &2.784006\cdot 10^{-1} \\[0.3cm]
 3.310612\cdot 10^{-6}& -6.335872\cdot 10^{-7} & 2.784006\cdot 10^{-1}  &5.175472\cdot 10^{-1}
     \end{pmatrix} \\[0.3cm]
\left[\boldsymbol{A}^{(H)}_{\beta}\right]_A &= \begin{pmatrix}
2.0161083\cdot 10^{-1} & 1.8993027\cdot 10^{-6} & 9.1697254\cdot 10^{-7} & 3.1613815\cdot 10^{-1}\\[0.3cm]
1.8993027\cdot 10^{-6} & 1.7892644\cdot 10^{-11}  & 8.6384666\cdot 10^{-12} & 2.9782232\cdot 10^{-6}\\[0.3cm]
9.1697254\cdot 10^{-7}  & 8.6384666\cdot 10^{-12} & 4.1706025\cdot 10^{-12}  & 1.4378692\cdot 10^{-6}\\[0.3cm]
3.1613815\cdot 10^{-1} & 2.9782232\cdot 10^{-6} & 1.4378692\cdot 10^{-6} & 4.9572402\cdot 10^{-1}
     \end{pmatrix},  
     \label{eq:app:A_OH+_4}
\end{align}
where the subscript \emph{A} indicates the active space part of the  $\boldsymbol{A}^{(K)}_{\sigma}$  matrices. Figure \ref{fig:app:OH+_MOs} shows the oo-qubit-ADAPT(4,4) spin-orbitals used to compute the  $\boldsymbol{A}^{(K)}_{\sigma}$  matrices \eqref{eq:app:A_OH+_1}-\eqref{eq:app:A_OH+_4}. The molecular orbital diagram is to be expected for OH$^{+}$ from a (4,4) active space calculation. The molecular orbitals consist of the bonding sigma orbital $\sigma_{1s-2p_z}$ formed by a $1s$ orbital from the hydrogen and a $2p_z$ orbital from the oxygen. The next two orbitals are the $2p_x$ and $2p_y$ orbitals from oxygen. The highest energy orbital is the anti-bonding sigma orbital $\sigma^*_{1s-2p_z}$. 

Note that the spin-orbitals in the $\boldsymbol{A}^{(K)}_{\alpha}$  matrix are not ordered by increasing energy (as in Figure \ref{fig:app:OH+_MOs}). The first, second, third, and fourth row in the  $\boldsymbol{A}^{(K)}_{\alpha}$ matrices correspond to the $2p_x$, $2p_y$, $\sigma_{2s-2p_z}$, and $\sigma^*_{2s-2p_z}$ spin-orbitals, respectively. The spin-orbitals in the $\boldsymbol{A}^{(K)}_{\beta}$ is ordered by increasing energy.

\section{Circuits}
\label{subsec:app:circuits}

The active part of the wave function $\ket{A(\vec{\theta})}$ (Eq.~\eqref{eq:active_space_wf} in the main text) is obtained from the qubit-ADAPT method~\cite{tang_qubit_ADAPT_VQE_2021} which approximates the zeroth-order wave function $\ket{\Psi^{(0)}_0}$. In qubit-ADAPT, the fermionic generators are the single and double excitation operators, and in the  Jordan-Wigner mapping  representation read~\cite{romero_strategies_2019} 
\begin{align}
\hat{a}^\dagger_a \hat{a}_i - \text{h.c.} =& \frac{i}{2} \bigotimes^{a-1}_{k=i+1} Z_k \left( Y_i X_a - X_i Y_a \right) \label{eq:app:one_elec_jw} \\[0.3cm]
\hat{a}^\dagger_b \hat{a}^\dagger_a \hat{a}_j   \hat{a}_i - \text{h.c.} =& \frac{i}{8} \bigotimes^{j-1}_{k=i+1} Z_k \bigotimes^{b-1}_{l=a+1}Z_l \left( X_i X_j Y_a X_b +  Y_i X_j Y_a Y_b +  X_i Y_j Y_a Y_b  + X_i X_j X_a Y_b  \right. \nonumber \\
&\left.- Y_i X_j X_a X_b - X_i Y_j X_a X_b - Y_i Y_j Y_a X_b  - Y_i Y_j X_a Y_b  \right) \label{eq:app:two_elec_jw} 
\end{align}
where $b>a>j>i$. We then break down the  fermionic operators after the Jordan-Wigner mapping and choose the individual Pauli strings given in Eqs. \eqref{eq:app:one_elec_jw} and  \eqref{eq:app:two_elec_jw} as the operator pool. Note that we remove the Pauli-\emph{Z} chains from the operators, thus the size of the Pauli-strings are either 2 or 4 depending if the Pauli-string is from a single or double excitation operator, respectively. The qubit-ADAPT methods may potentially break particle number, resulting in nonphysical states with different particle number in the determinants, however, setting a strict convergence criterion ensures that the resulting optimized state will be a $N_e$-particle state. The reason is the Hamiltonian does not couple states with different particle numbers, and thus the lowest energy cannot be a mix of particle numbers.

The circuit representation of the exponential operators obtained from the single and double excitation operator pool are shown in Figs. \ref{fig:app:circuit_blocks_one_body_exci} and \ref{fig:app:circuit_blocks_two_body_exci}. These are the circuit building blocks used in qubit-ADAPT.

\begin{figure}[H]
\centering
\begin{tikzpicture}
\node[scale=0.8] {
\begin{quantikz}[row sep={0.9cm,between origins}]
\lstick{$\ket{q_1}$}&\gate[2]{\exp(-i \frac{\theta}{2}\sigma_1 \sigma_2)}  & \\
\lstick{$\ket{q_2}$}& \qw &
\end{quantikz} = \begin{quantikz}[row sep={0.9cm,between origins}]
\lstick{$\ket{q_1}$}& \gate[1]{R_1} &  \qw &  \targ{} &  \gate{R_z(\theta)} &  \targ{} & \qw &  \gate[1]{R^\dagger_1} &   \qw  \\
\lstick{$\ket{q_2}$}& \gate[1]{R_2} &\qw&  \ctrl{-1}&  \qw &  \ctrl{-1} &   \qw&  \gate[1]{R^\dagger_2}&   \qw 
\end{quantikz} };
\end{tikzpicture} 
\caption{The circuit representation of the  exponential operators obtained from the single excitation operator pool in Eq. \eqref{eq:app:one_elec_jw}. Here, $\sigma_i \in \{X,Y\}$ and $R_i = \{H, HS^\dagger\}$. If  $\sigma_i = X$ then $R_i=H$ and if $\sigma_i = Y$ then   $R_i=HS^\dagger$.}
\label{fig:app:circuit_blocks_one_body_exci}
\end{figure}
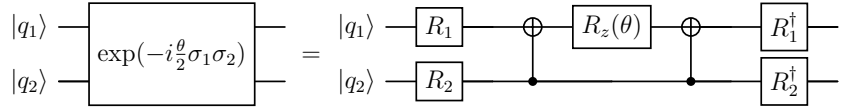
\begin{figure}[H]
\centering
\begin{tikzpicture}
\node[scale=0.8] {
\begin{quantikz}[row sep={0.9cm,between origins}]
\lstick{$\ket{q_1}$}&\gate[4]{\exp(-i \frac{\theta}{2}\sigma_1 \sigma_2 \sigma_3 \sigma_4)}  & \\
\lstick{$\ket{q_2}$}& \qw &\\
\lstick{$\ket{q_3}$}& \qw &\\
\lstick{$\ket{q_4}$}& \qw &
\end{quantikz} = \begin{quantikz}[row sep={0.9cm,between origins}]
\lstick{$\ket{q_1}$}& \gate[1]{R_1} &  \qw &  \qw &  \targ{} &  \gate{R_z(\theta)} &  \targ{} & \qw & \qw & \gate[1]{R^\dagger_1} &   \qw  \\
\lstick{$\ket{q_2}$}& \gate[1]{R_2} &\qw& \targ{}&  \ctrl{-1}&  \qw &  \ctrl{-1} & \targ{}&  \qw&  \gate[1]{R^\dagger_2}&   \qw \\
\lstick{$\ket{q_3}$}& \gate[1]{R_3} & \targ{} & \ctrl{-1}& \qw&  \qw &  \qw & \ctrl{-1}&  \targ{} &  \gate[1]{R^\dagger_3} &   \qw  \\
\lstick{$\ket{q_4}$}& \gate[1]{R_4} &\ctrl{-1} & \qw &   \qw &  \qw &  \qw & \qw &  \ctrl{-1} & \gate[1]{R^\dagger_4}&   \qw 
\end{quantikz} };
\end{tikzpicture} 
\caption{The circuit representation of the exponential operators obtained from the double excitation operator pool in Eq. \eqref{eq:app:two_elec_jw}. Here, $\sigma_i \in \{X,Y\}$ and $R_i = \{H, HS^\dagger\}$. If  $\sigma_i = X$ then $R_i=H$ and if $\sigma_i = Y$ then   $R_i=HS^\dagger$.}
\label{fig:app:circuit_blocks_two_body_exci}
\end{figure}
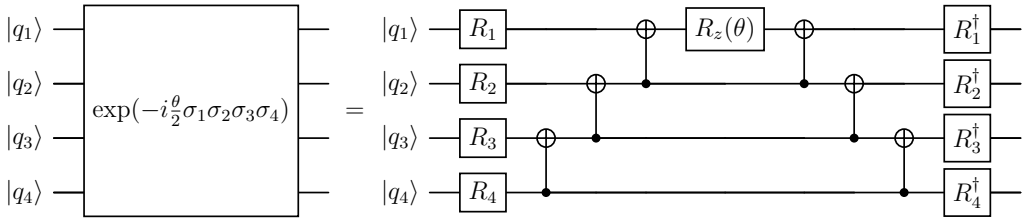


\subsection{Hydroxyl radical}

Figure \ref{fig:app:OH_circuit} shows the qubit-ADAPT(3,3) circuit for the OH$^{\bullet}$ molecule that approximates the zeroth-order wave function $\ket{\Psi^{(0)}_0}$, with the experimental internuclear distance of 0.9697 $\text{\AA}$ employing the 6-311++G**-J basis~\cite{kjaer_pople_2011}. The circuit and parameters were determined using a state-vector simulator \emph{Aurora}~\cite{aurora_2023}.

\begin{figure}[H]
\centering
\begin{tikzpicture}
\node[scale=0.7] {
\begin{quantikz}[row sep={0.7cm,between origins}]
\lstick{$\ket{0}$}& \qwbundle{6}  & \gate{U_{\text{HF}}} &  \gate[1]{\text{exp}(-i \frac{\theta_0 }{2}X_0X_2Y_3X_5)} & \gate[1,style={inner xsep=4pt}]{\text{exp}(-i\frac{\theta_1 }{2} X_0Y_2Y_3Y_5)} & \gate[1,style={inner xsep=3pt}]{\text{exp}(-i\frac{\theta_2 }{2} X_0Y_2Y_3Y_5)} & \gate[1]{\text{exp}(-i\frac{\theta_3 }{2} X_1X_2Y_3X_4)}  & \qw \\[0.5cm]
& \qw & \gate[1]{\text{exp}(-i \frac{\theta_4 }{2}X_3Y_5)} & \gate[1,style={inner xsep=14pt}]{\text{exp}(-i\frac{\theta_5 }{2} X_0 Y_2)} & \gate[1,style={inner xsep=4pt}]{\text{exp}(-i\frac{\theta_6 }{2} Y_0Y_1X_4Y_5)} & \gate[1]{\text{exp}(-i\frac{\theta_7 }{2} X_1Y_2X_4X_5)} & \gate[1]{\text{exp}(-i\frac{\theta_8 }{2} X_0X_1X_3Y_4)} & \qw \\[0.5cm]
& \qw & \gate[1]{\text{exp}(-i \frac{\theta_9 }{2}X_3Y_5)} & \gate[1,style={inner xsep=12pt}]{\text{exp}(-i\frac{\theta_{10} }{2} X_3Y_4)} & \gate[1]{\text{exp}(-i\frac{\theta_{11} }{2} X_1X_2X_3Y_5)} & \qw
\end{quantikz} };
\end{tikzpicture} 
\caption{The qubit-ADAPT(3,3) circuit for the hydroxyl radical (OH$^{\bullet}$) molecule.  The circuit representation of the exponential operators are shown in Figs. \ref{fig:app:circuit_blocks_one_body_exci} and \ref{fig:app:circuit_blocks_two_body_exci}, and $U_{\text{HF}}=X_0X_1X_3$ prepares the   Hartree-Fock state. The circuit parameters are given in \eqref{eq:app:cir_params_OH}.
}
\label{fig:app:OH_circuit}
\end{figure}

\begin{figure}[ht]
\centering  
\includegraphics[width=0.8\textwidth]{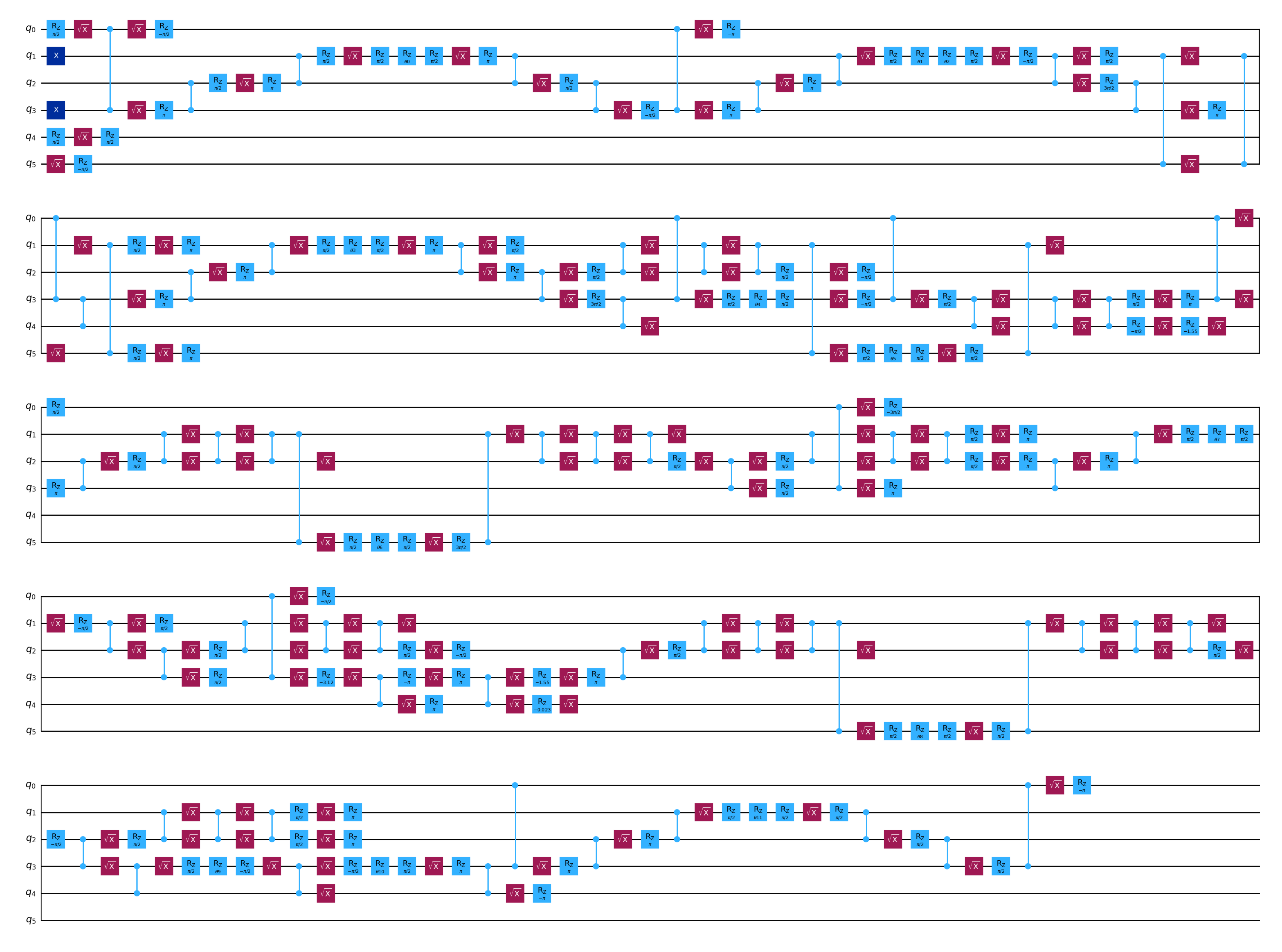}
\caption{ The transpiled circuit for the hydroxyl radical (OH$^{\bullet}$) molecule executed on IBM’s Torino device for the qubit-ADAPT circuit shown in Figure \ref{fig:app:OH_circuit}. The circuit parameters are given in \eqref{eq:app:cir_params_OH}.  }
\label{fig:app:OH_transpiled_circuit}
\end{figure}

The optimized circuit parameters in \ref{fig:app:OH_circuit} are given by
\begin{align}
\begin{tabular}{ lllll } 
  $\theta_0 =0.20271028$ & $\theta_1 = 2.097688\cdot 10^{-2} $  & $\theta_2 =  -2.781581 \cdot 10^{-2}$&  $\theta_3 =6.78393 \cdot 10^{-2}$  \\ 
$\theta_4 =-3.885881 \cdot 10^{-2}$  &$\theta_5 = -1.998574 \cdot 10^{-2}$ & $\theta_6 =  7.12922\cdot 10^{-3}$  & $\theta_7 =  1.2466\cdot 10^{-3}$  \\ 
$\theta_8 =-5.3797 \cdot 10^{-4}$  &$\theta_9 = 3.7292  \cdot 10^{-4}$ & $\theta_{10} =  -2.6724\cdot 10^{-4}$  & $\theta_{11} =  -5.3667764\cdot 10^{-5}$.  
\end{tabular} \label{eq:app:cir_params_OH}
\end{align}
The state-vector, generated from circuit \ref{fig:app:OH_circuit}, is 
\begin{align}
\left|\Psi^{(0)}_0 (\vec{\theta}_{\text{opt}})\right> =&0.993734
\left|110100\right>    
+0.104570
\left|011001\right>   
+1.82855\cdot 10^{-2}
\left|110001\right>   \nonumber \\  
+&3.40835\cdot 10^{-2}
\left|101010\right> 
+7.89272\cdot 10^{-3}
\left|011100\right>   
+3.18078\cdot 10^{-4}
\left|110010\right>  \nonumber \\   
+&2.66658\cdot 10^{-5}
\left|101001\right> 
-1.14002\cdot 10^{-5}
\left|101100\right>, \label{eq:app_OH_statevector}
\end{align}
with the notation $\ket{\alpha_{\sigma_{2s-2p_z}},\alpha_{2p_x},\alpha_{\sigma^*_{2s-2p_z}},\beta_{\sigma_{2s-2p_z}},\beta_{2p_x},\beta_{\sigma^*_{2s-2p_z}}}$ (see Section \ref{sec:app:spin_orbs_OH}). The energy is

\begin{align}
\braket{\Psi^{(0)}_0 (\vec{\theta}_{\text{opt}})|\hat{H}^{(0)}|\Psi^{(0)}_0 (\vec{\theta}_{\text{opt}})} &=-75.43387721 E_H.
\end{align}
The circuit that was transpiled and executed on IBM’s Torino device is shown in Figure \ref{fig:app:OH_transpiled_circuit}.  Entanglement is achieved through the two-qubit CZ gate and one-qubit operations are implemented using $R_z(\theta)$, $\sqrt{X}$, $X$ gates. In total, the transpiled circuit consists of 77 CZ gates, 136 $\sqrt{X}$ gates, 118 $R_z(\theta)$ gates, and 2 $X$ gates. 

\subsection{Nitric oxide}
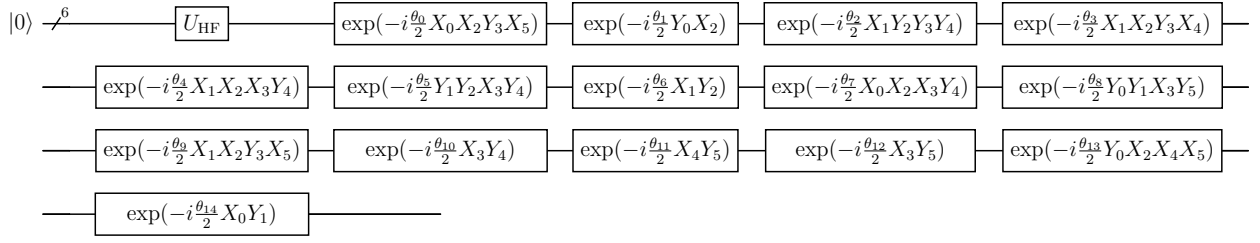
\begin{figure}[H]
\centering
\begin{tikzpicture}
\node[scale=0.7] {
\begin{quantikz}[row sep={0.7cm,between origins}]
\lstick{$\ket{0}$}& \qwbundle{6}  & \gate{U_\text{HF}}&  \gate[1]{\text{exp}(-i \frac{\theta_0 }{2}X_0X_2Y_3X_5)} & \gate[1,style={inner xsep=2pt}]{\text{exp}(-i\frac{\theta_1 }{2} Y_0X_2)} & \gate[1,style={inner xsep=3pt}]{\text{exp}(-i\frac{\theta_2 }{2} X_1Y_2Y_3Y_4)} & \gate[1,style={inner xsep=2pt}]{\text{exp}(-i\frac{\theta_3 }{2} X_1X_2Y_3X_4)} &  \qw \\[0.5cm]
& \qw& \gate[1]{\text{exp}(-i\frac{\theta_4 }{2} X_1X_2X_3Y_4)}&\gate[1,style={inner xsep=3pt}]{\text{exp}(-i \frac{\theta_5 }{2}Y_1Y_2X_3Y_4)} & \gate[1,style={inner xsep=2pt}]{\text{exp}(-i\frac{\theta_6 }{2} X_1Y_2)} & \gate[1]{\text{exp}(-i\frac{\theta_7 }{2} X_0X_2X_3Y_4)}& \gate[1,style={inner xsep=5pt}]{\text{exp}(-i\frac{\theta_8 }{2} Y_0Y_1X_3Y_5)}&\qw  \\[0.5cm]
& \qw&  \gate[1]{\text{exp}(-i\frac{\theta_9 }{2} X_1X_2Y_3X_5)}  &   \gate[1,style={inner xsep=13pt}]{\text{exp}(-i\frac{\theta_{10} }{2} X_3 Y_4)}  & \gate[1]{\text{exp}(-i\frac{\theta_{11} }{2} X_4 Y_5)} &  \gate[1,style={inner xsep=12pt}]{\text{exp}(-i\frac{\theta_{12} }{2} X_3Y_5)}& \gate[1]{\text{exp}(-i\frac{\theta_{13} }{2} Y_0 X_2 X_4 X_5)}&  \qw\\[0.5cm]
& \qw& \gate[1,style={inner xsep=13pt}]{\text{exp}(-i\frac{\theta_{14} }{2} X_0 Y_1)}&  \qw
\end{quantikz} };
\end{tikzpicture} 
\caption{The  qubit-ADAPT(3,3) circuit for the nitric oxide (NO$^{\bullet}$) molecule. The circuit representation of the exponential operators are
shown in Figs. \ref{fig:app:circuit_blocks_one_body_exci} and \ref{fig:app:circuit_blocks_two_body_exci}, and $U_{\text{HF}}=X_0X_1X_3$ prepares the   Hartree-Fock state.
}
\label{fig:app:NO_circuit}
\end{figure}

Figure \ref{fig:app:NO_circuit} shows the qubit-ADAPT(3,3) circuit for the NO$^{\bullet}$ molecule that approximates the zeroth-order wave function $\ket{\Psi^{(0)}_0}$, with the experimental
internuclear distance of 1.1508 $\text{\AA}$ employing the 6-311++G**-J basis~\cite{kjaer_pople_2011}. The circuit and parameters were determined using a state-vector simulator Aurora~\cite{aurora_2023}.

\begin{figure}[ht]
\centering  
\includegraphics[width=0.8\textwidth]{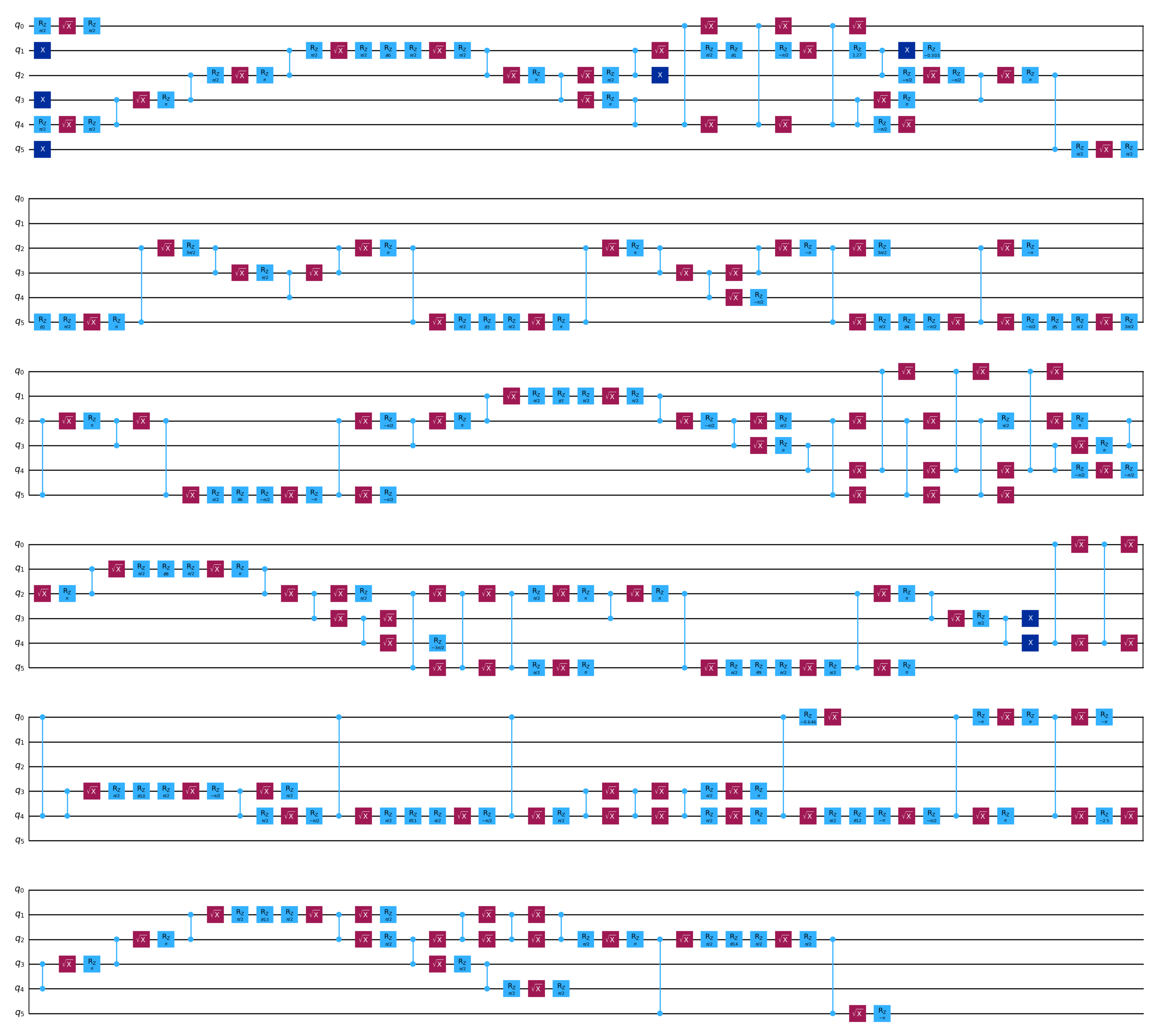}
\caption{ The transpiled circuit for the nitric oxide (NO$^{\bullet}$)  molecule executed on IBM’s Torino device for the qubit-ADAPT(3,3) circuit shown in Figure \ref{fig:app:NO_circuit}. The circuit parameters are given in \eqref{eq:app:cir_params_NO}.  }
\label{fig:app:NO_transpiled_circuit}
\end{figure}

The optimized circuit parameters in \ref{fig:app:NO_circuit} are given by
\begin{align}
\begin{tabular}{ lllll } 
  $\theta_0 = 0.422643$ & $\theta_1 = -0.14015686$  & $\theta_2 =1.486188  \cdot 10^{-2}$& $\theta_3 = -1.485007 \cdot 10^{-2}$  \\ 
$\theta_4 = 1.559565\cdot 10^{-2}$ & $\theta_5 = 1.560275\cdot 10^{-2}$ & $\theta_6 =2.84955\cdot 10^{-3}$ & $\theta_7 =-1.31488\cdot 10^{-3}$ \\ 
 $\theta_8 =-6.8496 \cdot 10^{-4}$& $\theta_9 =-1.13918 \cdot 10^{-3}$ & $\theta_{10} =8.4636 \cdot 10^{-4}$ & $\theta_{11} =2.4635 \cdot 10^{-4}$  \\ 
 $\theta_{12} =2.86484\cdot 10^{-3}$&  $\theta_{13} =1.6687\cdot 10^{-4}$&  $\theta_{14} = -9.45245626\cdot 10^{-5}$. 
\end{tabular} \label{eq:app:cir_params_NO}
\end{align}
The state-vector, generated from circuit \ref{fig:app:NO_circuit}, is 
\begin{align}
\left|\Psi^{(0)}_0 (\vec{\theta}_{\text{opt}})\right> =&0.974921 \left|110100\right> 
+0.209334
\left|011001\right> 
-6.81633\cdot 10^{-2}
\left|011100\right> \nonumber \\
-&2.96996\cdot 10^{-2}
\left|101010\right> 
+1.32915\cdot 10^{-2}
\left|110001\right> 
-1.46251\cdot 10^{-3}
\left|101100\right> \nonumber \\
+&6.95278\cdot 10^{-4}
\left|011010\right> 
-6.03809\cdot 10^{-4}
\left|101001\right> 
-5.15543\cdot 10^{-4}
\left|110010\right>, 
\end{align}
with the notation $\ket{\alpha_{\pi_{2\pi_y}},\alpha_{\sigma_{2s-2p_z}},\alpha_{\pi^*_{2\pi_y}},\beta_{\pi_{2\pi_y}},\beta_{\sigma_{dz^2-2p_z}},\beta_{\pi^*_{2\pi_y}}}$ (see Section \ref{sec:app:spin_orbs_no}). The energy is

\begin{align}
\braket{\Psi^{(0)}_0 (\vec{\theta}_{\text{opt}})|\hat{H}^{(0)}|\Psi^{(0)}_0 (\vec{\theta}_{\text{opt}})}=-129.31953239 E_H.
\end{align}
The state $\ket{101111}$ was also included in the state-vector but with a small amplitude of $3.24225\cdot 10^{-5}$. Setting a more strict convergence criterion would remove this state entirely. The circuit that was transpiled and executed on IBM’s Torino device is shown in Figure \ref{fig:app:NO_transpiled_circuit}.  In total, the transpiled circuit consists of 78 CZ gates, 127 $\sqrt{X}$ gates, 131 $R_z(\theta)$ gates, and 7 $X$ gates. 

\subsection{Hydroxyl cation}

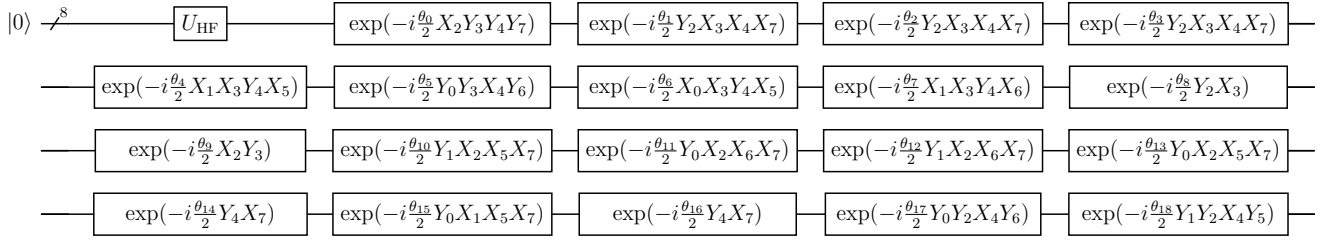
\begin{figure}[H]
\centering
\begin{tikzpicture}
\node[scale=0.7] {
\begin{quantikz}[row sep={0.7cm,between origins}]
\lstick{$\ket{0}$}& \qwbundle{8}  & \gate{U_{\text{HF}}} &  \gate[1,style={inner xsep=4pt}]{\text{exp}(-i \frac{\theta_0 }{2}X_2Y_3Y_4Y_7)} & \gate[1,style={inner xsep=2pt}]{\text{exp}(-i\frac{\theta_1 }{2} Y_2X_3X_4X_7)} & \gate[1,style={inner xsep=2pt}]{\text{exp}(-i\frac{\theta_2 }{2}  Y_2X_3X_4X_7)} & \gate[1,style={inner xsep=2pt}]{\text{exp}(-i\frac{\theta_3 }{2} Y_2X_3X_4X_7)}  & \qw \\[0.5cm]
& \qw& \gate[1]{\text{exp}(-i\frac{\theta_4 }{2} X_1X_3Y_4X_5)} & \gate[1,style={inner xsep=4pt}]{\text{exp}(-i\frac{\theta_5 }{2} Y_0Y_3X_4Y_6)} & \gate[1,style={inner xsep=2pt}]{\text{exp}(-i\frac{\theta_6 }{2} X_0X_3Y_4X_5)} & \gate[1,style={inner xsep=2pt}]{\text{exp}(-i\frac{\theta_7 }{2} X_1X_3Y_4X_6)} & \gate[1,style={inner xsep=16pt}]{\text{exp}(-i\frac{\theta_8 }{2} Y_2X_3)} & \qw \\[0.5cm]
& \qw&   \gate[1,style={inner xsep=14pt}]{\text{exp}(-i\frac{\theta_9 }{2} X_2Y_3)} &\gate[1]{\text{exp}(-i\frac{\theta_{10} }{2} Y_1X_2X_5X_7)}&\gate[1]{\text{exp}(-i\frac{\theta_{11} }{2} Y_0X_2X_6X_7)} &\gate[1]{\text{exp}(-i\frac{\theta_{12} }{2} Y_1X_2X_6X_7)} &\gate[1]{\text{exp}(-i\frac{\theta_{13} }{2} Y_0X_2X_5X_7)}& \qw \\[0.5cm]
& \qw&  \gate[1,style={inner xsep=13pt}]{\text{exp}(-i\frac{\theta_{14} }{2} Y_4X_7)} &\gate[1]{\text{exp}(-i\frac{\theta_{15} }{2} Y_0X_1X_5X_7)} &\gate[1,style={inner xsep=14pt}]{\text{exp}(-i\frac{\theta_{16} }{2} Y_4X_7)} &\gate[1,style={inner xsep=2pt}]{\text{exp}(-i\frac{\theta_{17} }{2} Y_0Y_2X_4Y_6)} &\gate[1,style={inner xsep=3pt}]{\text{exp}(-i\frac{\theta_{18} }{2} Y_1Y_2X_4Y_5)}& \qw 
\end{quantikz} };
\end{tikzpicture} 
\caption{The qubit-ADAPT(4,4) circuit for the hydroxyl cation (OH$^{+}$) molecule.  The circuit representation of the exponential operators are shown in Figs. \ref{fig:app:circuit_blocks_one_body_exci} and \ref{fig:app:circuit_blocks_two_body_exci}, and $U_{\text{HF}}=X_0X_1X_2X_4$ prepares the   Hartree-Fock state. The circuit parameters are given in \eqref{eq:app:cir_params_OH+}.
}
\label{fig:app:OH+_circuit}
\end{figure}

Figure \ref{fig:app:OH+_circuit} shows the qubit-ADAPT(4,4) circuit for the OH$^{+}$ molecule that approximates the zeroth-order wave function $\ket{\Psi^{(0)}_0}$, with the experimental internuclear distance of 1.0289 $\text{\AA}$ employing the 6-311++G**-J basis~\cite{kjaer_pople_2011}. The circuit and parameters were determined using a state-vector simulator \emph{Aurora}~\cite{aurora_2023}.

The optimized circuit parameters in \ref{fig:app:OH+_circuit} are given by
\begin{align}
\begin{tabular}{ lllll } 
  $\theta_0 =-3.43274 \cdot 10^{-2}$ & $\theta_1 =  0.107651 $  & $\theta_2 = 2.23381 \cdot 10^{-2}$& $\theta_3 =3.93659\cdot 10^{-2}$  \\ 
$\theta_4 = 6.48614\cdot 10^{-2}$ & $\theta_5 = 6.47689\cdot 10^{-2}$ & $\theta_6 =3.25836\cdot 10^{-2}$ & $\theta_7 = 3.25376\cdot 10^{-2}$ \\ 
 $\theta_8 =-1.95053 \cdot 10^{-2}$& $\theta_9 =1.95903 \cdot 10^{-2}$ & $\theta_{10} =-6.59998 \cdot 10^{-3}$ & $\theta_{11} =6.59044 \cdot 10^{-3}$  \\ 
 $\theta_{12} =-3.34069\cdot 10^{-3}$&  $\theta_{13} =-3.34468\cdot 10^{-3}$&  $\theta_{14} =2.45060\cdot 10^{-3}$  &$\theta_{15} =1.54535\cdot 10^{-3}$ \\ 
 $\theta_{16} =8.93870\cdot 10^{-4}$ &$\theta_{17} =-1.72548\cdot 10^{-5}$&$\theta_{18} =1.71730\cdot 10^{-5}$.
\end{tabular} \label{eq:app:cir_params_OH+}
\end{align}
The state-vector, generated from circuit \ref{fig:app:OH+_circuit}, is 
\begin{align}
\left|\Psi^{(0)}_0 (\vec{\theta}_{\text{opt}})\right> =&
0.993337 
\left|11101000\right> 
-0.101276
\left|11010001\right> +
3.25532\cdot 10^{-2}
\left|01110010\right> \nonumber \\ 
-&3.25494\cdot 10^{-2} 
\left|10110100\right> +
1.92507\cdot 10^{-2}
\left|11011000\right> 
-1.63637\cdot 10^{-2}
\left|10110010\right> \nonumber \\
-&1.63614\cdot 10^{-2}
\left|01110100\right> +
3.23463\cdot 10^{-4}
\left|11100001\right> +
1.57621\cdot 10^{-4}
\left|00010111\right> \nonumber \\
+& 2.12627\cdot 10^{-5}
\left|00101110\right> +
1.42300\cdot 10^{-5}
\left|01001011\right> 
-1.41685\cdot 10^{-5}
\left|10001101\right> 
\end{align}
with the notation $\ket{\alpha_{2p_x},\alpha_{2p_y},\alpha_{\sigma_{2s-2p_z}},\alpha_{\sigma^*_{2s-2p_z}},\beta_{\sigma_{2s-2p_z}},\beta_{2p_x},\beta_{2p_y},\beta_{\sigma^*_{2s-2p_z}}}$ (see Section \ref{sec:app:spin_orbs_oh+}). The energy is

\begin{align}
\braket{\Psi^{(0)}_0 (\vec{\theta}_{\text{opt}})|\hat{H}^{(0)}|\Psi^{(0)}_0 (\vec{\theta}_{\text{opt}})}=-75.0178478249 E_H.
\end{align}
 The circuit that was transpiled and executed on IBM’s Torino device is shown in Figure \ref{fig:app:NO_transpiled_circuit}.  In total, the transpiled circuit consists of 144 CZ gates, 242 $\sqrt{X}$ gates, 182 $R_z(\theta)$ gates, and 4 $X$ gates. 
\begin{figure}[ht]
\centering  
\includegraphics[width=1.0\textwidth]{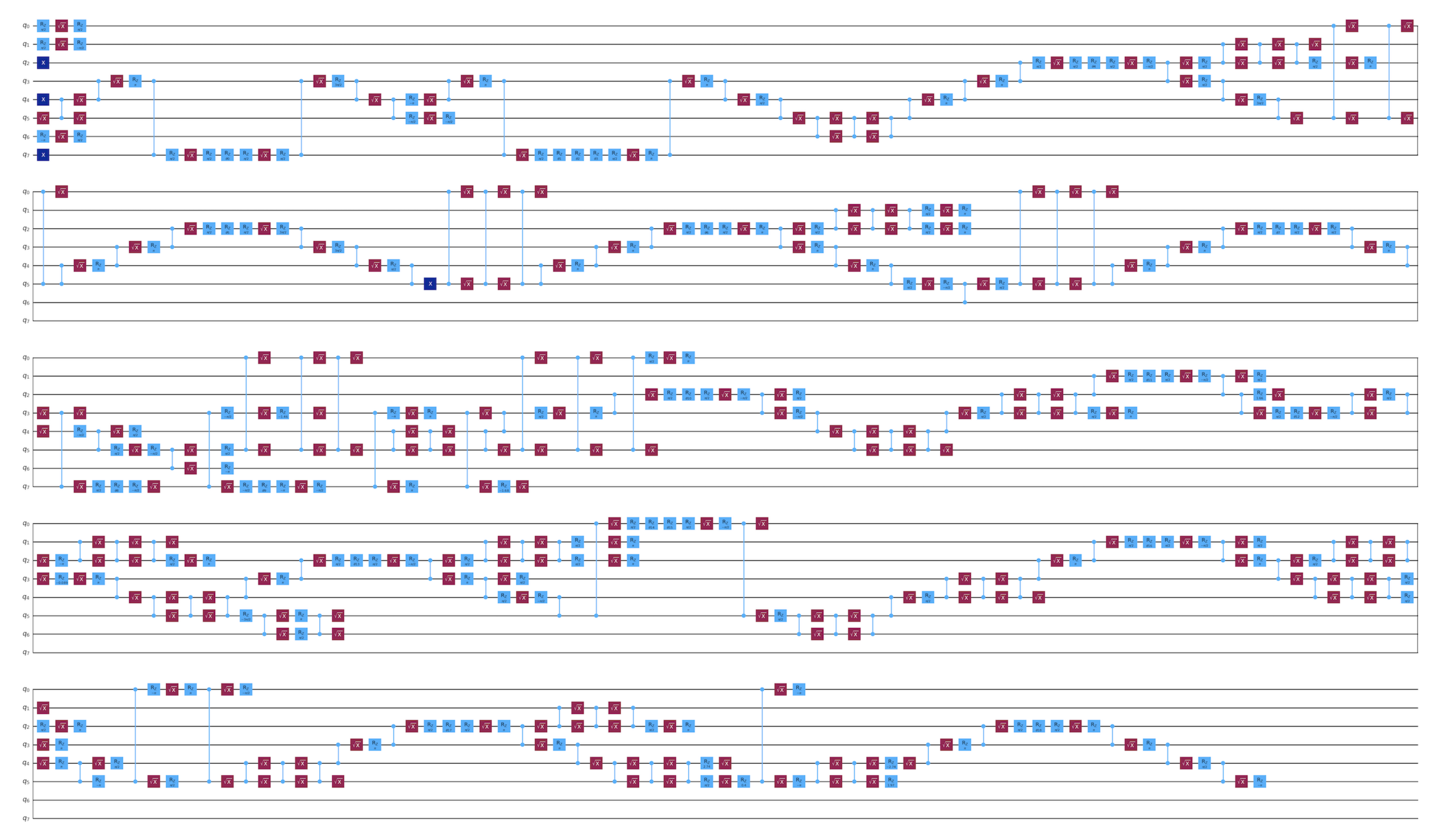}
\caption{ The transpiled circuit for the hydroxyl cation (OH$^{+}$) executed on IBM’s Torino device for the qubit-ADAPT(4,4) circuit shown in Figure \ref{fig:app:OH+_circuit}. The circuit parameters are given in \eqref{eq:app:cir_params_OH+}.  }
\label{fig:app:OH+_transpiled_circuit}
\end{figure}

\section{Hardware calculations}
\label{app:hardware_calculations}

On the quantum hardware, we prepare the circuits shown in Figure \ref{fig:main_circuits} in the main text, where the transpiled circuits are shown in Section \ref{subsec:app:circuits}, and calculate the 1-RDM elements as

\begin{align}
D_{v\sigma, w\sigma}=\left< A (\vec{\theta}_{\text{opt}}) \right|\hat{a}^\dagger_{v\sigma}\hat{a}_{w\sigma} \left|A (\vec{\theta}_{\text{opt}})\right> \label{eq:app_D_psqs_hardware}
\end{align}
where $\ket{A (\vec{\theta}_{\text{opt}})}$, defined in Eq.~\eqref{eq:active_space_wf} in the main text, is prepared using the qubit-ADAPT circuits in Figure \ref{fig:main_circuits}, and the fermionic excitation operators are mapped into qubit operators via the Jordan-Wigner transformation. In section \ref{sec:app:1_rdm}, we show the quantum hardware experiments for computing the 1-RDM elements in Eq. \ref{eq:app_D_psqs_hardware}, which are then used to calculate the HFCs.  In section \ref{sec:app:occ_nums} we show the occupation numbers, which correspond to the eigenvalues of the 1-RDMs, obtained from the quantum hardware experiments.

\subsection{One-electron reduced density matrix}
\label{sec:app:1_rdm}

\begin{figure}[t!]
    \centering
   \includegraphics[width=0.8\textwidth]{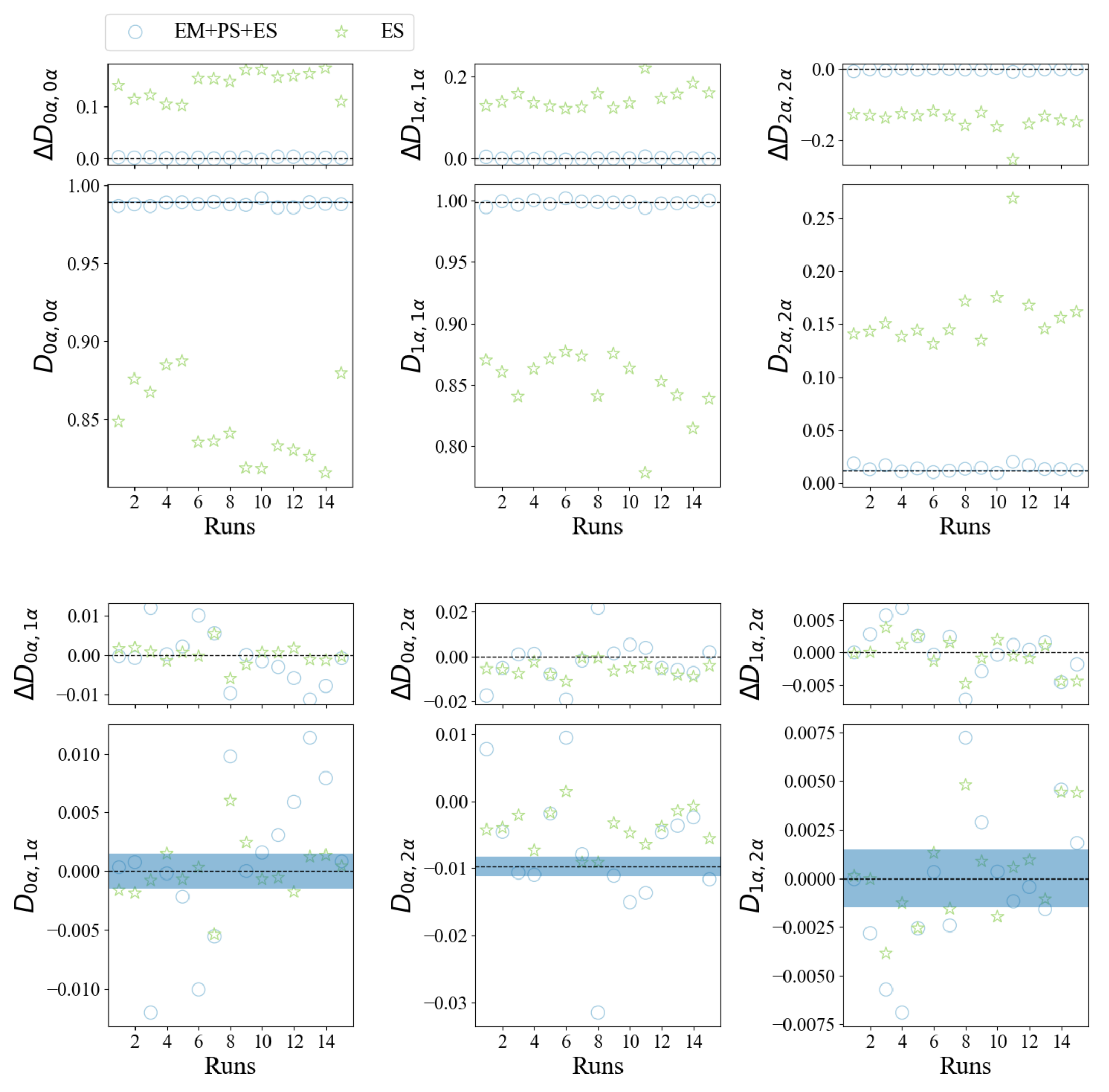}
    \caption{$\alpha$-spin one-electron reduced density matrix (1-RDM) elements obtained from the quantum hardware experiments  (IBM Torino device) for  the hydroxyl radical (OH$^{\bullet}$) using the circuit in Figure \ref{fig:app:OH_circuit}. The tables in the main text (Tables \ref{tbl:overview_processing} and \ref{tbl:overview_QPU_time}) show an overview of the various tools used in the hardware experiments and the number of measurements and QPU times.  The circles (EM+PS+ES) include error mitigation ($\boldsymbol{M}_{U_0}$), post-selection, and error suppression, and the stars (ES) includes only error suppression. The horizontal dashed line shows the exact (UCASSCF) value of the 1-RDM elements. The blue area centered around the exact value shows the standard deviation for shot-noise only.   }
    \label{fig:app:OH_rdms_alpha}
\end{figure}

\begin{figure}[t!]
    \centering
   \includegraphics[width=0.8\textwidth]{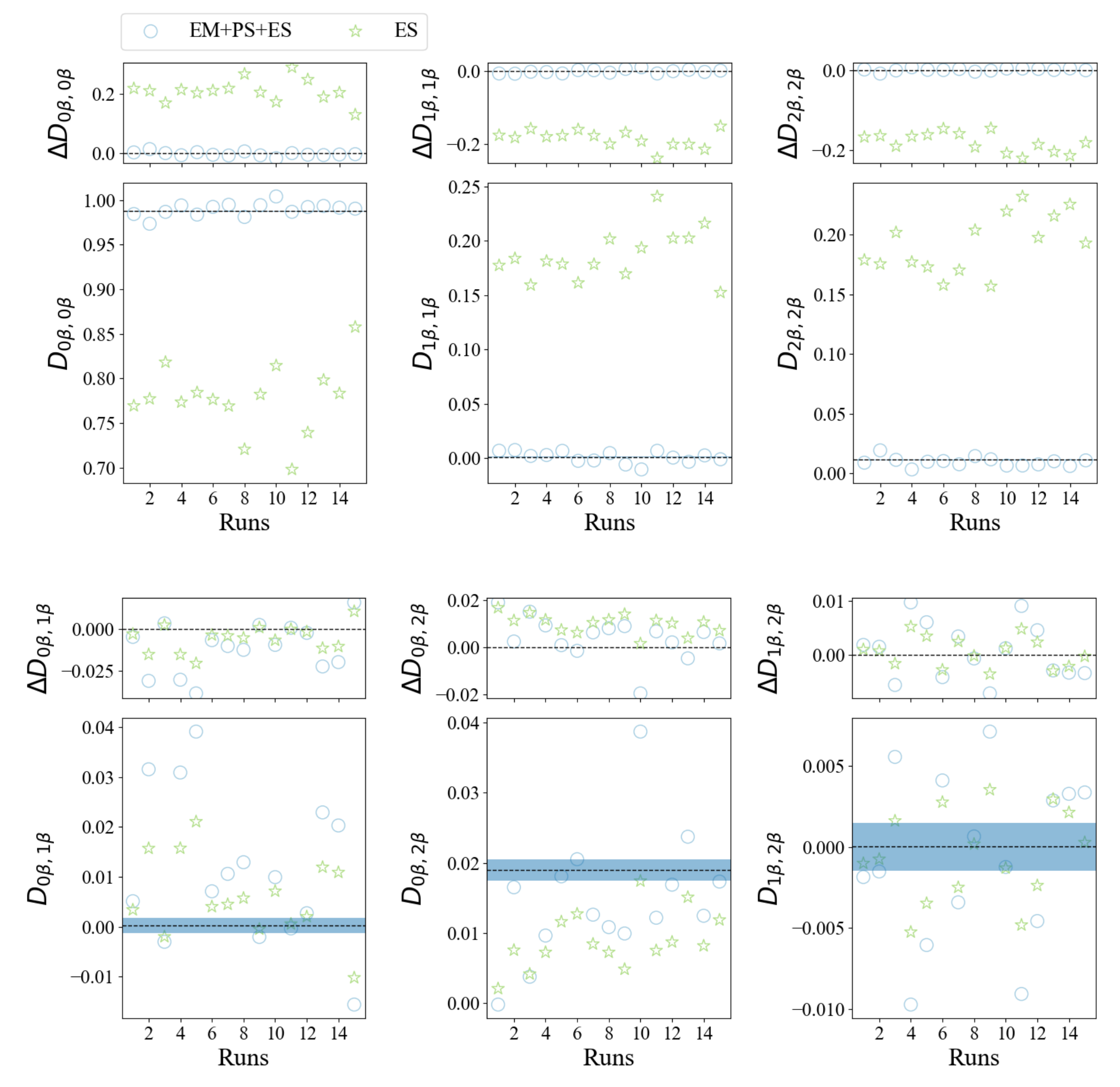}
    \caption{$\beta$-spin one-electron reduced density matrix (1-RDM) elements obtained from the quantum hardware experiments  (IBM Torino device) for  the hydroxyl radical (OH$^{\bullet}$) using the circuit in Figure \ref{fig:app:OH_circuit}.   The tables in the main text (Tables \ref{tbl:overview_processing} and \ref{tbl:overview_QPU_time}) show an overview of the various tools used in the hardware experiments and the number of measurements and QPU times.  The circles (EM+PS+ES) include error mitigation ($\boldsymbol{M}_{U_0}$), post-selection, and error suppression, and the stars (ES) includes only error suppression. The horizontal dashed line shows the exact (UCASSCF) value of the 1-RDM elements. The blue area centered around the exact value shows the standard deviation for shot-noise only. }
    \label{fig:app:OH_rdms_beta}
\end{figure}

\begin{figure}[H]
    \centering
   \includegraphics[width=0.8\textwidth]{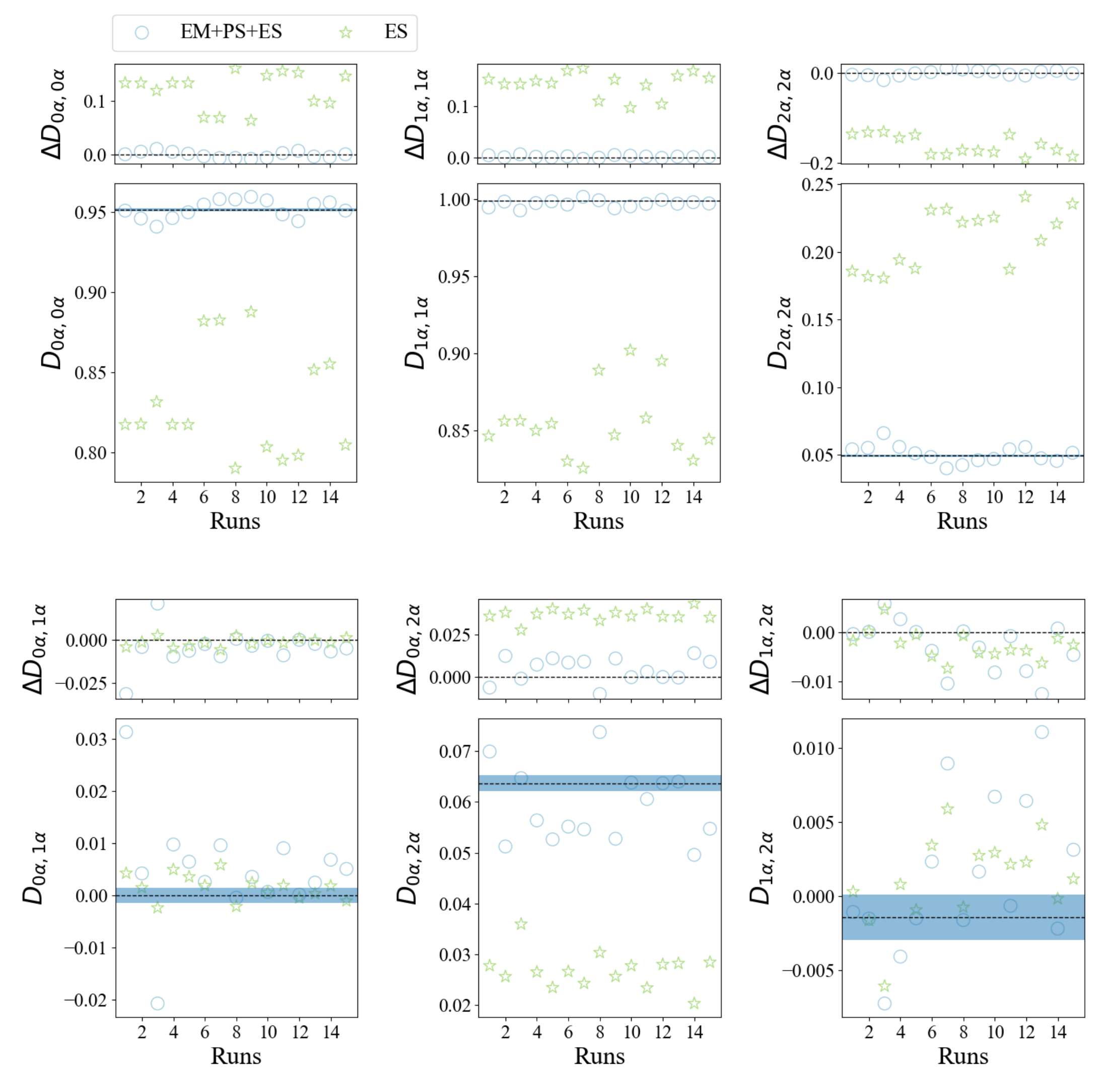}
    \caption{$\alpha$-spin one-electron reduced density matrix (1-RDM) elements obtained from the quantum hardware experiments  (IBM Torino device) for  the  nitric oxide (NO$^{\bullet}$) molecule using the circuit in Figure \ref{fig:app:NO_circuit}.   The tables in the main text (Tables \ref{tbl:overview_processing} and \ref{tbl:overview_QPU_time}) show an overview of the various tools used in the hardware experiments and the number of measurements and QPU times.  The circles (EM+PS+ES) include error mitigation ($\boldsymbol{M}_{U_0}$), post-selection, and error suppression, and the stars (ES) includes only error suppression. The horizontal dashed line shows the exact (UCASSCF) value of the 1-RDM elements. The blue area centered around the exact value shows the standard deviation for shot-noise only. }
    \label{fig:app:NO_rdms_alpha}
\end{figure}

\begin{figure}[H]
    \centering
   \includegraphics[width=0.8\textwidth]{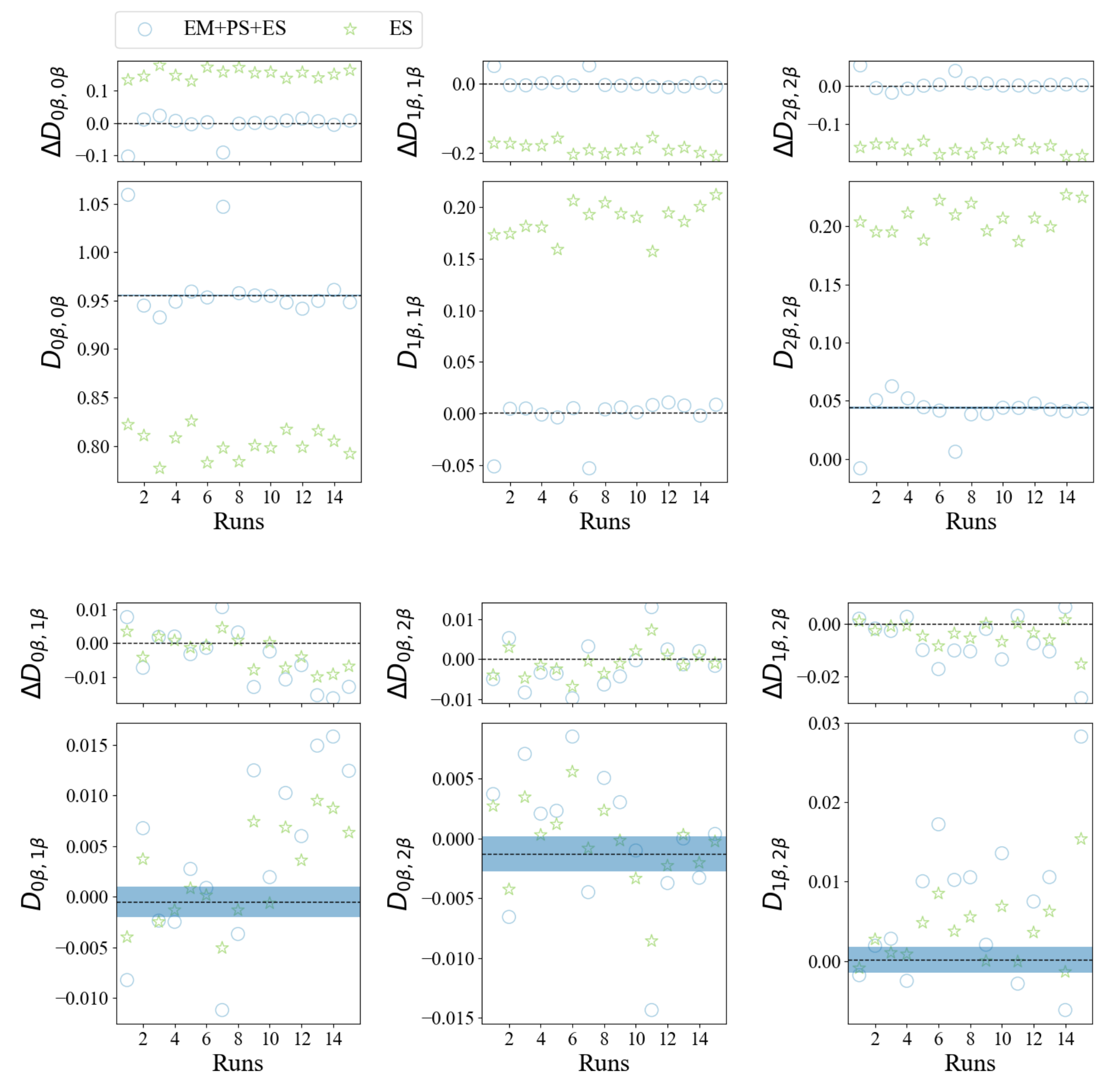}
    \caption{$\beta$-spin  one-electron reduced density matrix (1-RDM) elements obtained from the quantum hardware experiments  (IBM Torino device) for  the  nitric oxide (NO$^{\bullet}$) molecule using the circuit in Figure \ref{fig:app:NO_circuit}.   The tables in the main text (Tables \ref{tbl:overview_processing} and \ref{tbl:overview_QPU_time}) show an overview of the various tools used in the hardware experiments and the number of measurements and QPU times.  The circles (EM+PS+ES) include error mitigation ($\boldsymbol{M}_{U_0}$), post-selection, and error suppression, and the stars (ES) includes only error suppression. The horizontal dashed line shows the exact (UCASSCF) value of the 1-RDM elements. The blue area centered around the exact value shows the standard deviation for shot-noise only.  }
    \label{fig:app:NO_rdms_beta}
\end{figure}

\begin{figure}[H]
    \centering
   \includegraphics[width=0.8\textwidth]{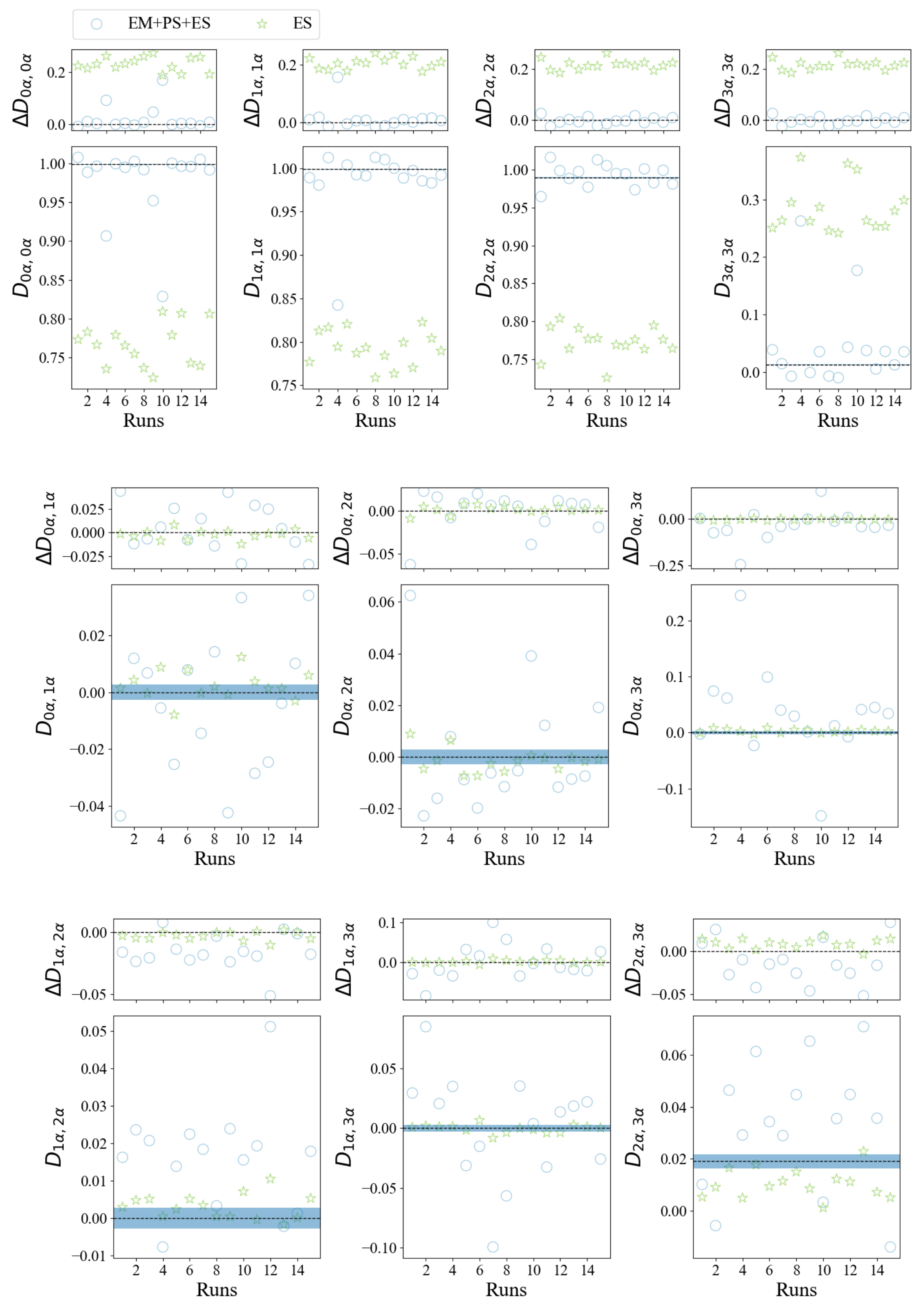}
    \caption{$\alpha$-spin one-electron reduced density matrix (1-RDM) diagonal elements obtained from the quantum hardware experiments  (IBM Torino device) for  the  hydroxyl cation (OH$^+$) molecule using the circuit in Figure \ref{fig:app:OH+_circuit}.   The tables in the main text (Tables \ref{tbl:overview_processing} and \ref{tbl:overview_QPU_time}) show an overview of the various tools used in the hardware experiments and the number of measurements and QPU times.  The circles (EM+PS+ES) include error mitigation ($\boldsymbol{M}_{U_0}$), post-selection, and error suppression, and the stars (ES) includes only error suppression. The horizontal dashed line shows the exact (UCASSCF) value of the 1-RDM elements. The blue area centered around the exact value shows the standard deviation for shot-noise only. }
    \label{fig:app:OH+_rdms_alpha}
\end{figure}

\begin{figure}[H]
    \centering
   \includegraphics[width=0.8\textwidth]{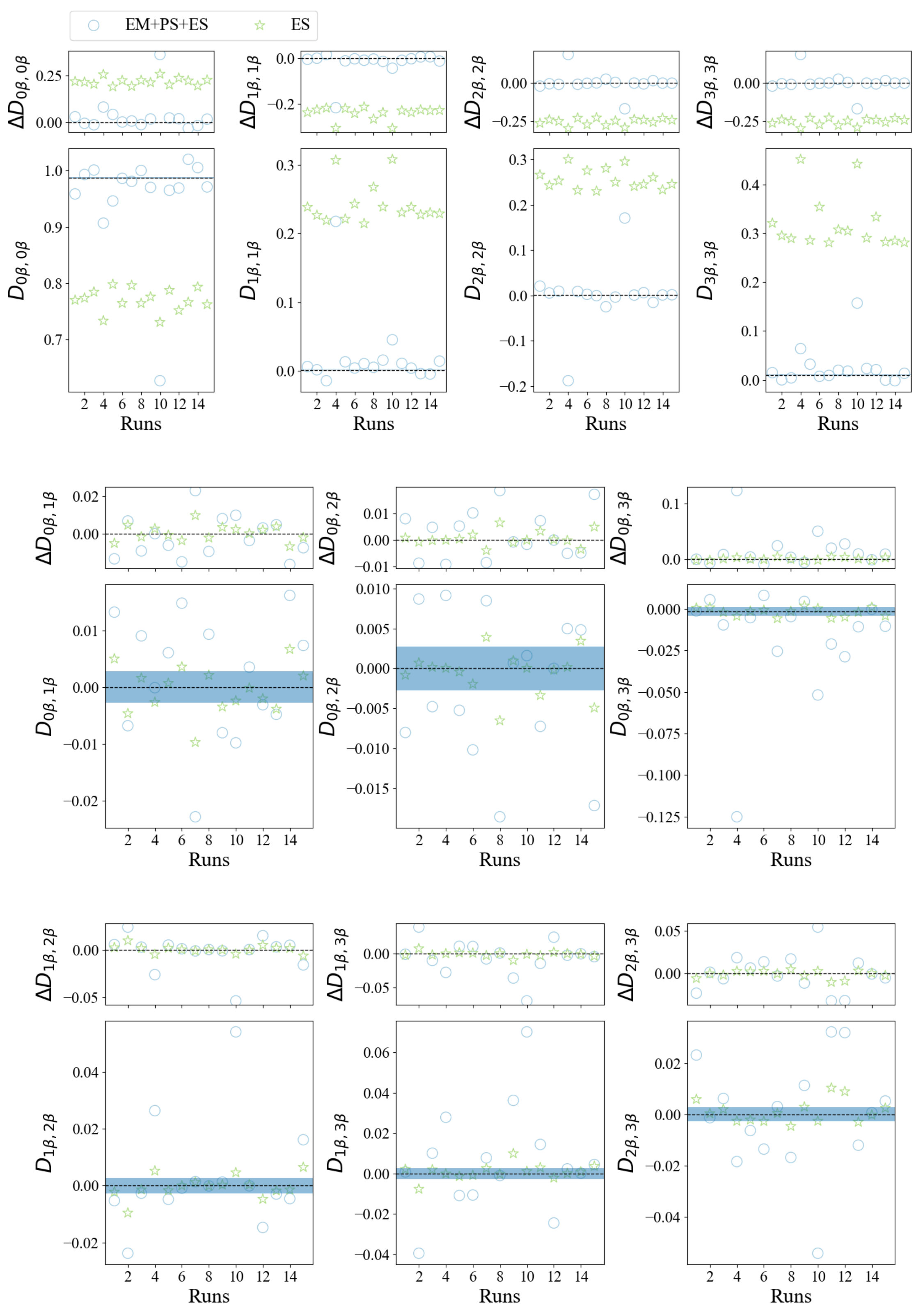}
    \caption{$\beta$-spin one-electron reduced density matrix (1-RDM) off-diagonal elements obtained from the quantum hardware experiments  (IBM Torino device) for  the  hydroxyl cation (OH$^+$) molecule using the circuit in Figure \ref{fig:app:OH+_circuit}.   The tables in the main text (Tables \ref{tbl:overview_processing} and \ref{tbl:overview_QPU_time}) show an overview of the various tools used in the hardware experiments and the number of measurements and QPU times.  The circles (EM+PS+ES) include error mitigation ($\boldsymbol{M}_{U_0}$), post-selection, and error suppression, and the stars (ES) includes only error suppression. The horizontal dashed line shows the exact (UCASSCF) value of the 1-RDM elements. The blue area centered around the exact value shows the standard deviation for shot-noise only. }
    \label{fig:app:OH+_rdms_beta}
\end{figure}

Figures \ref{fig:app:OH_rdms_alpha} and  \ref{fig:app:OH_rdms_beta} show the 1-RDM elements obtained from the quantum hardware experiments for OH$^{\bullet}$. Figures  \ref{fig:app:NO_rdms_alpha} and  \ref{fig:app:NO_rdms_beta} show the quantum hardware experiments for NO$^{\bullet}$. Figures \ref{fig:app:OH+_rdms_alpha} and  \ref{fig:app:OH+_rdms_beta}  show the quantum hardware experiments for  OH$^+$.  We use their experimental bond lengths of 0.9697 $\text{\r{A}}$, 1.1508 $\text{\r{A}}$, and 1.0289 $\text{\r{A}}$ for OH$^{\bullet}$, NO$^{\bullet}$, and OH$^+$, respectively. The 6-311++G**-J basis set~\cite{kjaer_pople_2011}, which is  optimized for ESR properties, was adopted in all calculations.

\subsection{Occupation number}
\label{sec:app:occ_nums}

\begin{figure}[H]
    \centering
   \includegraphics[width=1.0\textwidth]{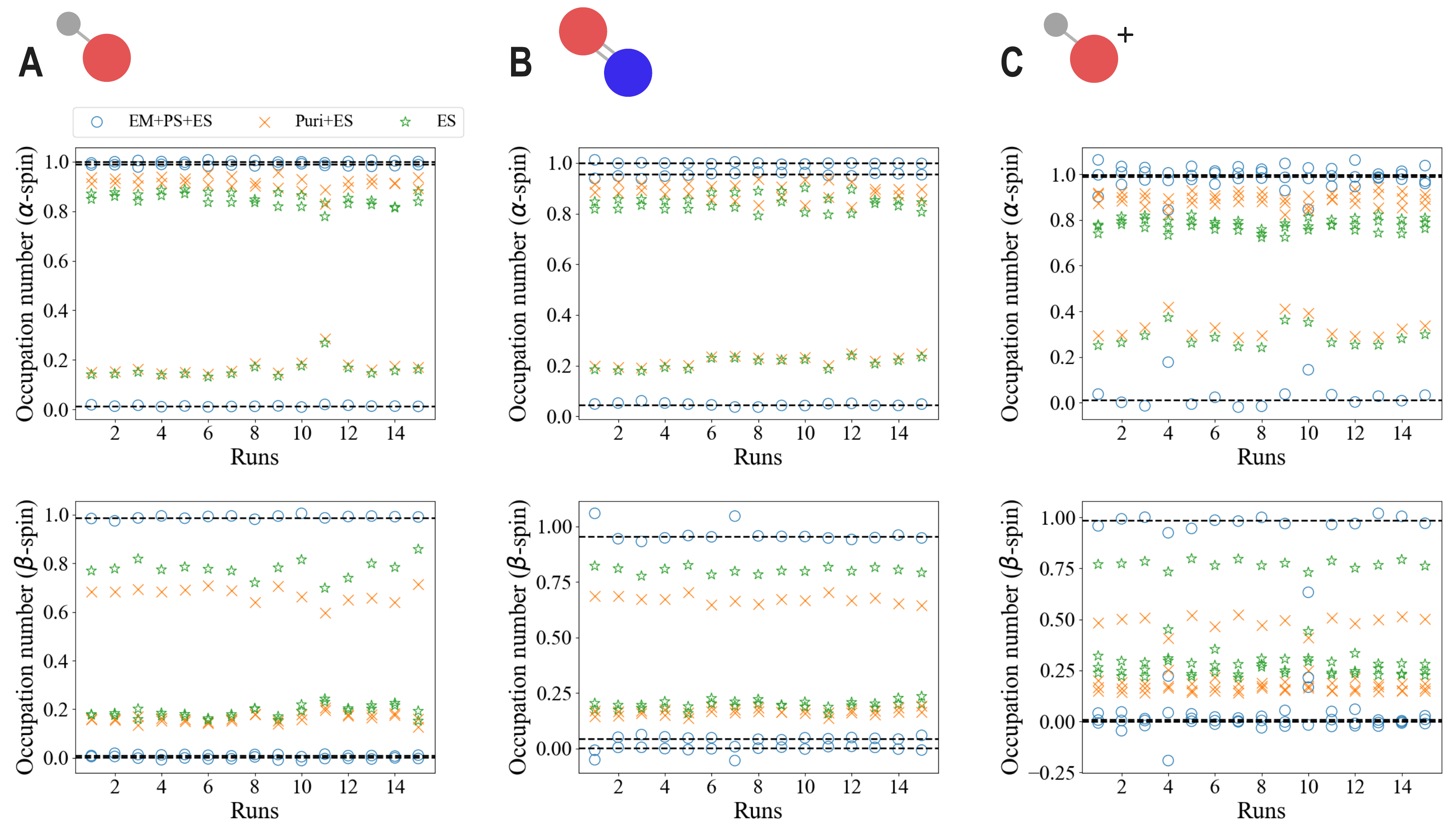}
    \caption{ Occupation numbers obtained from quantum hardware experiments for 15 independent runs on the IBM Torino device for OH$^{\bullet}$ (A),  NO$^{\bullet}$ (B), and OH$^{+}$ (C)  using the circuits shown in Figure \ref{fig:main_circuits}. An overview of the tools used in hardware experiments is provided in Table \ref{tbl:overview_processing}, while Table \ref{tbl:overview_QPU_time} summarizes the number of measurements and QPU times. The circles (EM+PS+ES) include error mitigation ($\boldsymbol{M}_{U_0}$), post-selection, and error suppression, the crosses (Puri+ES) include the 1-RDM purification method and error suppression, and the stars (ES) includes only error suppression. The horizontal dashed lines correspond to the exact (UCASSCF) occupation numbers.   }
    \label{fig:app:occ_nums}
\end{figure}